\documentclass[aip,reprint,jcp,floatfix]{revtex4-1}

\usepackage[colorlinks=true,linkcolor=blue,citecolor=blue,urlcolor=blue]{hyperref}
\usepackage{setspace} 
\usepackage{graphicx}
\usepackage{amsmath}
\usepackage{color}
\usepackage{amsmath}
\usepackage{amssymb}
\usepackage{verbatim}
\usepackage{latexsym}
\usepackage{enumerate} 
\usepackage{bm} 
\usepackage{chemfig} 

\begin{document}

\title{Temperature effects in first-principles solid state calculations of the chemical shielding tensor made simple}

\author{Bartomeu Monserrat}

\email{bm418@cam.ac.uk}

\affiliation{TCM Group, Cavendish Laboratory, University of Cambridge,
  J.\ J.\ Thomson Avenue, Cambridge CB3 0HE, United Kingdom}

\author{Richard J.\ Needs}

\affiliation{TCM Group, Cavendish Laboratory, University of Cambridge,
  J.\ J.\ Thomson Avenue, Cambridge CB3 0HE, United Kingdom}

\author{Chris J.\ Pickard}

\affiliation{Department of Physics and Astronomy, University College London, Gower Street, London WC1E 6BT, United Kingdom}

\date{\today}

\begin{abstract}
We study the effects of atomic vibrations on the solid-state chemical shielding tensor using first principles density functional theory calculations. At the harmonic level, we use a Monte Carlo method and a perturbative expansion. The Monte Carlo method is accurate but computationally expensive, while the perturbative method is computationally more efficient, but approximate. We find excellent agreement between the two methods for both the isotropic shift and the shielding anisotropy. The effects of zero-point quantum mechanical nuclear motion are important up to relatively high temperatures: at $500$~K they still represent about half of the overall vibrational contribution. We also investigate the effects of anharmonic vibrations, finding that their contribution to the zero-point correction to the chemical shielding tensor is small. We exemplify these ideas using magnesium oxide and the molecular crystals L-alanine and $\beta$-aspartyl-L-alanine. We therefore propose as the method of choice to incorporate the effects of temperature in solid state chemical shielding tensor calculations the perturbative expansion within the harmonic approximation. This approach is accurate and requires a computational effort that is about an order of magnitude smaller than that of dynamical or Monte Carlo approaches, so these effects might be routinely accounted for.
\end{abstract}


\maketitle

\section{Introduction}

Nuclear magnetic resonance (NMR) is a central technique in the study of the structure and dynamics of solid-state systems,\cite{exp_nmr_review,exp_nmr_review2,gipaw_review} providing direct access to the microscopic local environment of solids. Interestingly, NMR can be used to study systems which lack periodicity and it can probe light elements such as hydrogen, both of which are challenging to study with other structure determination techniques such as X-ray diffraction.\cite{exp_nmr_review_disorder}

Experimental NMR spectra cannot usually be related to the underlying structure and dynamics in a straightforward manner. This difficulty can be addressed by theoretically predicting NMR spectra that may then be used in conjunction with experimental results to study the structure and dynamics of the systems of interest. For solid-state NMR, the use of first-principles calculations based on pseudopotential plane-wave density functional theory (DFT)\cite{PhysRev.136.B864,PhysRev.140.A1133,dft_rev_mod_phys} has proved to be a very powerful approach. These calculations are typically based on the gauge including projector augmented waves (GIPAW) theory of Pickard and Mauri,\cite{gipaw,gipaw_ultrasoft,gipaw_review} that permits the reconstruction of the electronic wave function near atomic nuclei when pseudopotentials are used, and has been successfully applied to the study of organic systems,\cite{gipaw_example_organic} inorganic systems,\cite{gipaw_example_inorganic} glasses,\cite{gipaw_example_glass} polymers,\cite{gipaw_example_polymer1,gipaw_example_polymer2} and surfaces.\cite{gipaw_example_surface} However, first-principles calculations are typically performed in the static lattice approximation, in which the zero-point (ZP) and thermal motions of atomic nuclei are ignored due to the high computational cost of including them. The uncertainty associated with the static lattice approximation may be quite severe for light elements and, considering their importance in NMR, it is evident that a proper treatment of nuclear motion is desirable. 

The quantum-mechanical study of NMR parameters including nuclear motion has been extensively explored in the gas phase using a variety of methods.\cite{perturbative_shift,schulte_pimd_nmr_2,schulte_pimd_nmr_1,perturbative_jj,perturbative_shift_jj} In the solid phase there exists a smaller number of first-principles NMR studies that take into account nuclear motion.\cite{pickard_original_nmr_vibrational,pickard_nmr_vibrations_organic,md_quadrupolar_coupling,md_nmr_lee,nmr_force_fields,long_time_md,dracinsky_md_nmr,dracinsky_pimd_nmr,dracinsky_md_quadratic_nmr} In these studies, the vibrational phase space of the solid is explored either by Monte Carlo sampling of the harmonic vibrational wave function,\cite{pickard_original_nmr_vibrational,pickard_nmr_vibrations_organic} molecular dynamics,\cite{md_quadrupolar_coupling,md_nmr_lee,pickard_nmr_vibrations_organic,nmr_force_fields,long_time_md,dracinsky_md_nmr} path-integral molecular dynamics,\cite{dracinsky_pimd_nmr} or perturbation theory.\cite{dracinsky_md_quadratic_nmr} These approaches incorporate the effects of nuclear motion in solid state NMR calculations, but their computational expense prohibits their use in many applications.

In this work, we investigate the accuracy and computational expense of a method for NMR calculations for solids that treats the vibrations within the harmonic approximation and the coupling between vibrations and the chemical shielding tensor perturbatively. We expand the chemical shielding tensor about the static lattice value in terms of vibrational normal mode amplitudes. Truncating this expansion at second order leads to a formalism that requires the determination of the chemical shielding tensor at a number of points in the vibrational phase space that scales linearly with the system size. We test the accuracy and efficiency of the scheme using the simple crystal magnesium oxide (MgO) and the molecular crystals L-alanine and $\beta$-aspartyl-L-alanine (bDA), chosen because previous calculations including the effects of vibrations exist for each of them.

We also investigate the effects of anharmonic vibrations in L-alanine using a recently proposed method for solid-state anharmonic calculations.\cite{PhysRevB.87.144302,helium,prl_dissociation_hydrogen} We find that anharmonic corrections are small even for the hydrogen atoms.

Our main conclusion is that the perturbative approach within the harmonic approximation accurately describes the changes in the chemical shielding tensor induced by atomic motion. As this approach is about an order of magnitude computationally cheaper than Monte Carlo methods for the system sizes of interest in NMR, we believe it should be the method of choice for systematically incorporating temperature effects in first principles NMR calculations.

The rest of this paper is arranged as follows. In Sec.~\ref{sec:formalism} we describe the theoretical formalism and computational implementation, and in Sec.~\ref{sec:comp} we detail the computational parameters used in our calculations. In Sec.~\ref{sec:harmonic} we present our results obtained by treating vibrations at the harmonic level, with particular attention to the accuracy and computational expense of the harmonic method. In Sec.~\ref{sec:anharmonic} we describe the effects of including anharmonic terms for the coupling of vibrations to the chemical shielding tensor. Finally, in Sec.~\ref{sec:conclusions} we summarize our findings and draw our conclusions.

\section{Formalism} \label{sec:formalism}

\subsection{Theoretical formalism}

The expectation value of the chemical shielding tensor $\bm{\sigma}$ with respect to the vibrations of the atoms in a solid at temperature $T$ can be written as
\begin{equation}
\langle\bm{\sigma}\rangle = \frac{1}{\mathcal{Z}}\sum_{\mathbf{S}}\langle\Phi^{\mathbf{S}}(\mathbf{q})|\bm{\sigma}(\mathbf{q})|\Phi^{\mathbf{S}}(\mathbf{q})\rangle e^{-E_{\mathbf{S}}/k_{\mathrm{B}}T}. \label{eq:exp_value}
\end{equation}
In this expression, the vibrational state $\mathbf{S}$ has wave function $|\Phi^{\mathbf{S}}\rangle$ and energy $E_{\mathbf{S}}$, $\mathcal{Z}=\sum_{\mathbf{S}}e^{-E_{\mathbf{S}}/k_{\mathrm{B}}T}$ is the partition function, and $k_{\mathrm{B}}$ is Boltzmann's constant. The collective coordinate $\mathbf{q}$ is a $3N$-dimensional vector containing the vibrational amplitudes of the $N$ atoms in the solid. 

The expectation value in Eq.~(\ref{eq:exp_value}) has been evaluated in the past by sampling the vibrational density using Monte Carlo techniques,\cite{pickard_original_nmr_vibrational,pickard_nmr_vibrations_organic} molecular dynamics,\cite{md_quadrupolar_coupling,md_nmr_lee,pickard_nmr_vibrations_organic,nmr_force_fields,long_time_md,dracinsky_md_nmr} path-integral molecular dynamics,\cite{dracinsky_pimd_nmr} or perturbation theory.\cite{dracinsky_md_quadratic_nmr} Molecular dynamics samples phase space classically, and therefore neglects the effects of zero-point motion. Monte Carlo sampling provides information about the zero-point motion, which can be important for light elements and at low temperatures, but it is limited by the approximations used to describe $|\Phi^{\mathbf{S}}\rangle$, usually the harmonic approximation. Path integral methods sample phase space including the quantum nature of nuclear vibrations, and also terms beyond the harmonic approximation. Each of these approaches suffers from a significant computational cost due to the large number of sampling points required to reduce the statistical uncertainty. In the cases of molecular dynamics and path-integral molecular dynamics, this cost must be added to the intrinsic cost of the phase space sampling. Perturbation theory\cite{dracinsky_md_quadratic_nmr} has been used including terms up to fourth order in the expansion of the Born-Oppenheimer potential, and up to second order in the expansion of the chemical shielding tensor about its equilibrium value. The inclusion of high-order terms also makes perturbative expansions computationally intensive. 

In this work we evaluate the expectation value in Eq.~(\ref{eq:exp_value}) by approximating the coupling of the chemical shielding tensor to the vibrational state in an expansion in terms of the amplitudes of the vibrational modes:
\begin{equation}
\bm{\sigma}(\mathbf{q})=\bm{\sigma}(\mathbf{0})+\sum_{n,\mathbf{k}}\bm{c}_{n\mathbf{k}}^{(1)}q_{n\mathbf{k}} +\!\!\!\! \sum_{n,\mathbf{k},n'\mathbf{k}'}\!\!\bm{c}_{n\mathbf{k};n'\mathbf{k}'}^{(2)}q_{n\mathbf{k}}q_{n'\mathbf{k}'}, \label{eq:expansion}
\end{equation}
where we have retained terms up to second order. In Eq.~(\ref{eq:expansion}), $(n,\mathbf{k})$ denotes the branch $n$ and wave vector $\mathbf{k}$ of a vibrational normal mode, $q_{n\mathbf{k}}$ is the normal mode vibrational amplitude, and $\textbf{c}_{n\mathbf{k}}$ is the tensor coupling of vibrational mode $(n,\mathbf{k})$ to the chemical shielding tensor. Within the harmonic approximation, the vibrational wave function is even, and therefore the only non-zero terms in the expectation value of Eq.~(\ref{eq:exp_value}) when using the expression in Eq.~(\ref{eq:expansion}) are the quadratic diagonal terms $\textbf{c}_{n\mathbf{k};n\mathbf{k}}^{(2)}$, which from now on we simply write as $\textbf{c}_{n\mathbf{k}}$. Therefore, without loss of generality, we may write 
\begin{equation}
\bm{\sigma}(\mathbf{q})=\sum_{n,\mathbf{k}}\textbf{c}_{n\mathbf{k}}q_{n\mathbf{k}}^2. \label{eq:quadratic}
\end{equation}
The use of Eq.~(\ref{eq:quadratic}) to evaluate the expectation value in Eq.~(\ref{eq:exp_value}) within the harmonic approximation leads to a simple expression for the temperature dependence of the chemical shielding tensor:
\begin{equation}
\langle\bm{\sigma}\rangle_{\mathrm{Q}}=\sum_{n,\mathbf{k}}\frac{\mathbf{c}_{n\mathbf{k}}}{2\omega_{n\mathbf{k}}}[1+2n_{\mathrm{B}}(\omega_{n\mathbf{k}})], \label{eq:tdep}
\end{equation}
where $n_{\mathrm{B}}(\omega)=(e^{\omega/k_{\mathrm{B}}T}-1)^{-1}$ is a Bose-Einstein factor, and Q makes reference to \textit{quadratic}.

This method is computationally more efficient than other approaches because it typically requires a smaller number of individual NMR calculations to be performed. However, the gain in computational cost comes at the expense of accuracy, as the expression in Eq.~(\ref{eq:quadratic}) assumes that the coupling of vibrational mode $(n,\mathbf{k})$ to the chemical shielding tensor is independent of all other vibrational modes, and also assumes that the dependence of the coupling on the vibrational amplitude is quadratic. Below we provide numerical evidence to support the validity of this approximation when vibrations are treated at the harmonic level. Finally, we note that apart from the gain in computational efficiency, the expression in Eq.~(\ref{eq:quadratic}) is also attractive because it provides microscopic information about the vibrational coupling to the chemical shielding tensor, which is obscured in sampling methods.

\subsection{Computational implementation}

\subsubsection{Quadratic approximation}

Within the quadratic approximation, the coupling constants $c_{n\mathbf{k}}^{ij}$ of the tensor component labelled by Cartesian coordinates $(i,j)$ can be evaluated by a finite displacement method. We displace the atoms along the harmonic vibrational modes with an amplitude $\Delta q_{n\mathbf{k}}$ of order $\sqrt{\langle q_{n\mathbf{k}}^2\rangle}=1/\sqrt{2\omega_{n\mathbf{k}}}$, and average over positive and negative displacements (see the discussion in Sec.~\ref{subsubsec:coupling_convergence} below for details of the choice of normal mode amplitude). The coupling constants $c_{n\mathbf{k}}^{ij}$ are then obtained from a quadratic fit to the finite displacement data, which can be formulated as
\begin{equation}
c^{ij}_{n\mathbf{k}}\!=\!\frac{\sigma^{ij}(0,\ldots,\Delta q_{n\mathbf{k}},\ldots,0)\!+\!\sigma^{ij}(0,\ldots,-\Delta q_{n\mathbf{k}},\ldots,0)}{2\Delta q_{n\mathbf{k}}^2}.
\end{equation}


The largest computational expense in the methodology described above arises in the first principles calculation of the finite-displacement chemical shielding tensor at each phonon branch $n$ and Brillouin zone (BZ) point $\mathbf{k}$. In order to reduce the number of calculations required, we may use symmetry to restrict the calculation of the finite-displacement chemical shielding tensor to the irreducible wedge of the BZ.
The chemical shielding tensor transforms according to
\begin{equation}
\bm{\sigma}(q_{n\mathcal{R}\mathbf{k}})=\mathcal{R}^{-1}\bm{\sigma}(q_{n\mathbf{k}})\mathcal{R},
\end{equation}
where $\mathcal{R}$ denotes the rotations of the point group of the crystal.
The use of symmetry is advantageous when considering the convergence of the renormalised chemical shielding tensor as a function of the size of the simulation cell. We note that convergence with respect to the simulation cell size, which is equivalent to convergence with respect to the phonon BZ sampling density, is also required to explore phase space when using Monte Carlo sampling or molecular dynamics methods. 




The quadratic approximation described above is computationally efficient because a small number of data points is needed to sample the relevant vibrational phase space. A further benefit in terms of computational efficiency compared with molecular dynamics and path-integral molecular dynamics methods is that the sampling points in the quadratic approximation are independent of each other (this is also true in the Monte Carlo sampling approach). This means that the sampling calculations can be simply parallelized over a large number of processors, and the results can be obtained very rapidly in real time, naturally fitting with the expanding field of high throughput computation. This should be contrasted with the need to follow a correlated path in the dynamics methods, which means that they cannot make similar usage of modern computational architectures.

\subsubsection{Monte Carlo sampling}

In this work we also use a Monte Carlo sampling scheme in order to validate the approximations involved in the quadratic expansion and to include the effects of anharmonic vibrations. The expectation value of the chemical shielding tensor with respect to the harmonic density is
\begin{equation}
\langle\bm\sigma\rangle = \int d\mathbf{q}|\Phi_{\mathrm{har}}(\mathbf{q})|^2\bm{\sigma}(\mathbf{q}),
\end{equation}
which may be evaluated using Monte Carlo sampling as 
\begin{equation}
\langle\bm{\sigma}\rangle_{\mathrm{MC}}\simeq\frac{1}{M}\sum_{i=1}^{M}\bm{\sigma}(\mathbf{q}_i),
\end{equation}
where $M$ is the number of random sampling points, distributed according to the density $|\Phi_{\mathrm{har}}|^2$, and MC stands for \textit{Monte Carlo}. The statistical uncertainty may be estimated as
\begin{equation}
\Delta\langle\bm\sigma\rangle_{\mathrm{MC}}\!\simeq\!\left[\!\frac{1}{M\!(M-1)}\!\sum_{i=1}^M\!\left(\!\bm{\sigma}(\mathbf{q}_i)\!-\!\frac{1}{M}\!\sum_{j=1}^M{\sigma}(\mathbf{q}_j)\!\right)^2\right]^{1/2}\!\!.
\end{equation}

Within the harmonic approximation, the density $|\Phi_{\mathrm{har}}|^2$ is a product of Gaussian functions $(2\pi s^2)^{-1/2}\exp(-\frac{q^2}{2s^2})$ whose widths $s$ depend on the temperature according to
\begin{equation}
s^2(T)=\frac{1}{2\omega}\coth\left(\frac{\omega}{2k_{\mathrm{B}}T}\right). 
\end{equation}
The random sampling of a Gaussian distribution is a computationally efficient process, making this approach more appealing than the alternative dynamical approaches.

\section{Computational details} \label{sec:comp}

In this section we describe the computational details of our first-principles calculations. All our calculations are performed within plane-wave pseudopotential DFT\cite*{PhysRev.136.B864,PhysRev.140.A1133,dft_rev_mod_phys} as implemented in the {\sc castep} package.\cite{CASTEP} We have used the generalized gradient approximation of Perdew-Burke-Ernzerhof to approximate the exchange-correlation functional,\cite{PhysRevLett.77.3865} and we have described the electron-ion interaction by means of ultrasoft ``on the fly'' pseudopotentials.\cite{PhysRevB.41.7892} We have used a plane-wave energy cut-off of $800$~eV for MgO and $700$~eV for L-alanine and bDA, and a Monkhorst-Pack\cite{PhysRevB.13.5188} BZ sampling grid of density $2\pi\times0.04$ \AA$^{-1}$ for the three systems. These parameters lead to differences between frozen-phonon structures below $10^{-4}$~eV/atom for the energy, $10^{-3}$~eV/\AA\@ for the forces on atoms, $10^{-2}$~GPa for the stress on MgO, and below $10^{-2}$~ppm for the components of the chemical shielding tensor.

\subsection{Structures}

MgO has a cubic crystal structure of space group $Fm\overline{3}m$ with two atoms in the primitive cell. We have relaxed the volume of the primitive cell of MgO within DFT, finding an equilibrium lattice constant equal to $a=4.245$~\AA, which is a little larger than the experimental value of $4.212$~\AA.

L-alanine is a molecular orthorhombic crystal of space group $P2_12_12_1$, in which each molecule has $13$ atoms (C$_3$H$_7$NO$_2$) and the primitive cell has $4$ molecules. We use the structure reported in Ref.~\onlinecite{alanine_crystal}, determined by neutron diffraction at a temperature of $60$~K. The orthorhombic structure has lattice parameters $a=5.806$~\AA, $b=5.940$~\AA, and $c=12.274$~\AA. We have relaxed the internal atomic coordinates to reduce the forces on each atom below $10^{-3}$~eV/\AA.

bDA is an orthorhombic molecular crystal of space group $P2_12_12_1$, in which each molecule has $26$ atoms (C$_7$H$_{12}$N$_2$O$_5$), and the primitive cell has $4$ molecules. We use the structure reported in Ref.~\onlinecite{bDA_crystal}, determined by X-ray diffraction at a temperature of $120$~K. The orthorhombic structure has lattice parameters $a=4.845$~\AA, $b=9.409$~\AA, $c=19.170$~\AA. We have relaxed the internal atomic coordinates to reduce the forces on each atom below $10^{-3}$~eV/\AA.

\subsection{NMR calculations}

The first-principles calculations of the chemical shielding tensor are performed using the GIPAW theory\cite{gipaw,gipaw_ultrasoft,gipaw_review} as implemented in the {\sc castep} package.\cite{CASTEP} We investigate the isotropic chemical shift
\begin{equation}
\sigma_{\textrm{iso}}=\frac{1}{3}\mathrm{tr} \bm{\sigma}, 
\end{equation}
and the shielding anisotropy (SA)
\begin{equation}
\sigma_{\mathrm{SA}}=\sigma_{zz}-\frac{1}{2}(\sigma_{xx}+\sigma_{yy})
\end{equation}
where appropriate. In the latter, the principal components of the chemical shielding tensor are ordered according to $|\sigma_{zz}-\sigma_{\mathrm{iso}}|\geq|\sigma_{xx}-\sigma_{\mathrm{iso}}|\geq|\sigma_{yy}-\sigma_{\mathrm{iso}}|$.

\subsection{Vibrational calculations}

For the vibrational harmonic calculations, we have calculated the harmonic frequencies and eigenvectors at the $\Gamma$ point for L-alanine and bDA, and at the phonon BZ $\mathbf{k}$-points commensurate with the size of the supercell in MgO. We have used the finite-displacement method\cite{phonon_finite_displacement} in order to construct the matrix of force constants, with atomic displacements of magnitude $0.005$~\AA\@ and averaging over positive and negative displacements. We have then diagonalized the corresponding dynamical matrix to find the harmonic frequencies and eigenvectors.

For the anharmonic vibrational calculations, we have used the principal axes approximation including independent modes to map the Born-Oppenheimer energy surface beyond the harmonic approximation,\cite{PhysRevB.87.144302} including only $\Gamma$-point vibrational modes. We have solved the resulting Schr\"{o}dinger equation using an expansion of the vibrational wave function in a basis of simple harmonic oscillator eigenstates, and we have used a total of $40$ basis functions for the description of each vibrational degree of freedom.

\section{Harmonic vibrations} \label{sec:harmonic}

In this section we assess the validity of the quadratic approximation $\langle\bm\sigma\rangle_{\mathrm{Q}}$ by comparing it with Monte Carlo sampling calculations $\langle\bm\sigma\rangle_{\mathrm{MC}}$. The results in this section are within the harmonic approximation, and we use MgO, L-alanine, and bDA as test systems. We report the vibrational correction, defined as the difference between the vibrationally-averaged value and the static lattice value, $\Delta\sigma_{\mathrm{iso}}=\langle\sigma_{\mathrm{iso}}\rangle-\sigma_{\mathrm{iso}}$ for the isotropic shift and $\Delta\sigma_{\mathrm{SA}}=\langle\sigma_{\mathrm{SA}}\rangle-\sigma_{\mathrm{SA}}$ for the shielding anisotropy.

\subsection{Magnesium Oxide}

In MgO, the chemical shielding tensor is diagonal and isotropic. Therefore, we report the isotropic chemical shift only. At the static lattice level, we find $\sigma_{\textrm{iso}}^{\textrm{O}}=-197.99$~ppm and $\sigma_{\textrm{iso}}^{\textrm{Mg}}=-535.05$~ppm.
All of the calculations correspond to simulation cells containing $16$ atoms.

\subsubsection{Quadratic approximation}

\begin{figure}
\centering
\includegraphics[scale=0.335]{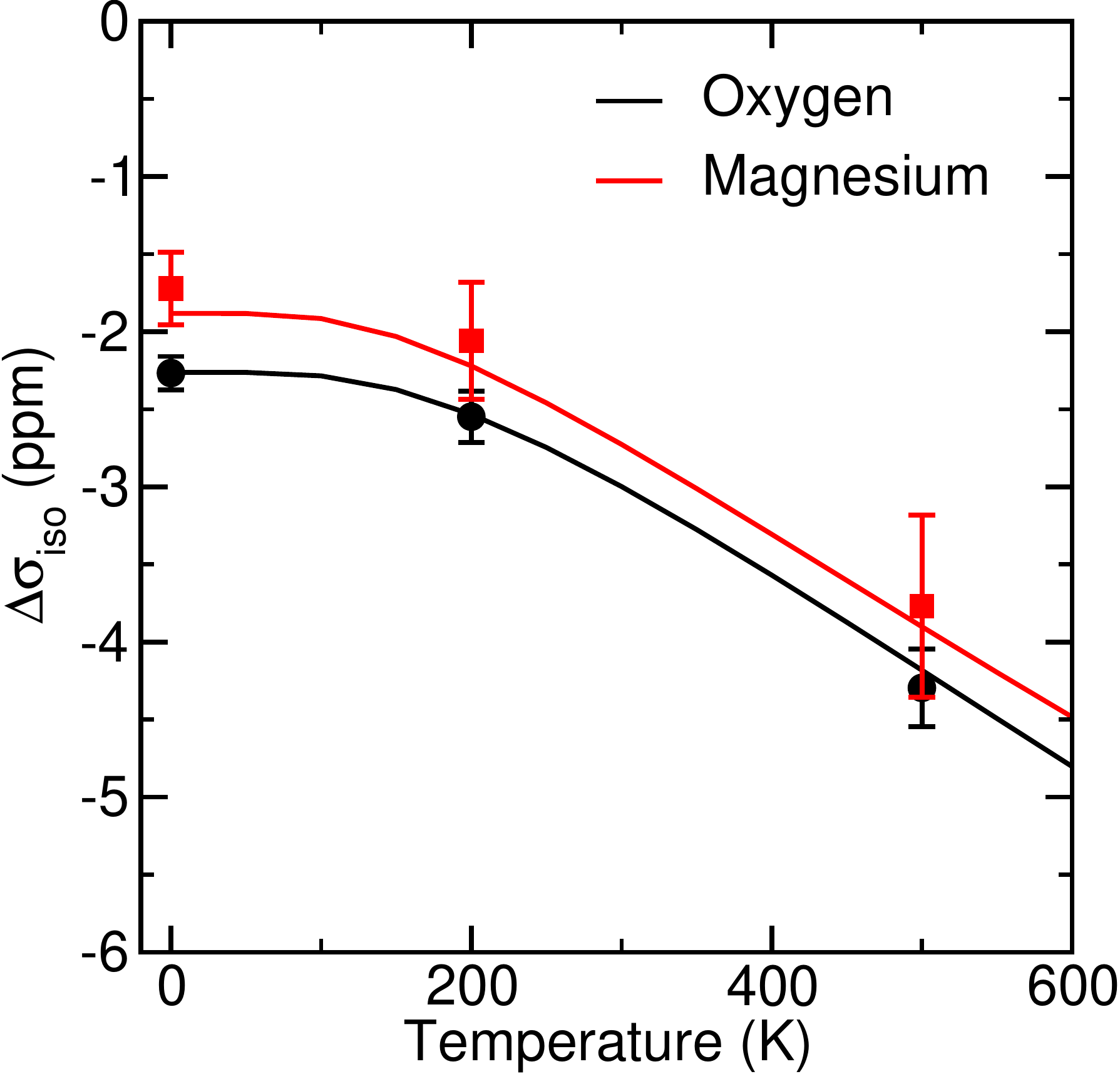}
\caption{Temperature dependence of the correction to the isotropic chemical shift from the static lattice value in MgO. The solid lines correspond to the results obtained using Eq.~(\ref{eq:quadratic}), and the circles and diamonds to the results obtained from Monte Carlo sampling. The statistical error bars give the standard deviation.}
\label{fig:tdependence}
\end{figure}

The temperature-dependent correction to the isotropic chemical shift $\Delta\sigma_{\textrm{iso}}$ evaluated by Monte Carlo sampling and by means of the quadratic expansion in Eq.~(\ref{eq:quadratic}) is given in Fig.~\ref{fig:tdependence}. The results show that, within the statistical uncertainty of the more accurate Monte Carlo sampling approach, the quadratic expansion leads to the same results, demonstrating the validity of the approximations involved in the latter.

The Monte Carlo sampling calculations reported include the sampling of $350$ points at each temperature ($0$~K, $200$~K, and $500$~K) and a total of $1,050$ points, and the statistical uncertainty is still $6\%$ of the total correction in oxygen at $T=500$~K, and $15\%$ in magnesium at the same temperature. In contrast, the quadratic expansion approach 
requires only $30$ sampling points, making the quadratic expansion approach more than an order of magnitude more efficient than the Monte Carlo sampling approach. Furthermore, these $30$ sampling points are sufficient to obtain the renormalization at any temperature by means of Eq.~(\ref{eq:tdep}).

Monte Carlo sampling is appropriate for the evaluation of high-dimensional integrals, as the number of sampling points required to obtain a given statistical uncertainty is, in principle, independent of the dimension of the integral.\footnote{This is true only if the variance of the function being sampled does not change when increasing the dimension of the integral.} Therefore, we expect that, for a given statistical uncertainty, Monte Carlo sampling will become the most efficient approach for the evaluation of Eq.~(\ref{eq:exp_value}) at some system size $N$. However, our results for MgO and the molecular crystals L-alanine and bDA presented below show that for all of these systems the quadratic approximation is significantly more efficient than Monte Carlo sampling. Therefore, it appears that, for most system sizes of interest in NMR, the quadratic approximation of Eq.~(\ref{eq:tdep}) is expected to lead to the smallest computational expense, and should therefore be the method of choice for such calculations. 

\subsubsection{Microscopic details}

Within the quadratic approximation, the couplings of the chemical shielding tensor to the vibrational state $\mathbf{c}_{n\mathbf{k}}$ are treated individually for each vibrational mode $(n,\mathbf{k})$. It is therefore possible to investigate the microscopic origin of the vibrational effects on the chemical shielding tensor by comparing the magnitudes of the different couplings.

The vibrational coupling to the chemical shielding tensor of oxygen atoms is dominated by short-wavelength vibrational modes in which an MgO pair oscillates against an adjacent MgO pair. The second largest contribution comes from vibrational modes in which oxygen atoms dominate the motion.

For magnesium atoms, the coupling to the chemical shielding is dominated by vibrational modes in which only the oxygen atoms vibrate. The second largest contributions come from vibrations of the oxygen and magnesium atoms against one another in a primitive cell.

%

\subsection{L-alanine}

\begin{figure}
\centering
\includegraphics[scale=0.85]{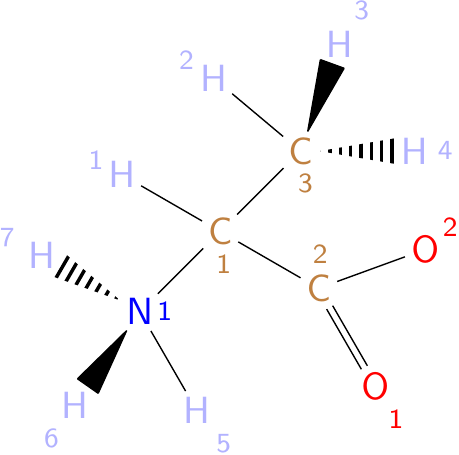}
\includegraphics[scale=0.85]{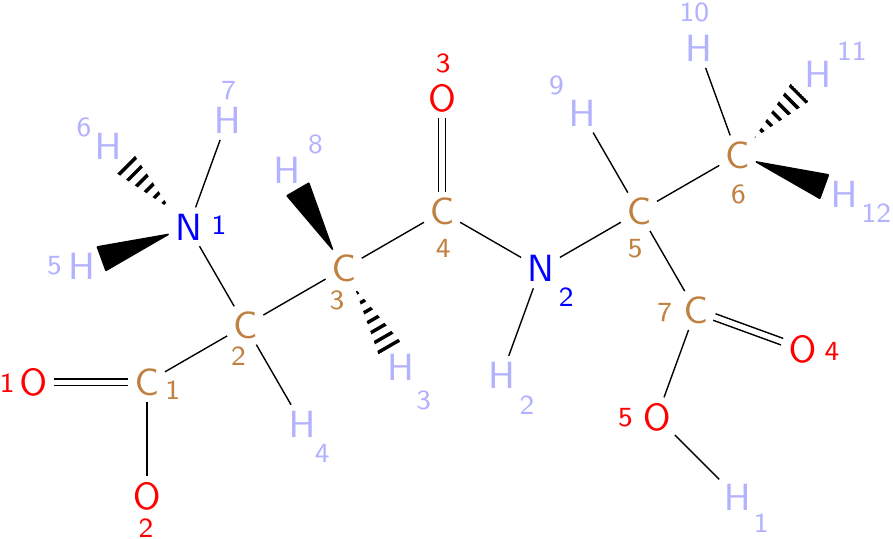}
\caption{Scheme used to label the atoms in L-alanine (top) and bDA (bottom).} 
\label{fig:bDA_structure}
\end{figure}

The scheme we use for labeling the atoms in the L-alanine molecule shown in the top diagram of Fig.~\ref{fig:bDA_structure} is used throughout this work.


\begin{figure}
\centering
\includegraphics[scale=0.35]{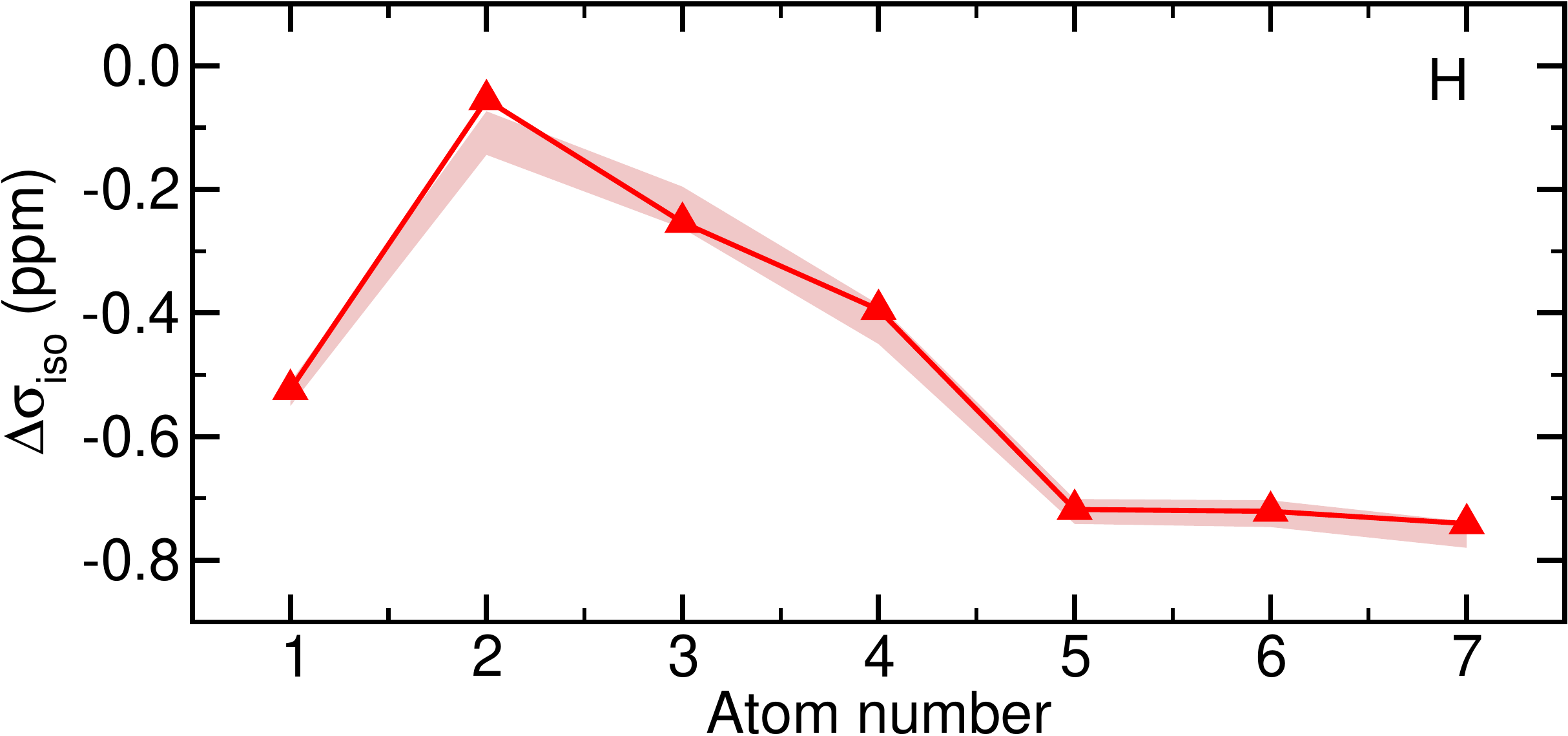}
\includegraphics[scale=0.35]{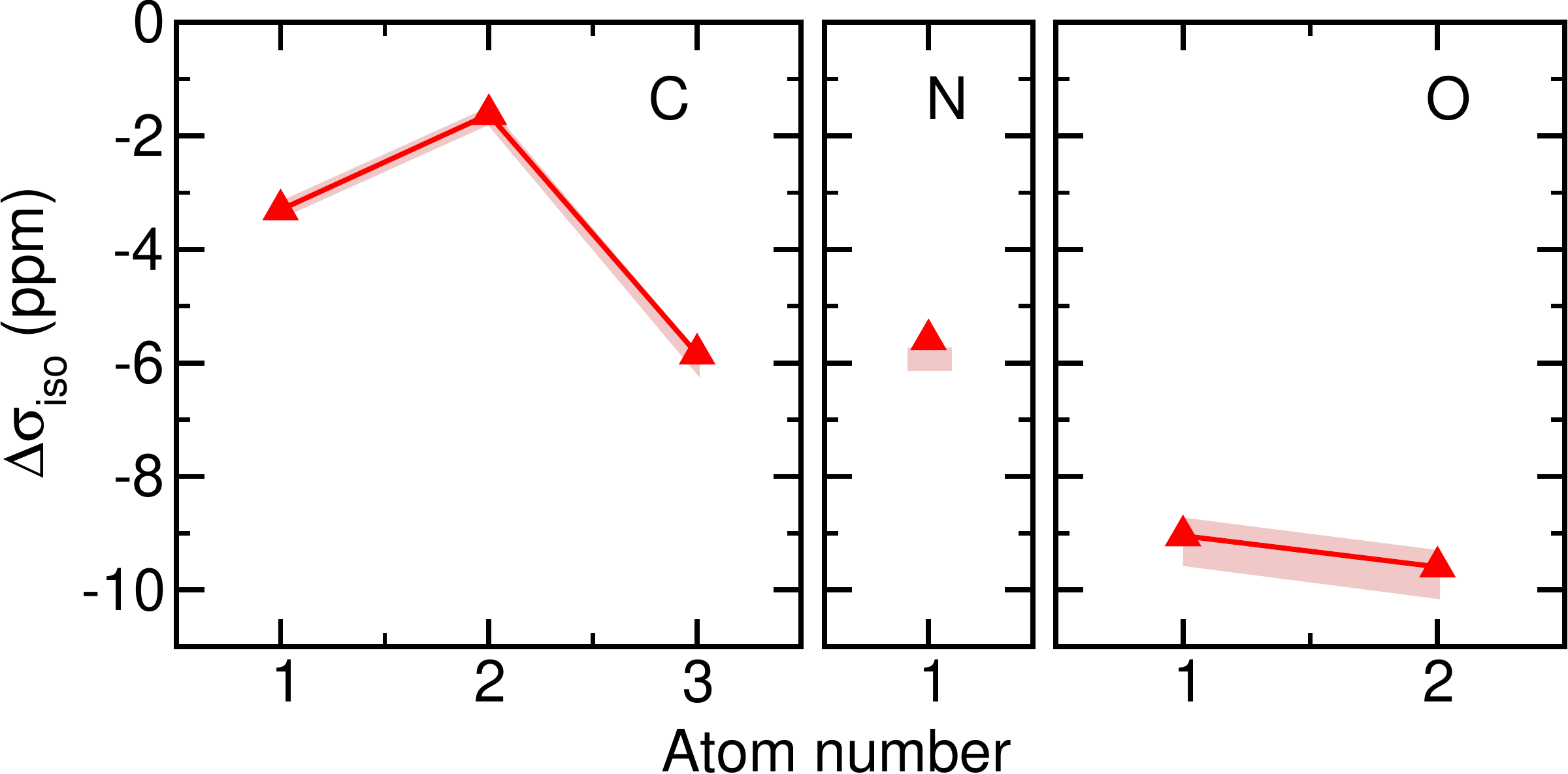}
\caption{ZP correction to the isotropic chemical shift from the static lattice value of L-alanine. The red triangles correspond to the results obtained using Eq.~(\ref{eq:quadratic}), and the light red bands to the results obtained from Monte Carlo sampling. The links between atom numbers are only an aid to the eye.}
\label{fig:alanine_iso}
\end{figure}

In Fig.~\ref{fig:alanine_iso} we report the ZP correction to the isotropic chemical shift of L-alanine, evaluated  by Monte Carlo sampling (light red bands) and using the quadratic expansion (red triangles). As already observed in the case of MgO, the two approaches are in agreement within the statistical uncertainty of the Monte Carlo sampling results for all atomic species and all atoms. We note that the uncertainty associated with the ZP correction of individual atoms in the Monte Carlo evaluation is highly correlated because for every structure sampled we calculate the chemical shielding tensor on all atoms in the crystal. This means that the statistical uncertainty is associated with a shift of the entire light red band rather than independent shifts of the values for individual atoms.

\begin{figure}
\centering
\includegraphics[scale=0.35]{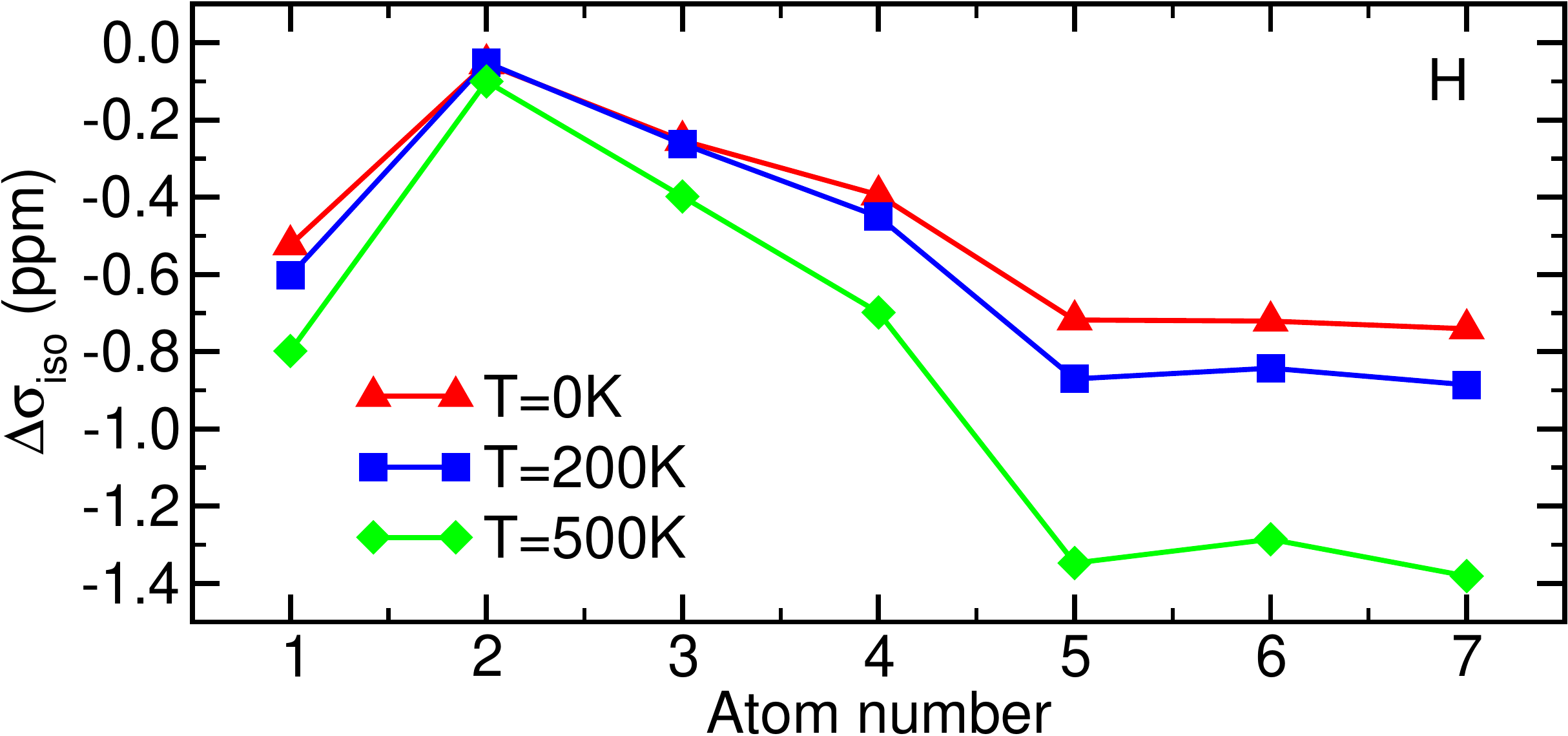}
\includegraphics[scale=0.35]{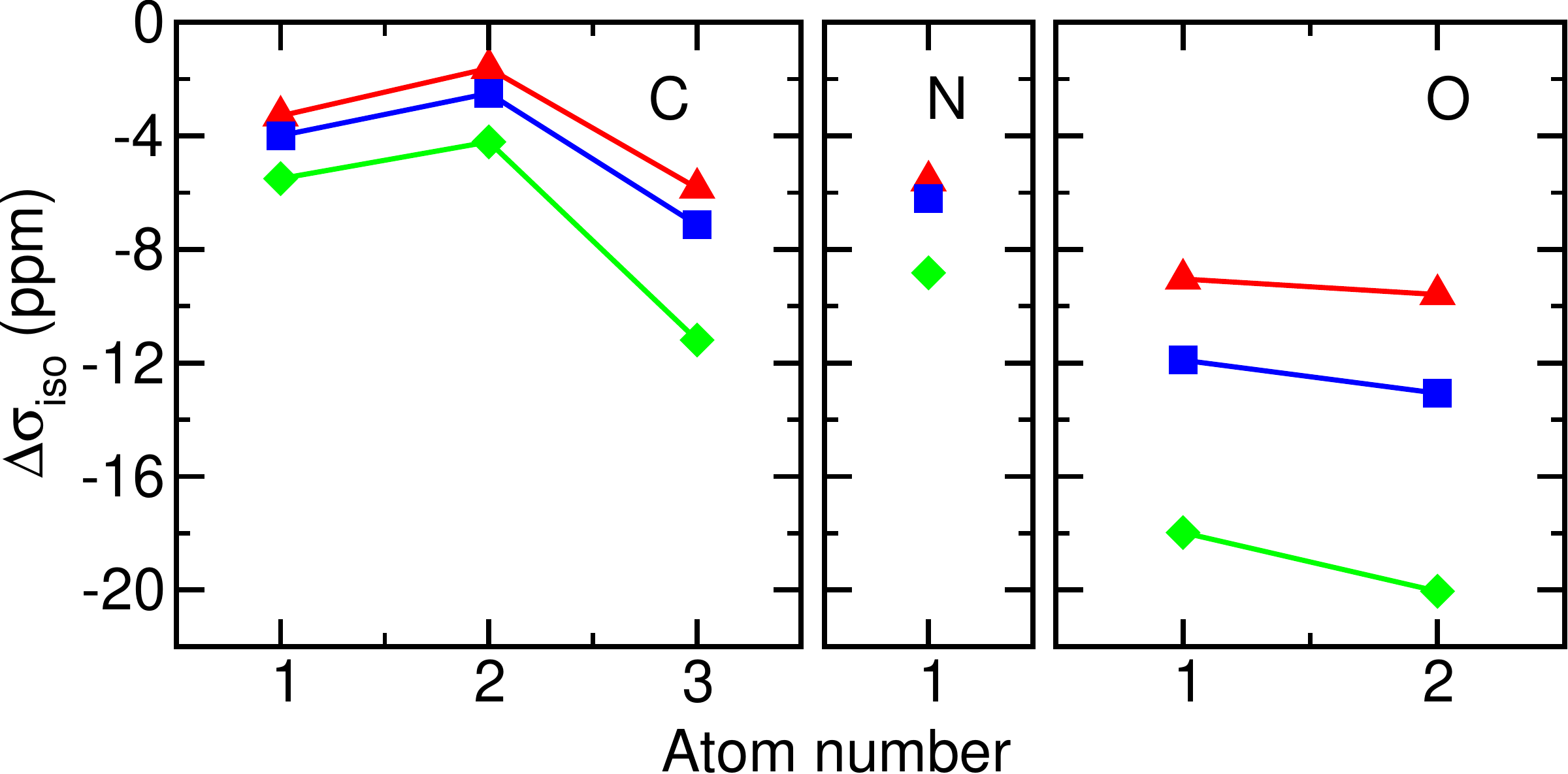}
\caption{Correction to the isotropic chemical shift from the static lattice value of L-alanine at temperatures of $T=0$~K (red triangles), $200$~K (blue squares), and $500$~K (green diamonds). The solid lines are an aid to the eye.}
\label{fig:alanine_iso_temp}
\end{figure}

We show the isotropic shift of L-alanine at temperatures of $T=0$~K, $200$~K, and $500$~K in Fig.~\ref{fig:alanine_iso_temp}. The temperature dependence is particularly strong in the NH$_3$ group, whose hydrogen atoms also have the strongest ZP correction. For all species and atoms, the ZP correction is quite large, and represents about $50$\% of the overall vibrational correction even at a temperature of $T=500$~K.

\begin{figure}
\centering
\includegraphics[scale=0.35]{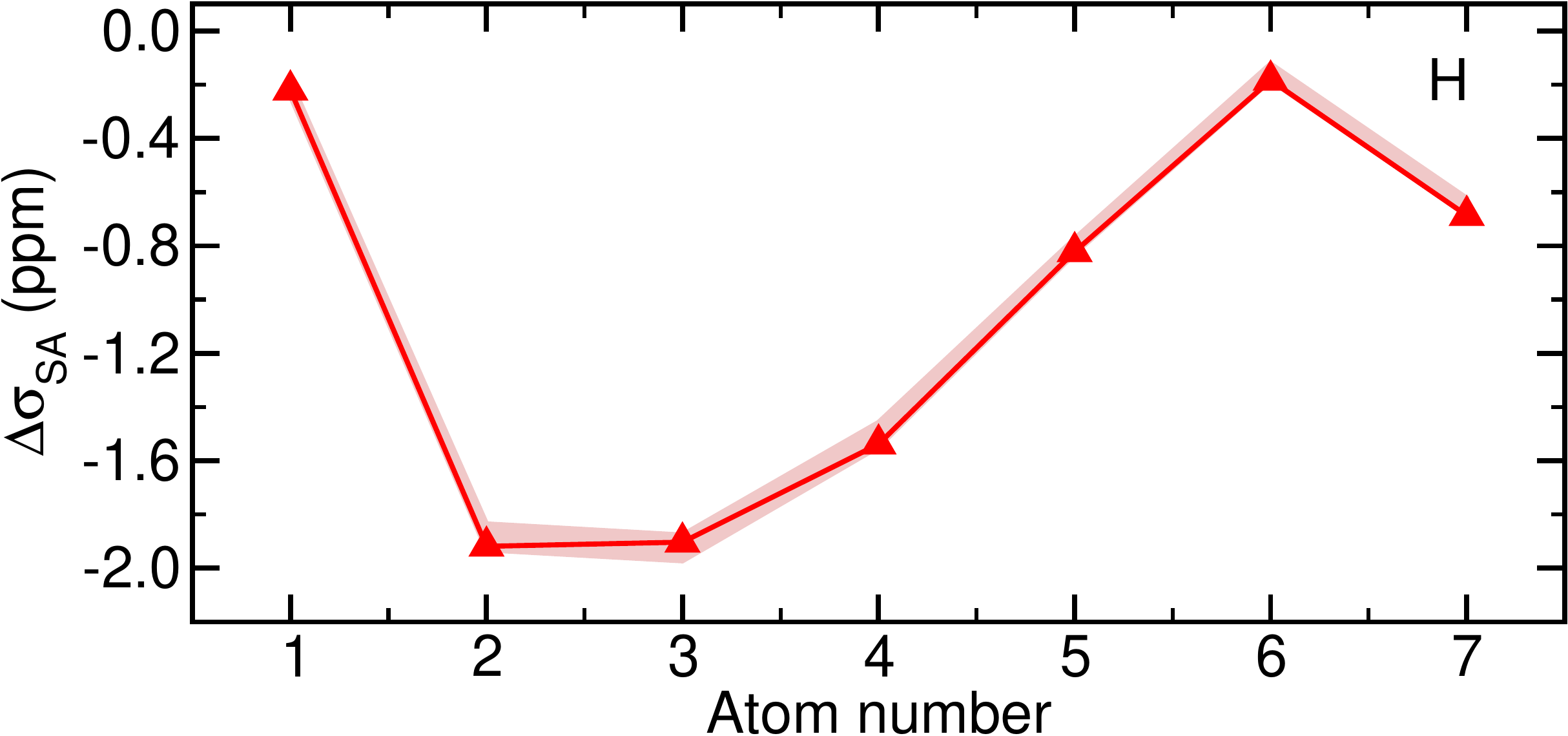}
\includegraphics[scale=0.35]{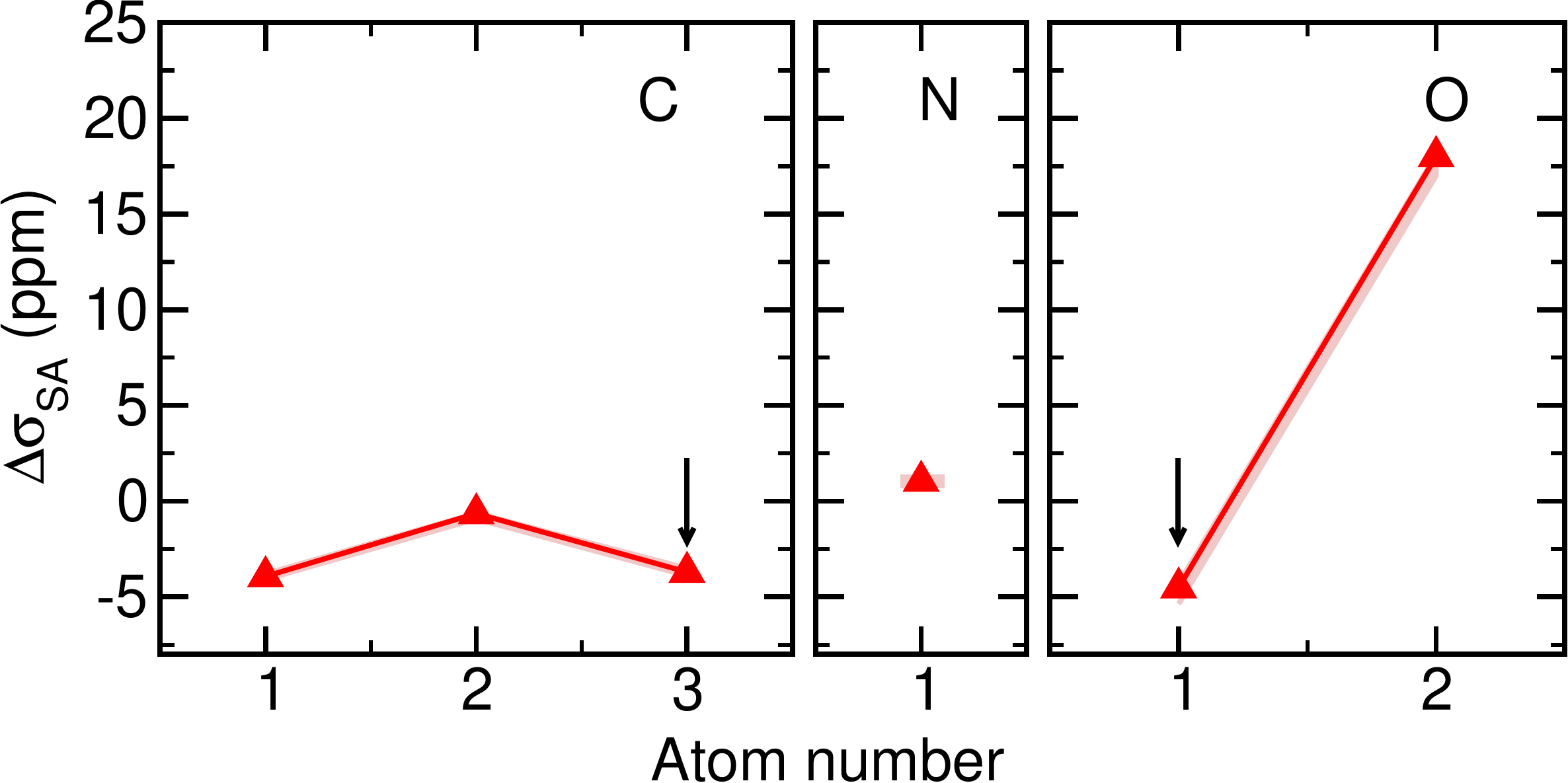}
\caption{ZP correction to the shielding anisotropy from the static lattice value of L-alanine. The red triangles correspond to the results obtained using Eq.~(\ref{eq:quadratic}), and the light red bands to the results from Monte Carlo sampling. The arrows indicate the atoms in which the shielding anisotropy changes sign from the static lattice value to the vibrationally averaged value. The links between atom numbers are only an aid to the eye.}
\label{fig:alanine_ani}
\end{figure}

In Fig.~\ref{fig:alanine_ani} we show the ZP correction to the shielding anisotropy of L-alanine, evaluated by Monte Carlo sampling (light red bands) and using the quadratic expansion (red triangles). For carbon atom $3$ and oxygen atom $1$, the ordering of the principal values of the chemical shielding tensor reverses between the static and vibrationally-averaged calculations, and this is indicated by the black arrows in Fig.~\ref{fig:alanine_ani}. In these cases, the ZP correction shown corresponds to $\Delta\sigma_{\mathrm{SA}}=|\langle\sigma_{\mathrm{SA}}\rangle|-|\sigma_{\mathrm{SA}}|$.

\begin{figure}
\centering
\includegraphics[scale=0.35]{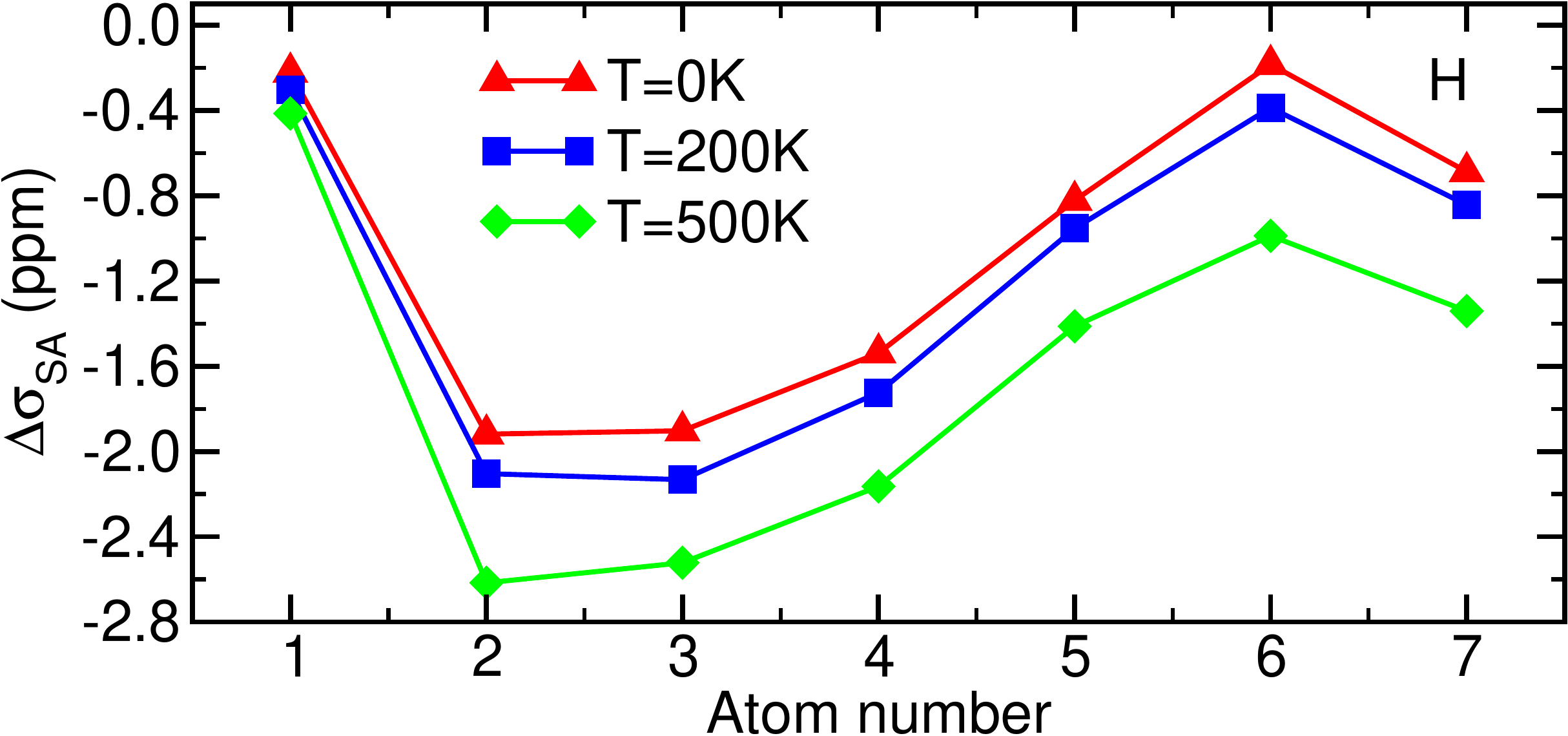}
\includegraphics[scale=0.35]{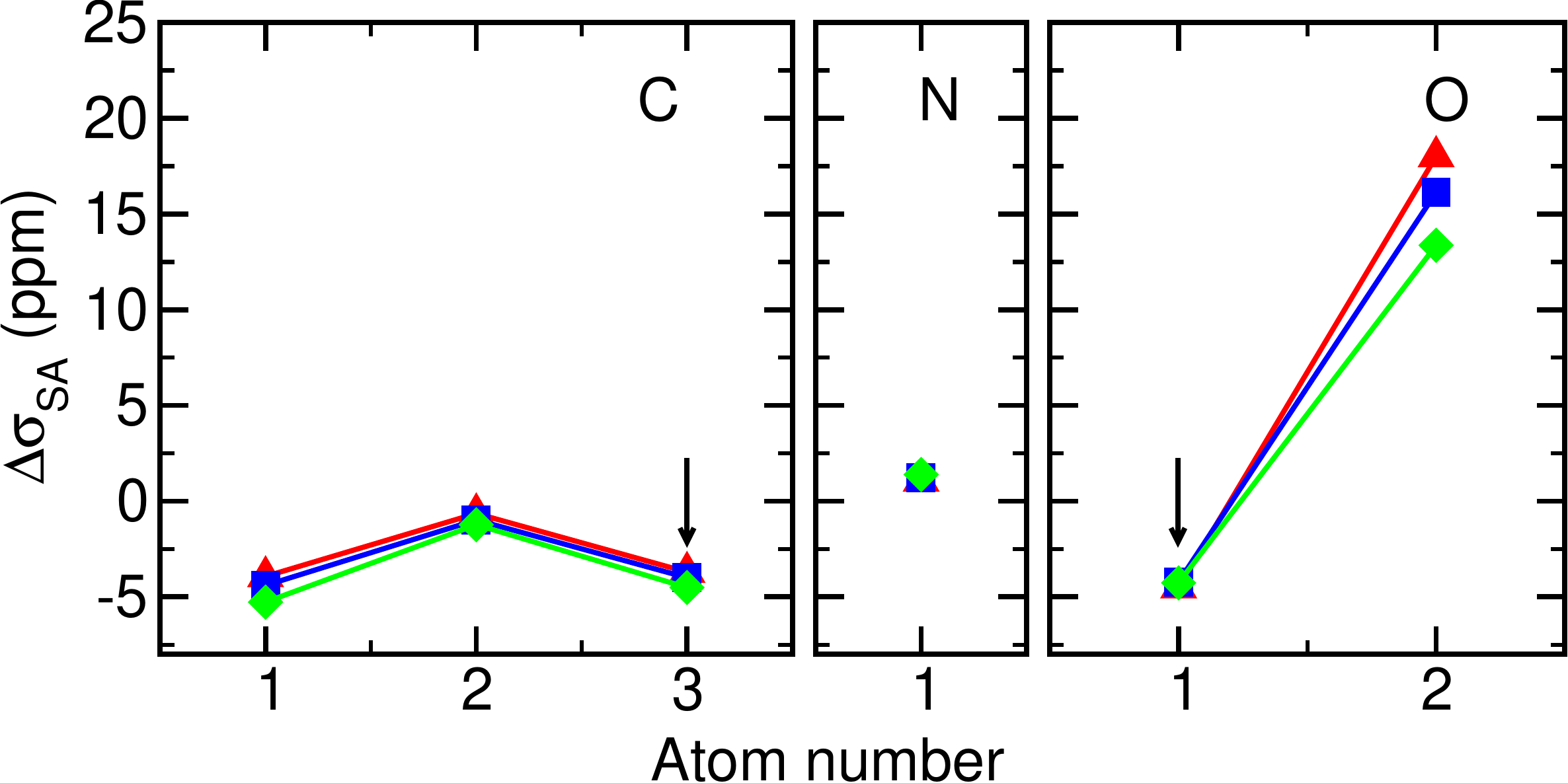}
\caption{Correction to the shielding anisotropy from the static lattice value of L-alanine at temperatures of $T=0$~K (red triangles), $200$~K (blue squares), and $500$~K (green diamonds). The solid lines are an aid to the eye.}
\label{fig:alanine_ani_temp}
\end{figure}

We show the temperature dependence of the shielding anisotropy at temperatures of $T=0$~K, $200$~K, and $500$~K in Fig.~\ref{fig:alanine_ani_temp}. The temperature dependence of the shielding anisotropy has a similar strength in the CH$_3$ and NH$_3$ groups, unlike the isotropic shift dependence. The magnitude of the correction due to temperature is smaller for the shielding anisotropy than for the isotropic shift. This feature is even more pronounced in the case of bDA (see Sec.~\ref{subsec:bDA} below). 

We have investigated the origin of the weak temperature dependence of the shielding anisotropy in the two molecular crystals considered, which contrasts with the strong temperature dependence of the isotropic chemical shift. We consider the change in each of the three eigenvalues of the chemical shielding tensor due to atomic vibrations from zero temperature, $\Sigma_{\alpha\alpha}=\sigma_{\alpha\alpha}(T)-\sigma_{\alpha\alpha}(0)$, where $\alpha$ stands for the three Cartesian directions $\alpha=x,y,z$. Although there are differences depending on the atom under consideration, in general the changes in the three eigenvalues for a given atom are similar $\Sigma_{xx}\simeq\Sigma_{yy}\simeq\Sigma_{zz}$, which leads to
\begin{eqnarray}
\sigma_{\mathrm{iso}}(T)\!-\!\sigma_{\mathrm{iso}}(0)\!&=&\!\frac{1}{3}\sum_{\alpha=x,y,z}\Sigma_{\alpha\alpha}\simeq\Sigma_{xx}, \\
\sigma_{\mathrm{SA}}(T)-\sigma_{\mathrm{SA}}(0)&=&\Sigma_{zz}-\frac{1}{2}(\Sigma_{xx}+\Sigma_{yy})\simeq0,
\end{eqnarray}
qualitatively explaining the observed temperature dependences of both the isotropic shift and the shielding anisotropy. In reality the eigenvalue with the largest absolute value has a somewhat larger correction, which ultimately leads to the non-vanishing temperature dependence of the shielding anisotropy.

The quadratic expansion requires the evaluation of $306$ data points, including the averaging over positive and negative displacements. For the Monte Carlo sampling evaluation, we have used a total of $10,000$ data points, which is about $30$ times larger than the set used in the quadratic calculation. We have used such a large number of sampling points because we reuse them for the anharmonic calculations (see Sec.~\ref{sec:anharmonic} below). The statistical error bars with this number of sampling points represent less than $10$\% of the overall ZP correction for all species and atoms, apart from the hydrogen atoms in the methyl group CH$_3$, for which they represent $10$--$30$\% of the ZP correction. If one is only interested in harmonic results, a more reasonable number may be of the order of $1000$ random sample points. This data set leads to statistical uncertainties in the ZP correction that represent about $10$\% of the overall ZP correction for hydrogen atoms $1$, $5$, $6$, and $7$, and about $50$\% for the hydrogen atoms of the methyl group CH$_3$. For carbon atoms the uncertainty is in the range $20$--$30$\% of the overall ZP correction, and about $10$\% for the nitrogen and oxygen atoms. In either case, the quadratic approximation is significantly more computationally efficient than the Monte Carlo sampling approach.

\subsubsection{Comparison with experiment}

The experimental shielding anisotropy at room temperature of the carbon atoms in L-alanine are reported in Ref.~\onlinecite{alanine_nmr_experiment}. In Table~\ref{tab:alanine_exp_sa} we show the experimental shielding anisotropies compared to the theoretical ones obtained within the static lattice approximation and the quadratic approximation at $T=293$~K. For carbon atom $1$ the vibrational contribution leads to a significant improvement in terms of the agreement between theory and experiment. For the other two atoms, we also find better agreement between theory and experiment when the effects of atomic vibrations are included, but the improvement is smaller.

\begin{table*}[t]
\setlength{\tabcolsep}{12pt} 
  \caption{Comparison of experimental and theoretical shielding anisotropies for the carbon atoms in L-alanine, in units of ppm. The experimental data is from Ref.~\onlinecite{alanine_nmr_experiment}.}
\label{tab:alanine_exp_sa}
\begin{tabular}{ccccc}
\hline
\hline
\textbf{Species} &  \textbf{Atom number} & \textbf{Experiment}  &  \textbf{Theory ($\mathbf{T=293}$~K)}  & \textbf{Theory (static)} \\
\hline
C  &  1   & $\phantom{0}29.5$   &  $\phantom{0}29.1$ & $\phantom{0}33.3$ \\
  &  2   &  $106.5$   &   $111.1$ & $111.8$ \\
  &  3   &  $\phantom{0}17.5$   &   $\phantom{0}19.0$ & $\phantom{0}15.1$ \\
\hline
\hline
\end{tabular}
\end{table*}


\subsection{$\beta$-aspartyl-L-alanine} \label{subsec:bDA}

The scheme we use for labeling the atoms in a bDA molecule is shown in the bottom diagram of Fig.~\ref{fig:bDA_structure}. This scheme is used throughout this work.

\subsubsection{Theoretical calculations}

\begin{figure}
\centering
\includegraphics[scale=0.35]{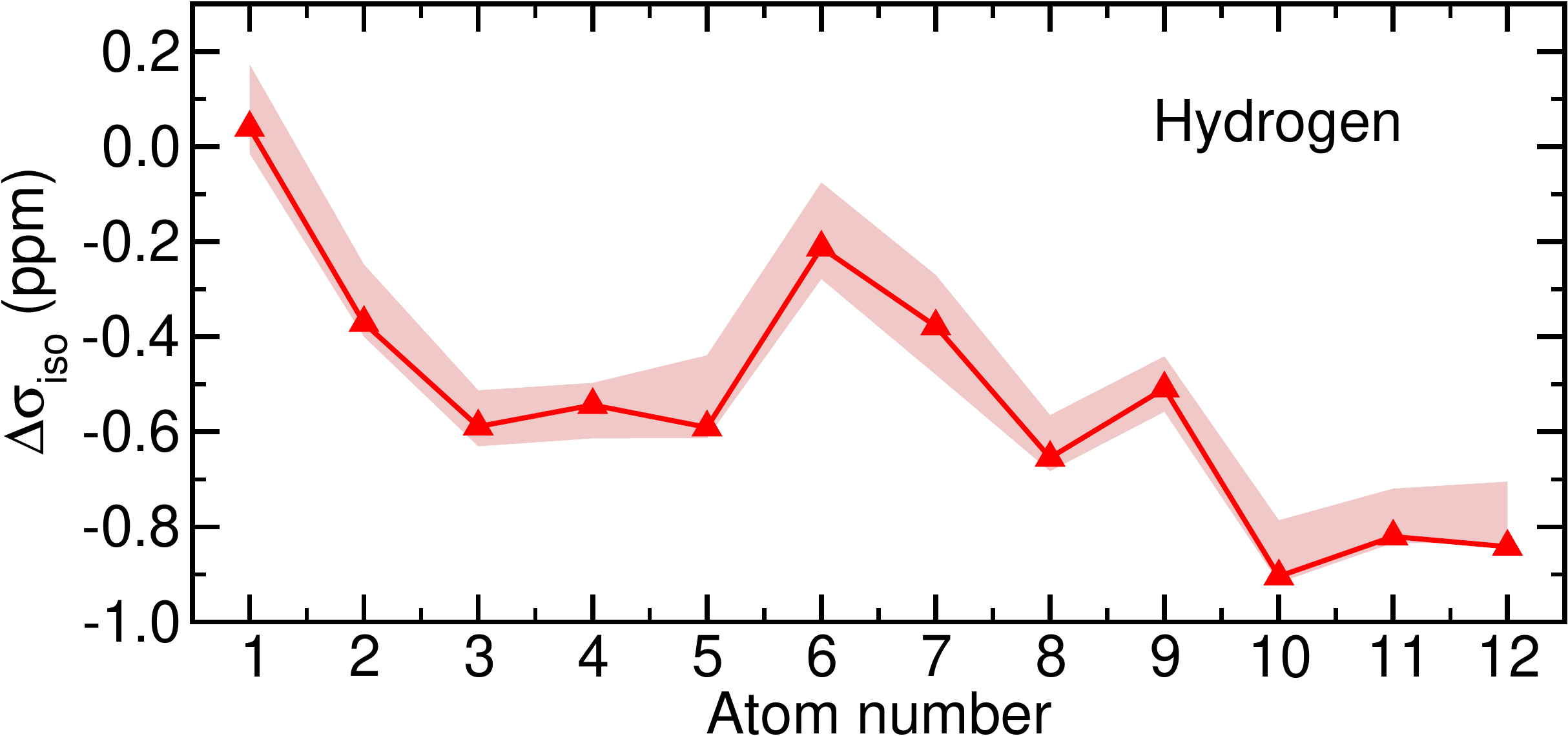}
\includegraphics[scale=0.35]{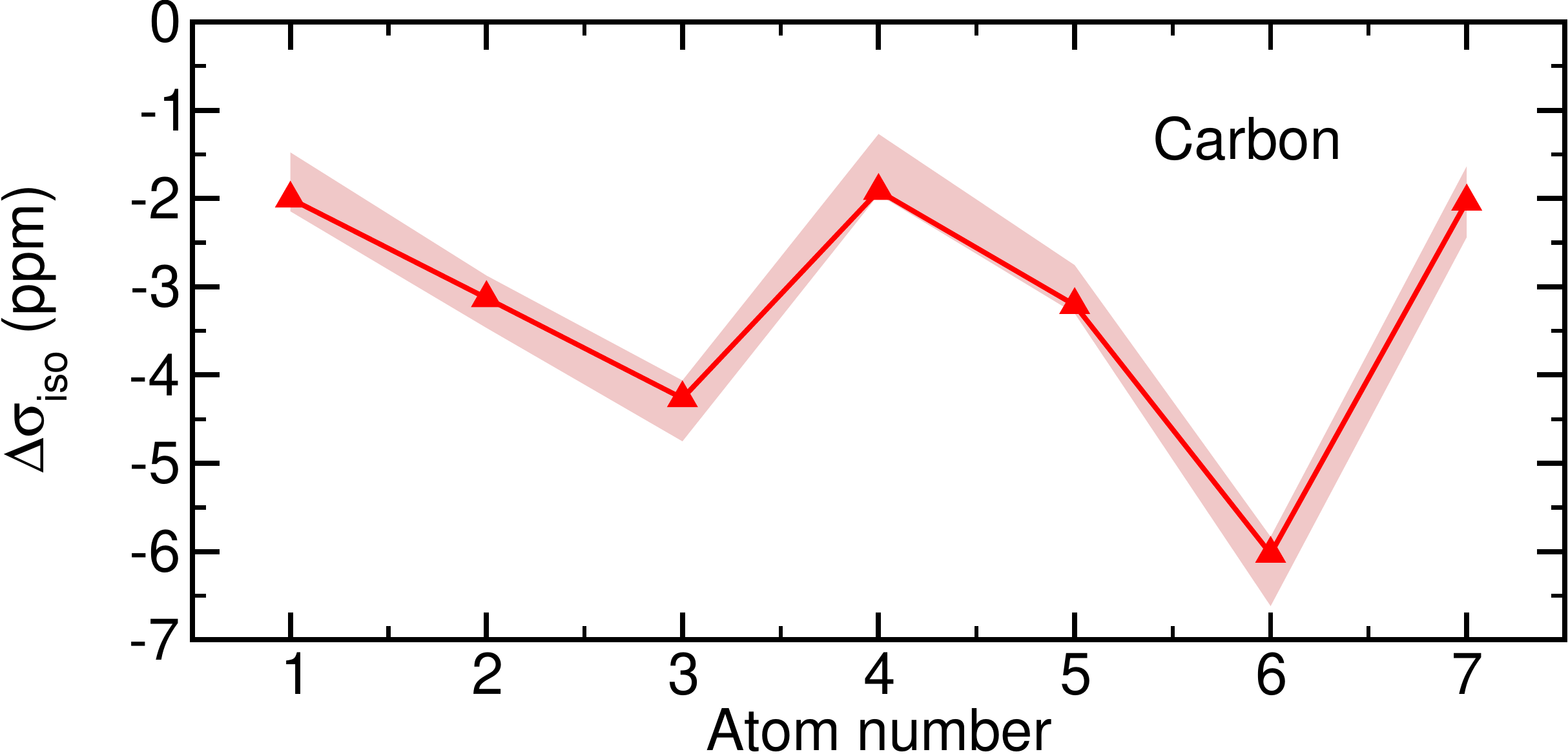}
\includegraphics[scale=0.35]{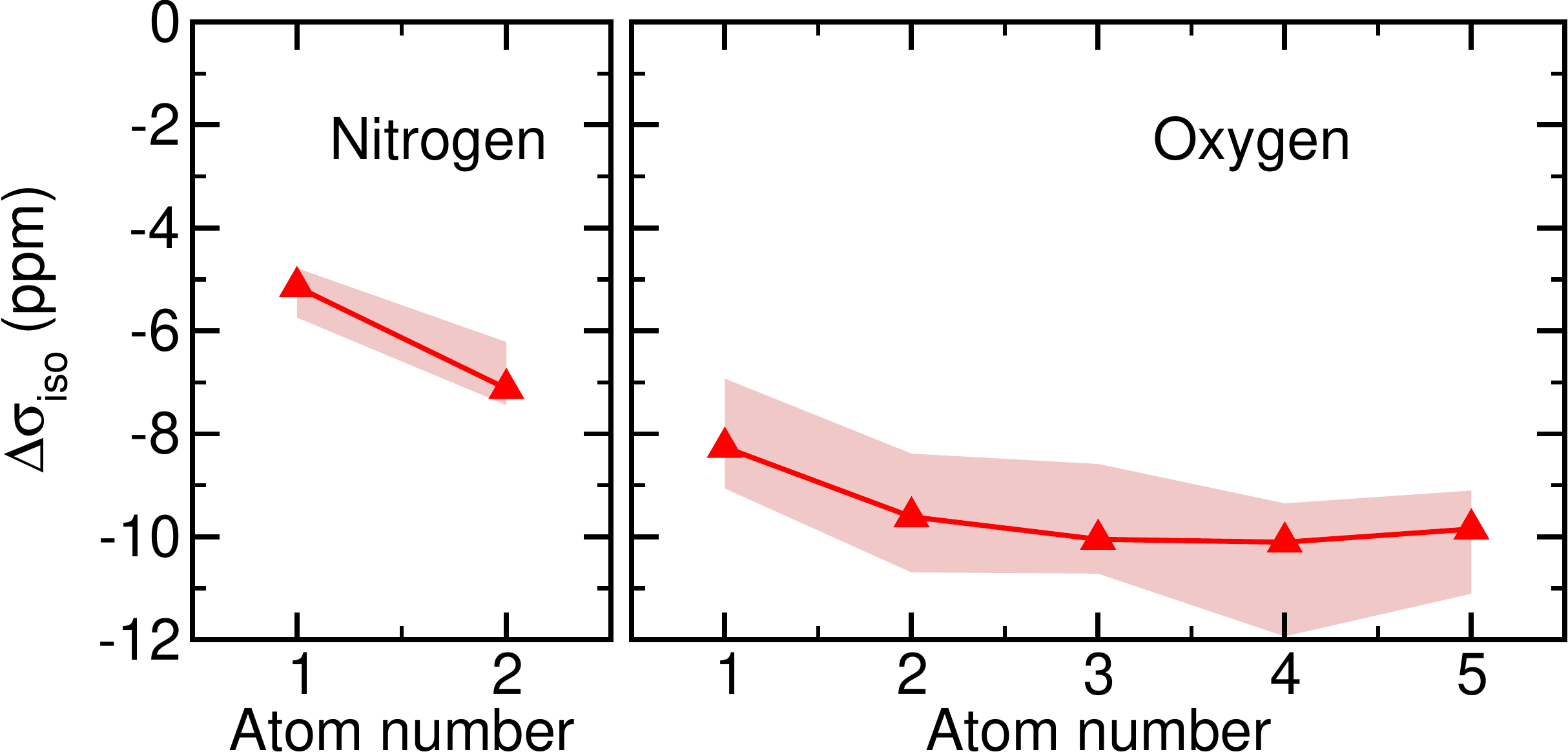}
\caption{ZP correction to the isotropic chemical shift from the static lattice value of bDA. The red triangles correspond to the results obtained using Eq.~(\ref{eq:quadratic}), and the light red bands to those from Monte Carlo sampling. The links between atom numbers are only an aid to the eye.}
\label{fig:bDA_iso}
\end{figure}

In Fig.~\ref{fig:bDA_iso} we show the ZP correction to the isotropic chemical shift of bDA, evaluated by Monte Carlo sampling (light red bands) and using the quadratic expansion (red triangles). These results further support the validity of the approximate Eq.~(\ref{eq:quadratic}), as the results from the quadratic expansion agree with those of the Monte Carlo sampling within the statistical uncertainty of the latter.

\begin{figure}
\centering
\includegraphics[scale=0.35]{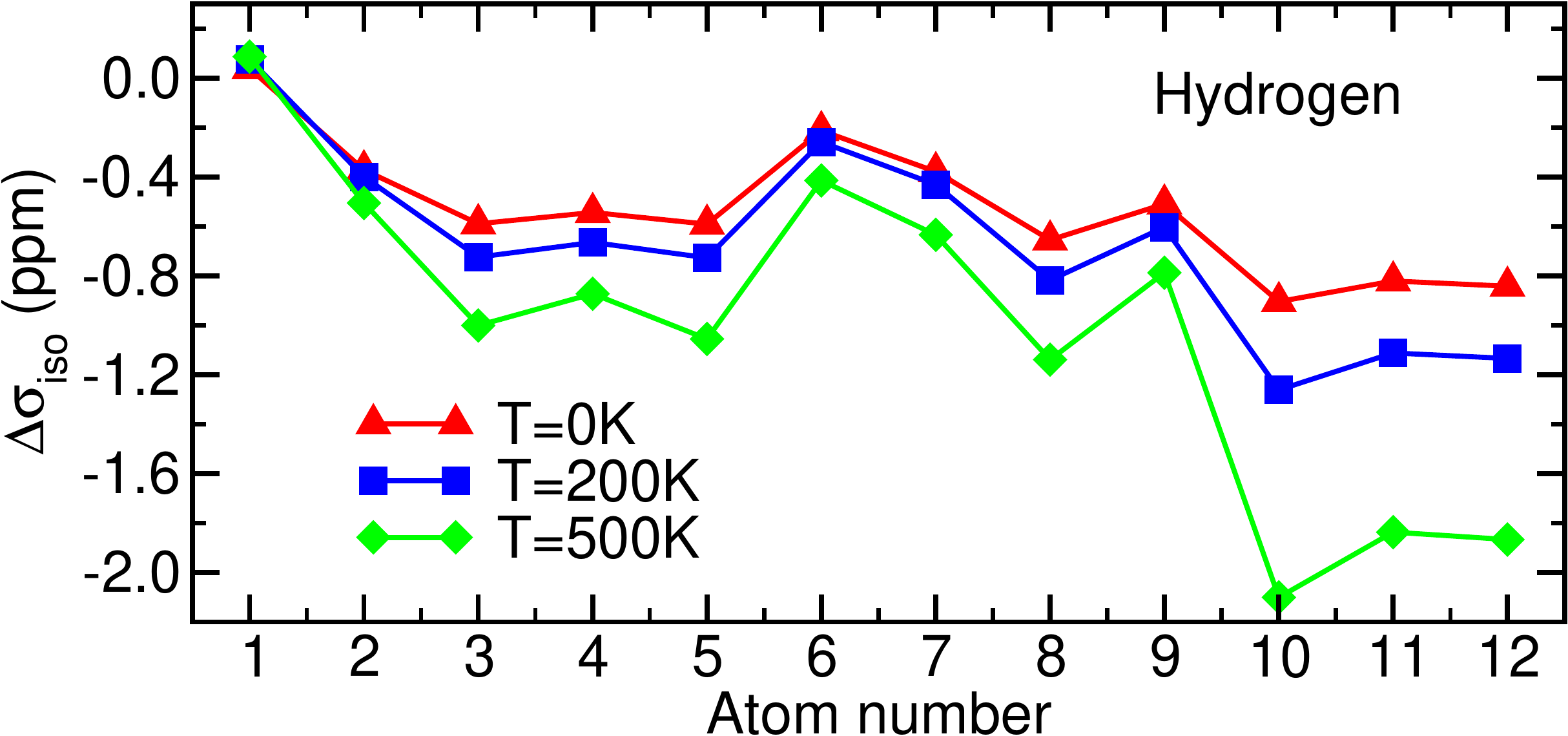}
\includegraphics[scale=0.35]{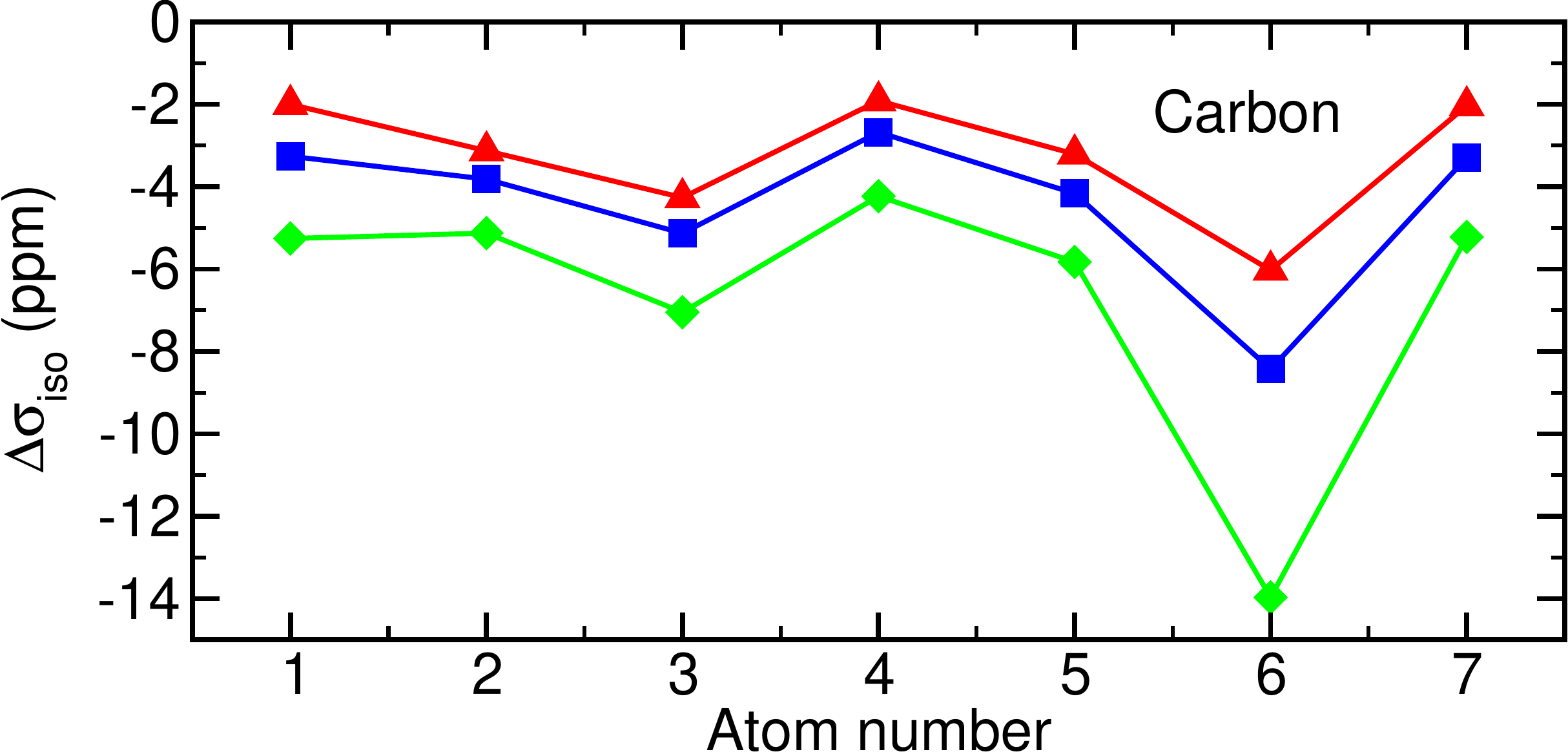}
\includegraphics[scale=0.35]{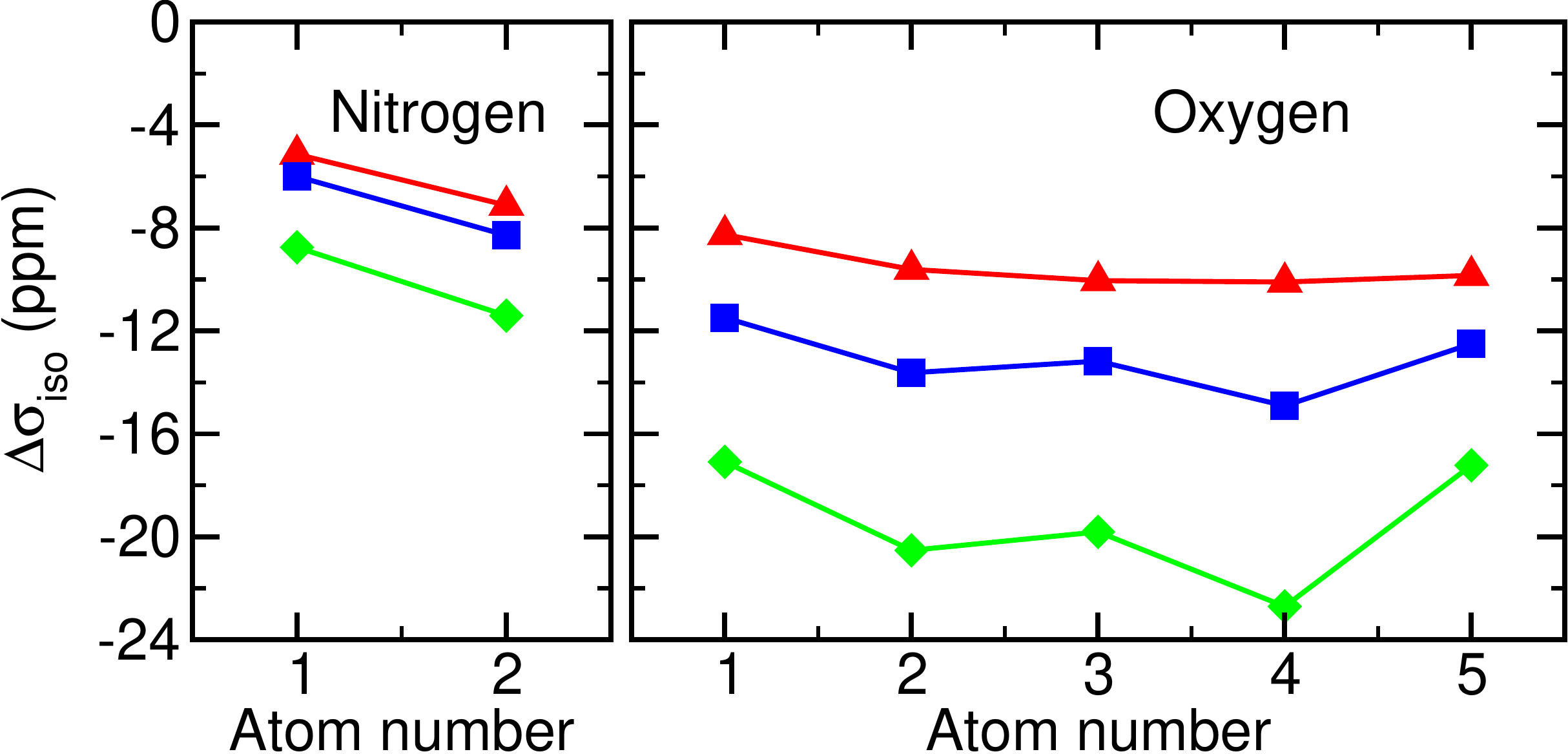}
\caption{Correction to the isotropic chemical shift from the static lattice value of bDA at temperatures of $T=0$~K (red triangles), $200$~K (blue squares), and $500$~K (green diamonds). The solid lines are an aid to the eye.}
\label{fig:bDA_iso_temp}
\end{figure}

In Fig.~\ref{fig:bDA_iso_temp} we show the isotropic shift of bDA at $T=0$~K, $200$~K, and $500$~K, calculated using the quadratic approximation of Eq.~(\ref{eq:tdep}). We observe a very strong temperature dependence of the isotropic chemical shift in the three hydrogen atoms of the CH$_3$ methyl group, and also in the corresponding carbon atom. Another noteworthy feature is the importance of the correction due to the quantum-mechanical ZP motion, which still represents about half of the overall correction at a temperature of $T=500$~K for all species and atoms. This suggests that methods of sampling phase space that neglect ZP motion, such as molecular dynamics methods, would lead to inaccurate results.

\begin{figure}
\centering
\includegraphics[scale=0.35]{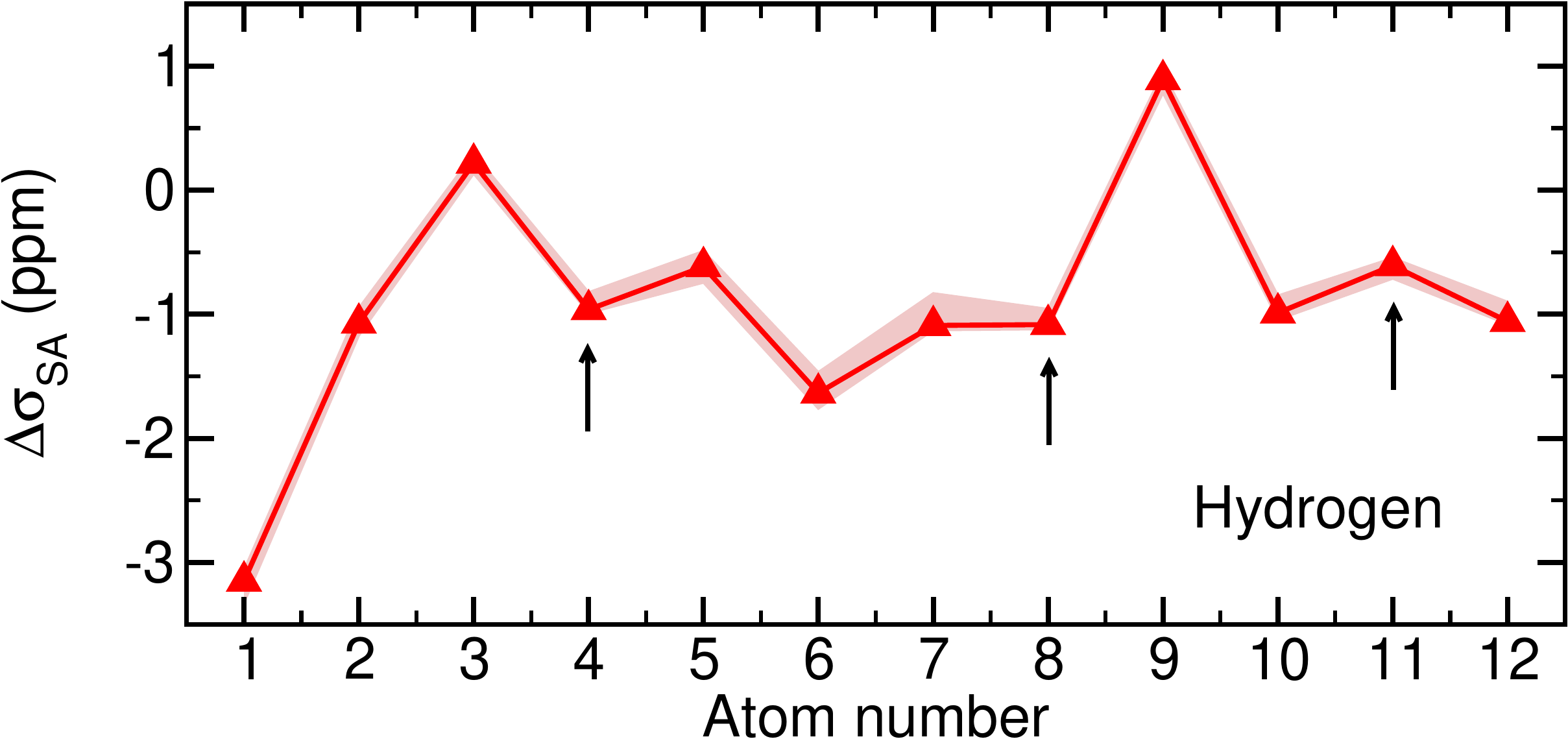}
\includegraphics[scale=0.35]{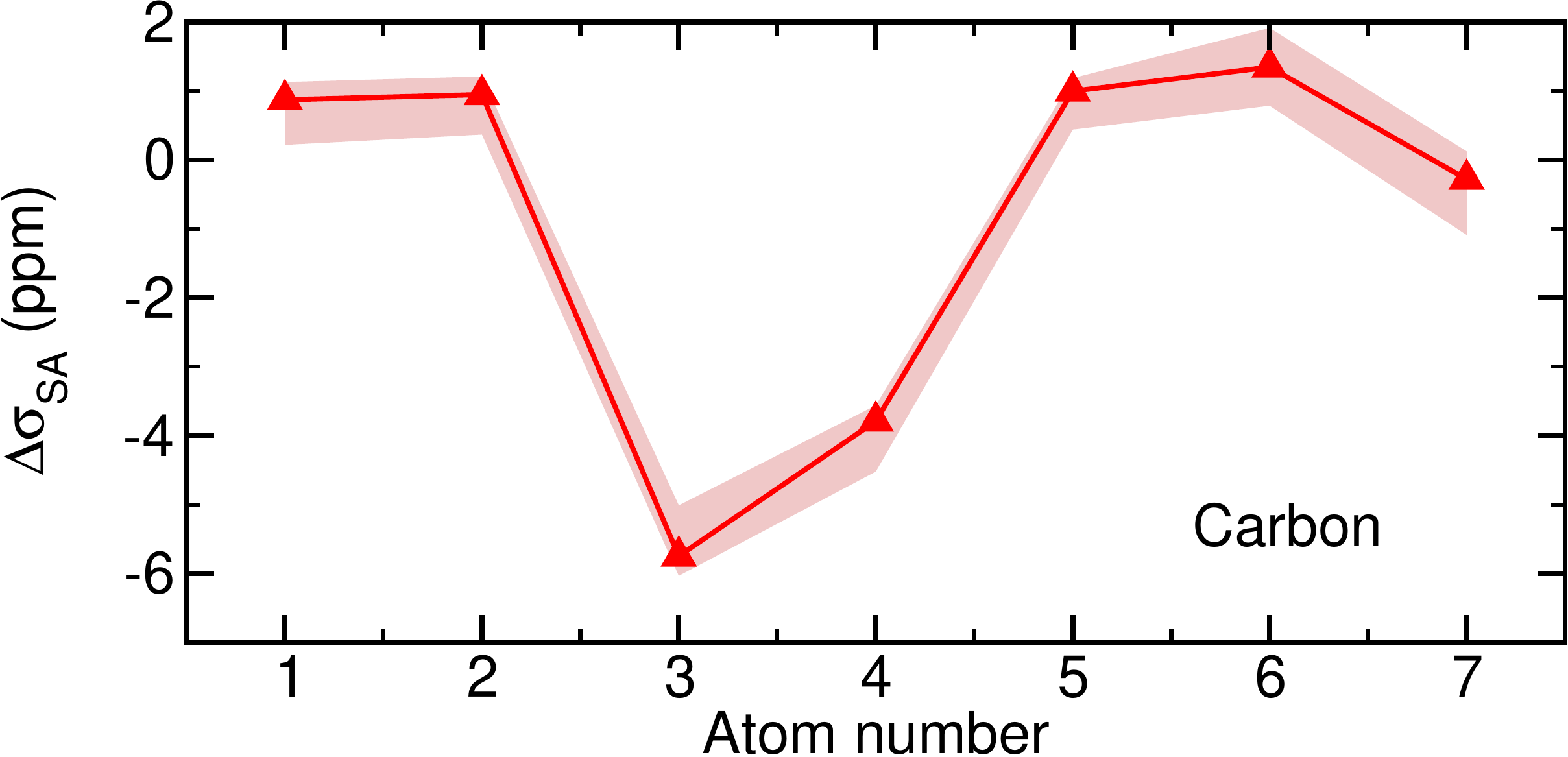}
\includegraphics[scale=0.35]{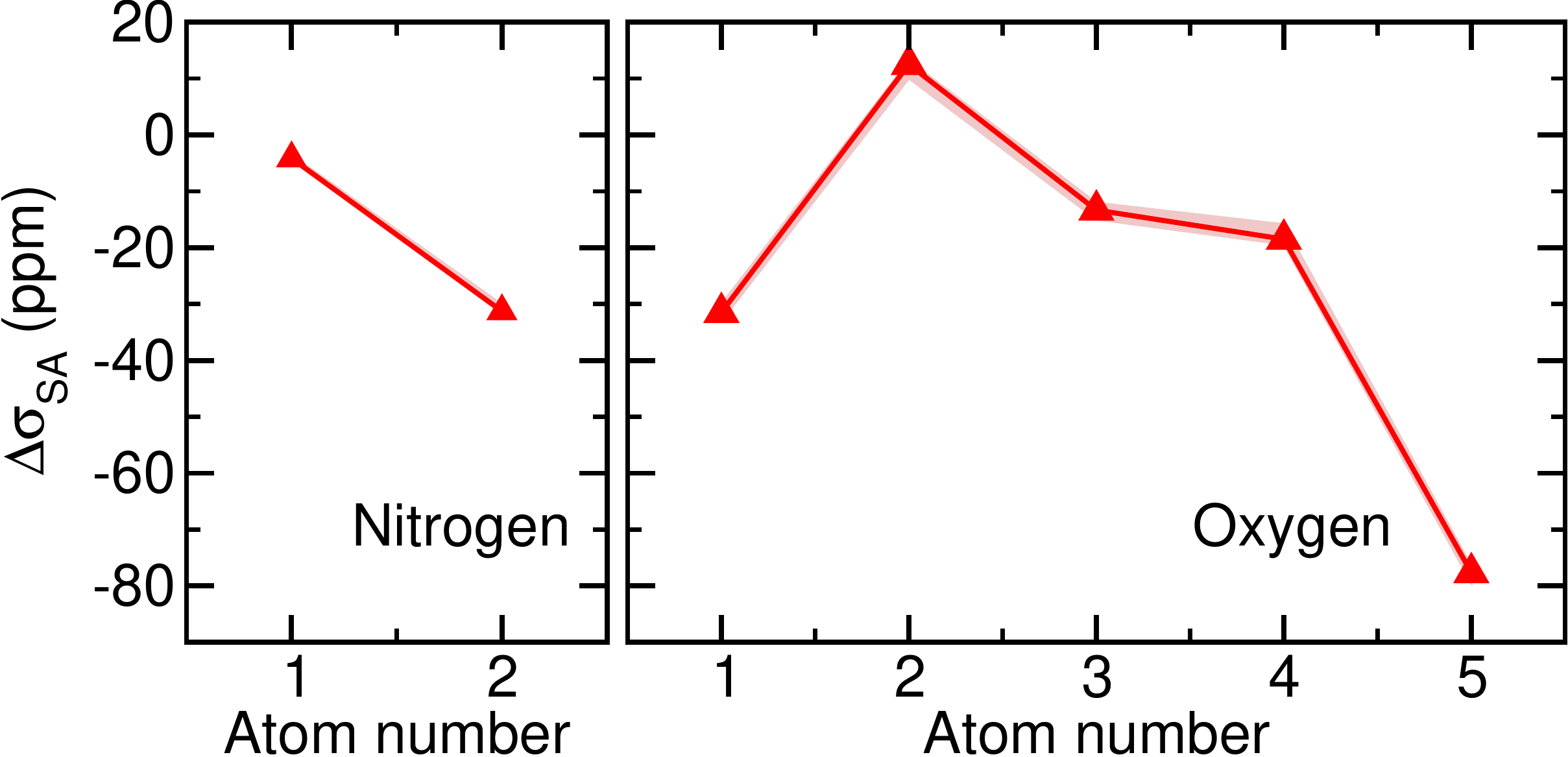}
\caption{ZP correction to the shielding anisotropy from the static lattice value for bDA. The red triangles correspond to the results obtained using Eq.~(\ref{eq:quadratic}), and the light red bands to the results obtained from Monte Carlo sampling. The arrows indicate the atoms for which the shielding anisotropy changes sign from the static lattice value to the vibrationally averaged value. The links between atom numbers are only an aid to the eye.}
\label{fig:bDA_ani}
\end{figure}

In Fig.~\ref{fig:bDA_ani} we show the ZP correction to the shielding anisotropy of bDA, evaluated by Monte Carlo sampling (light red bands) and using the quadratic expansion (red triangles). The agreement between the two methods is also very good in this case. The ZP correction to the shielding anisotropy is significantly larger than the ZP correction to the isotropic shift, and an accurate treatment of it including atomic vibrations could therefore be more important.

\begin{figure}
\centering
\includegraphics[scale=0.35]{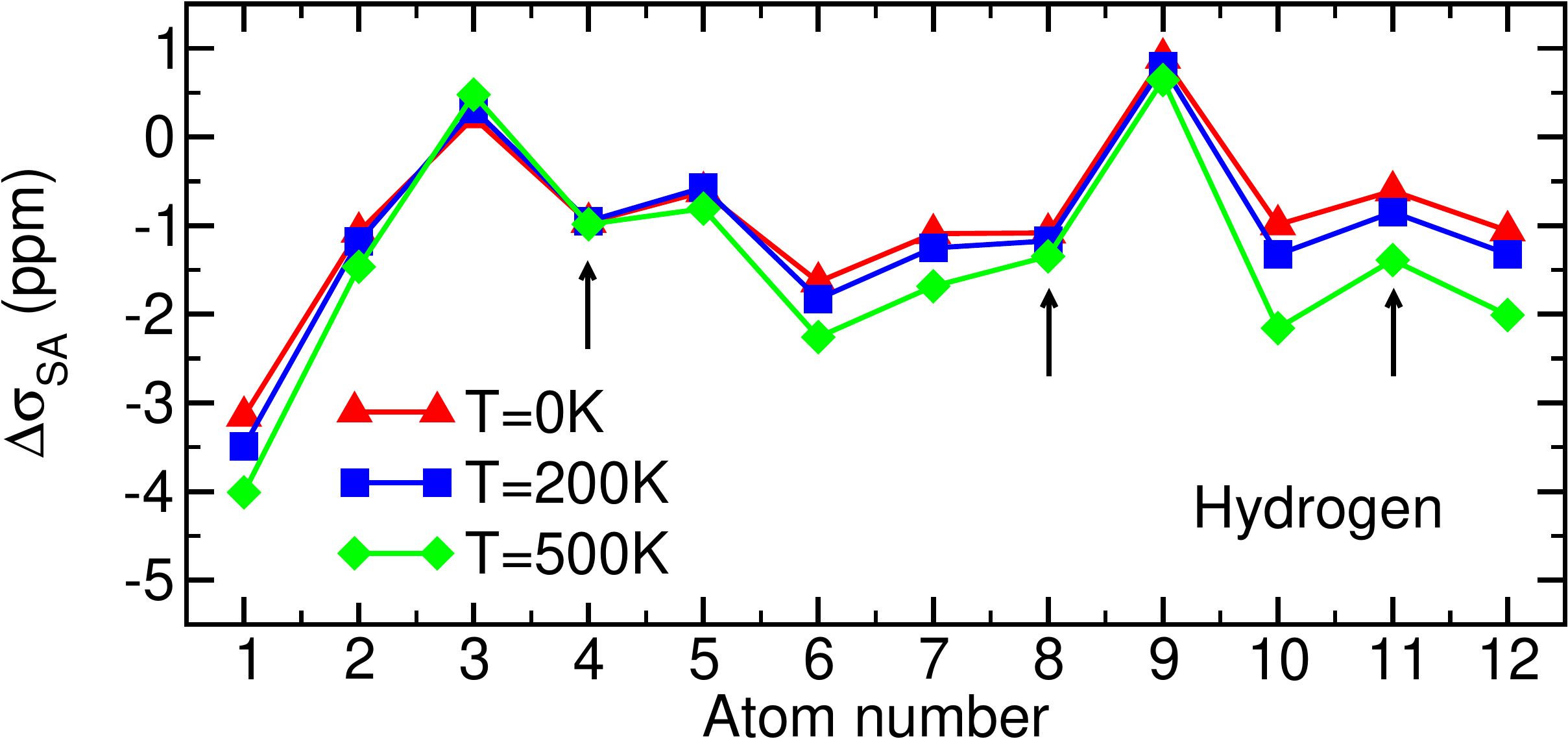}
\includegraphics[scale=0.35]{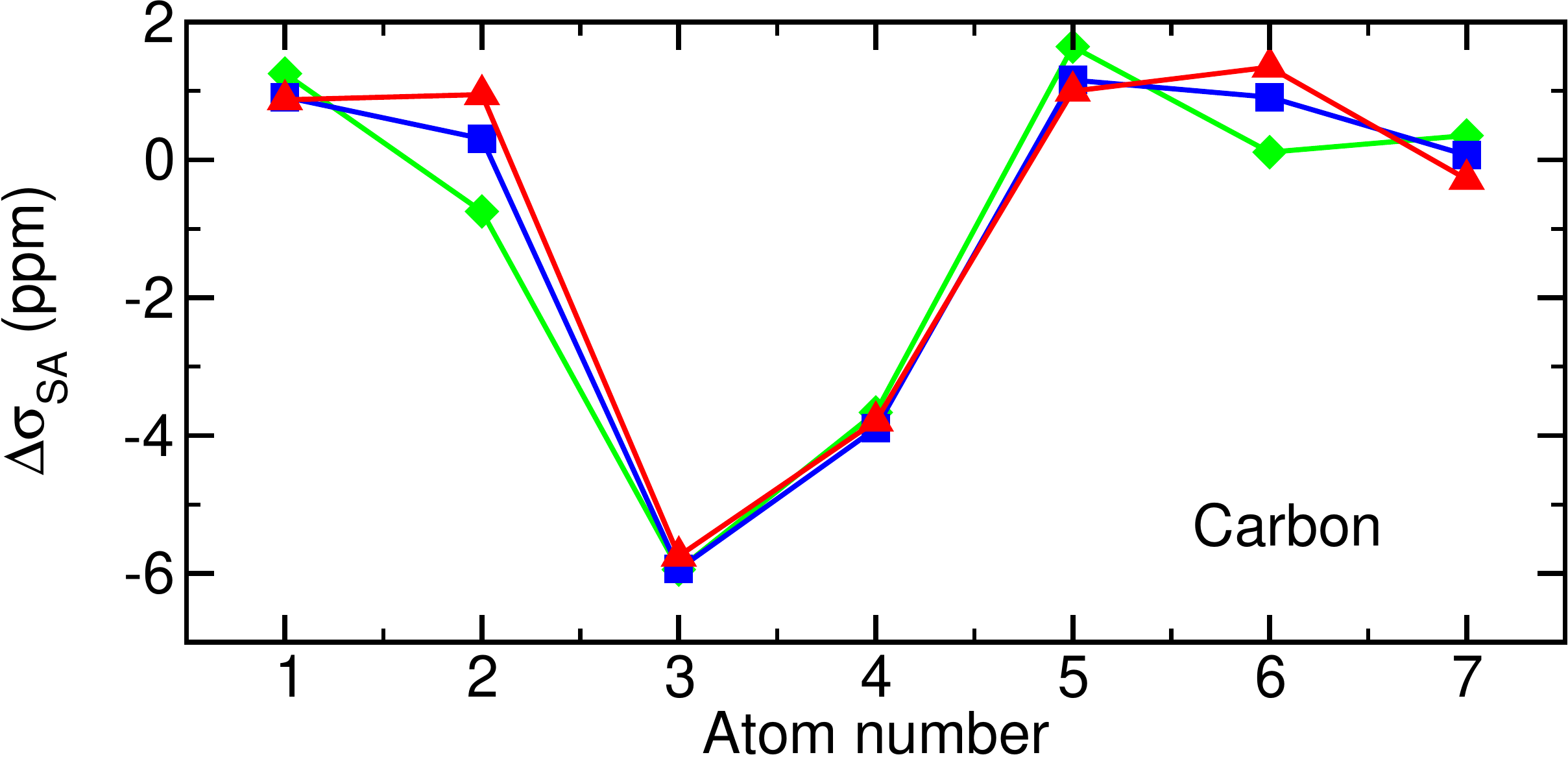}
\includegraphics[scale=0.35]{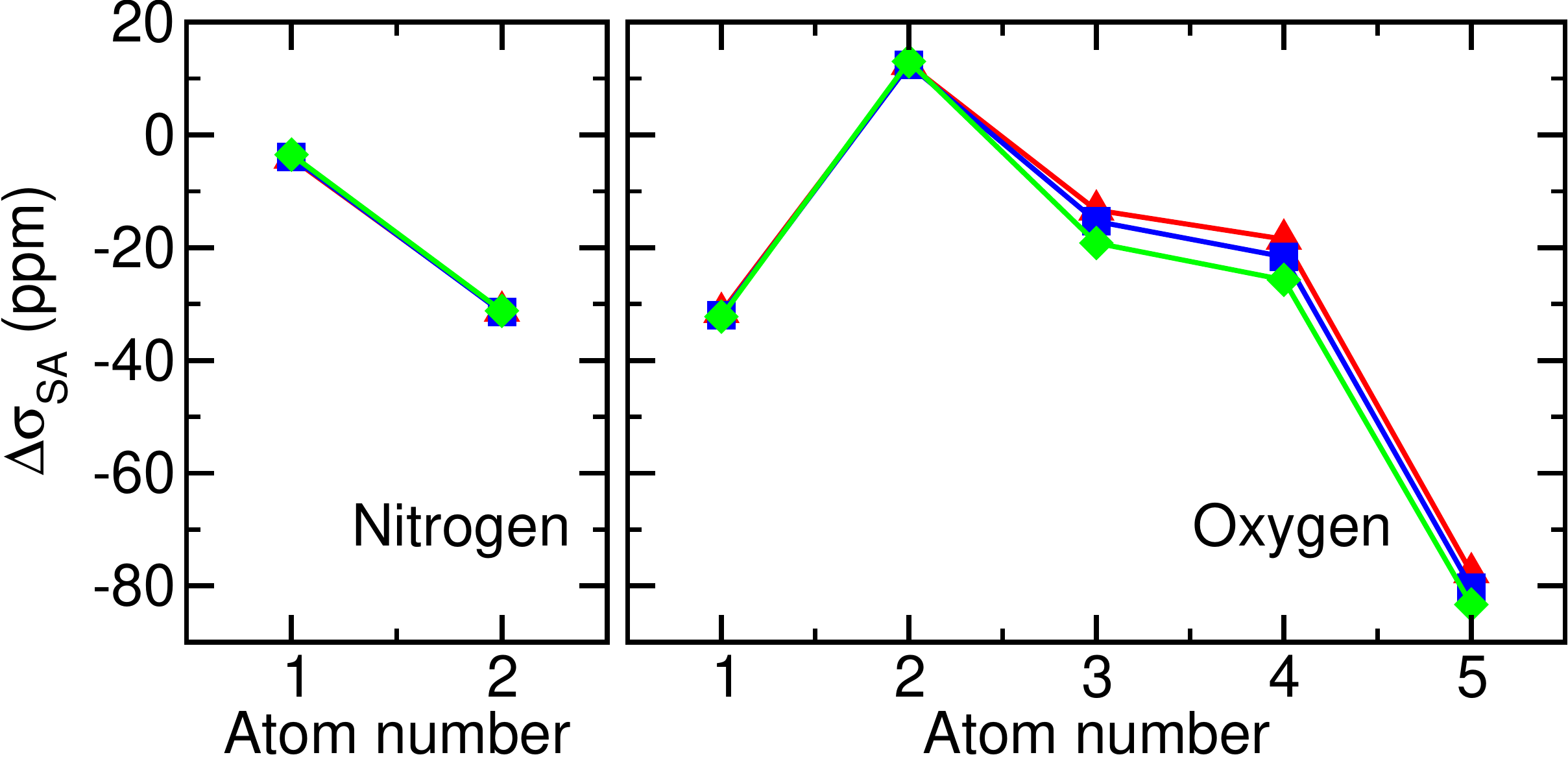}
\caption{Correction to the shielding anisotropy from the static lattice value for bDA at temperatures of $T=0$~K (red triangles), $200$~K (blue squares), and $500$~K (green diamonds).  The arrows indicate the atoms where the shielding anisotropy changes sign from the static lattice value to the vibrationally averaged value. The solid lines are an aid to the eye.}
\label{fig:bDA_ani_temp}
\end{figure}

We show the vibrational correction to the shielding anisotropy at temperatures of $T=0$~K, $T=200$~K, and $T=500$~K in Fig.~\ref{fig:bDA_ani_temp}. We observe a very weak temperature dependence of the shielding anisotropy for all species and atoms, in contrast to the strong temperature dependence found in the isotropic shift (see Fig.~\ref{fig:bDA_iso_temp}). This agrees with the equivalent observation for L-alanine, but in the case of bDA this feature is more prominent. The inclusion of ZP quantum motion is therefore central in calculating finite temperature shielding anisotropies, and sampling the vibrational phase space using molecular dynamics would fail to reproduce the effects of vibrations on the shielding anisotropy. In line with the observation from the temperature dependence of the isotropic shift, the CH$_3$ methyl group dominates the temperature dependence of the shielding anisotropy.

The Monte Carlo calculations reported used $1000$ sampling points. The quadratic expansion used $618$ data points, taking into account the need to average over positive and negative displacements. For this larger system, the computational gain obtained with the quadratic expansion is not as dramatic as for MgO and L-alanine, but it is nonetheless important. We note that the number of Monte Carlo sampling points required is determined by the desired size of the statistical uncertainty. For the isotropic shift, the use of $1000$ data points leads to statistical uncertainties in the range $10$--$30$\% of the full ZP correction for hydrogen atoms $2$--$4$ and $8$--$12$, for atom $6$ of about $60\%$, and for atom $1$ the statistical uncertainty is too large to be able to distinguish the ZP correction from zero. At finite temperature the uncertainties are expected to be even larger. These uncertainties suggest that the number of data points used here would still be inadequate for some applications. The statistical uncertainty in the shielding anisotropy has a similar behaviour across all species and atoms for the same number of data points. 

\subsubsection{Comparison with experiment}

\begin{figure}
\centering
\includegraphics[scale=0.35]{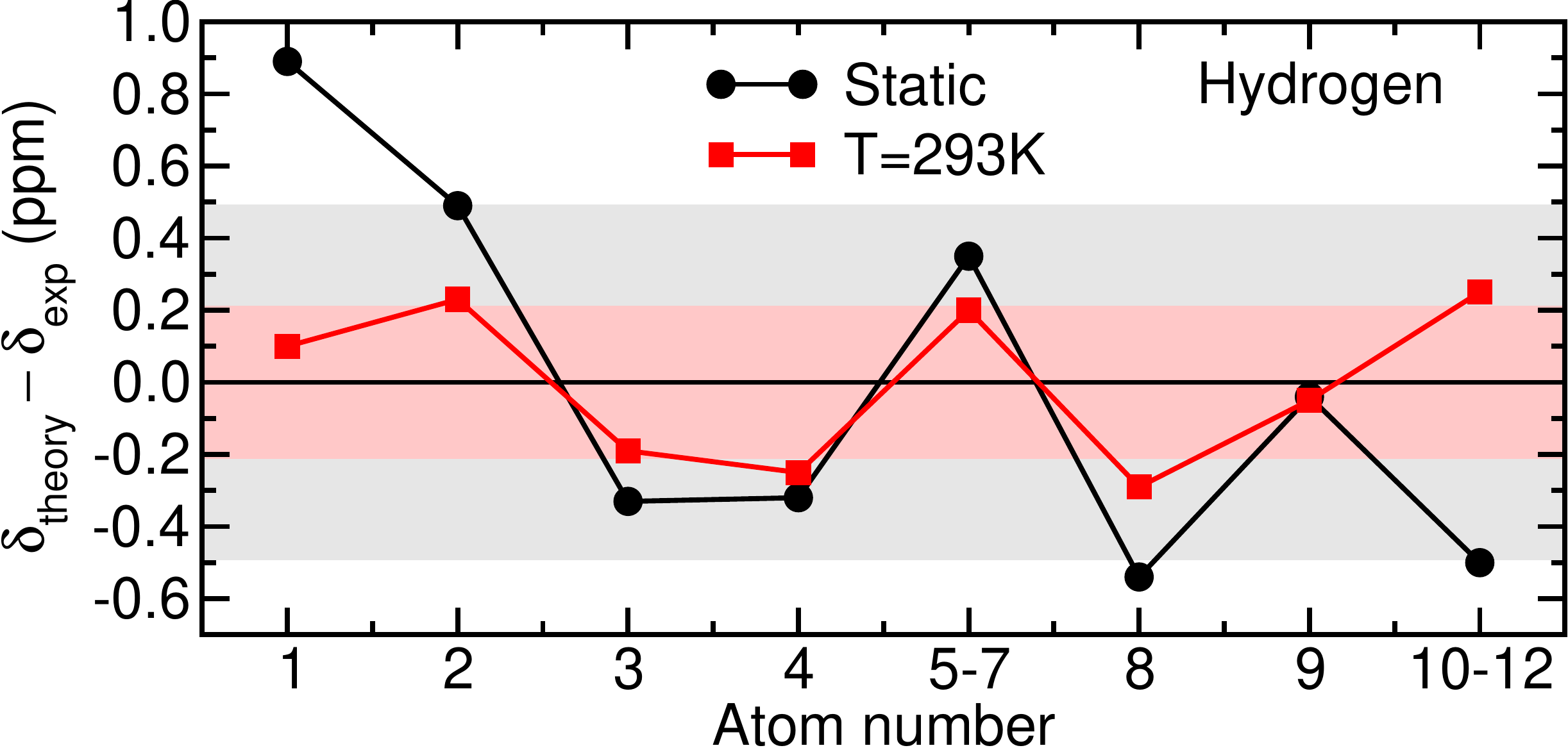}
\includegraphics[scale=0.35]{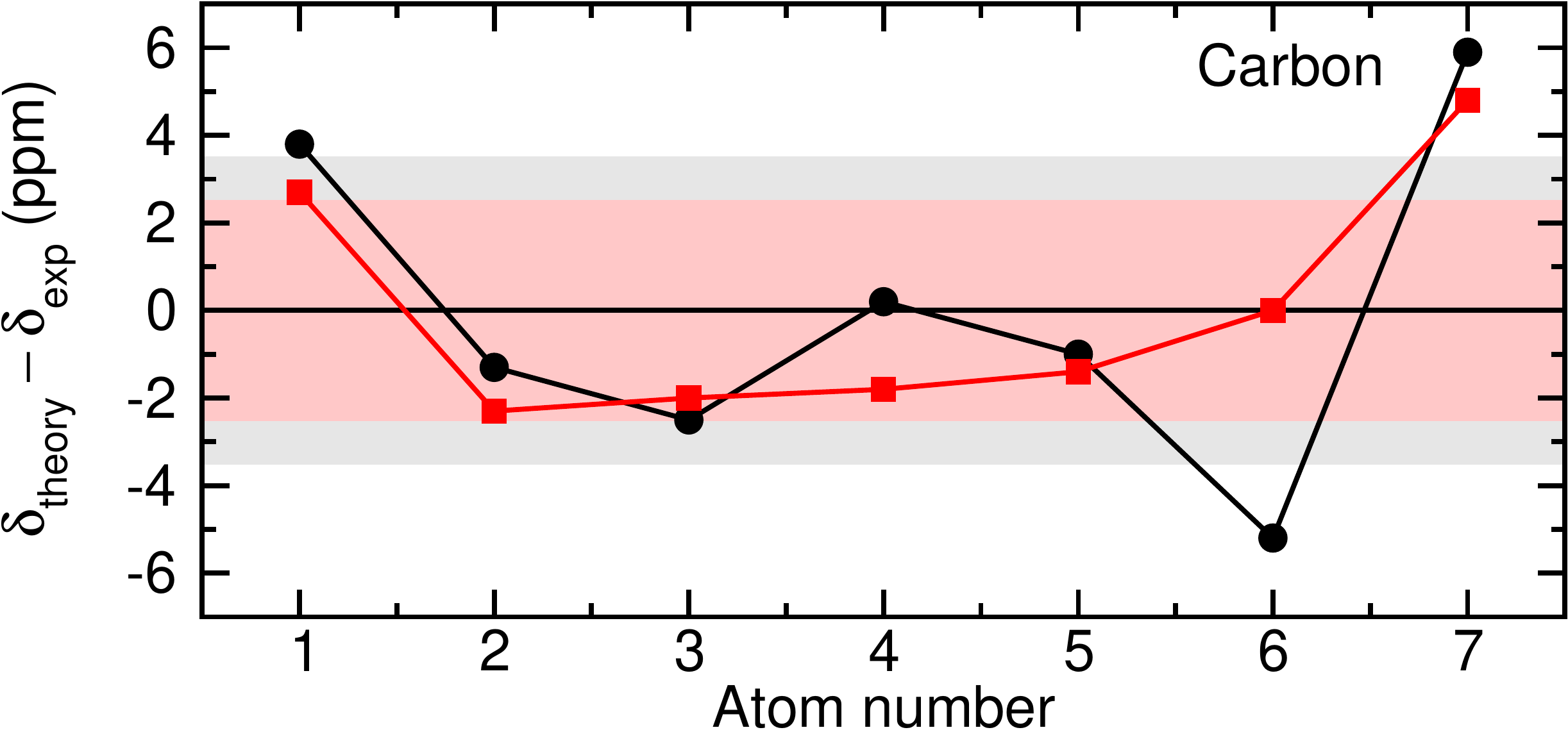}
\caption{Comparison between the theoretical and experimental chemical shifts of hydrogen and carbon in bDA. We report the theoretical shift calculated using the static lattice approximation (black circles) and using the quadratic approximation at $T=293$~K (red squares). The bands represent the root mean square deviation of the theoretical data compared to experiment. The solid lines are a guide to the eye only. Experimental results are from Ref.~\onlinecite{bDA_nmr_experiment}.}
\label{fig:bDA_exp}
\end{figure}

The experimental chemical shifts $\delta_{\mathrm{exp}}$ of hydrogen and carbon for bDA are reported in Ref.~\onlinecite{bDA_nmr_experiment}. The chemical shifts $\delta$, which are the quantities readily available experimentally, are defined with respect to a reference isotropic shift $\sigma_{\mathrm{ref}}$ according to
\begin{equation}
\delta=\sigma_{\mathrm{ref}}-\sigma_{\mathrm{iso}}.
\end{equation}
We have determined the value of $\sigma_{\mathrm{ref}}$ by fitting the function $f(x) = a - x$ to the experimental shifts against the theoretical isotropic shifts, with a single fitting parameter $a$. It follows that $\sigma_{\mathrm{ref}}=a$, and we have determined different references for the static lattice approximation and for the finite temperature quadratic approximation results. 

In Table~\ref{tab:bDA_exp_iso} we show experimental chemical shifts for bDA compared to the theoretical ones obtained using the static lattice approximation and the quadratic approximation at the experimental temperature of $T=293$~K. In each case we also report the reference shielding evaluated as described above. In Fig.~\ref{fig:bDA_exp} we show the difference between the theoretical and experimental chemical shifts for the calculations performed within the static lattice approximation (black circles) and at $T=293$~K within the quadratic approximation (red squares).  We also show the root mean square deviation (RMSD) of the theoretical data by coloured bands. For hydrogen, the RMSD decreases from $0.49$~ppm for the static lattice calculation to $0.21$~ppm when the effects of temperature are included. Therefore, the inclusion of temperature effects leads to a significantly better agreement with experiment. For carbon the improvement achieved by including the effects of temperature is more moderate: the RMSD is $3.5$~ppm for the static lattice calculations, and decreases to $2.5$~ppm when vibrational effects are taken into account.

\begin{table*}[t]
\setlength{\tabcolsep}{12pt} 
  \caption{Comparison of experimental and theoretical chemical shifts for bDA, in units of ppm. The experimental uncertainty is in the range $0.01$--$0.15$~ppm for H and $0.1$~ppm for C.}
\label{tab:bDA_exp_iso}
\begin{tabular}{ccccc}
\hline
\hline
\textbf{Species} &  \textbf{Atom number} & \textbf{Experiment}  &  \textbf{Theory ($\mathbf{T=293}$~K)}  & \textbf{Theory (static)} \\
\hline
H  &  1   & $14.01$   &  $14.11$ & $14.90$ \\
  &  2   &  $\phantom{0}8.34$   &   $\phantom{0}8.57$ & $\phantom{0}8.83$ \\
  &  3   &  $\phantom{0}3.23$   &   $\phantom{0}3.04$ & $\phantom{0}2.90$ \\
  &  4   &  $\phantom{0}4.40$   &   $\phantom{0}4.15$ & $\phantom{0}4.08$ \\
  &  5-7   &  $\phantom{0}7.94$   &   $\phantom{0}8.14$ & $\phantom{0}8.29$ \\
  &  8   &  $\phantom{0}2.70$   &   $\phantom{0}2.41$ & $\phantom{0}2.16$ \\
  &  9   &  $\phantom{0}5.32$   &   $\phantom{0}5.27$ & $\phantom{0}5.28$ \\
  &  10-12  &  $\phantom{0}1.27$   &   $\phantom{0}1.52$ & $\phantom{0}0.77$ \\
Reference & & & $29.85$ & $30.55$ \\
\hline
C  &  1   & $175.7$  & $178.4$  & $179.5$  \\
  &  2   & $\phantom{0}52.8$   & $\phantom{0}50.5$   & $\phantom{0}51.5$  \\
  &  3   & $\phantom{0}40.4$   & $\phantom{0}38.4$   & $\phantom{0}37.9$  \\
  &  4   & $171.2$  & $169.4$  & $171.4$  \\
  &  5   & $\phantom{0}47.5$   & $\phantom{0}46.1$   & $\phantom{0}46.5$  \\
  &  6   & $\phantom{0}16.9$   & $\phantom{0}16.9$   & $\phantom{0}11.7$  \\
  &  7   & $175.7$  & $180.5$  & $181.6$  \\
Reference  & & & $164.0$ & $169.3$  \\
\hline
\hline
\end{tabular}
\end{table*}

\subsection{Convergence details}

\subsubsection{Convergence with simulation cell size} 

In this section we discuss the convergence of the vibrational correction to the chemical shielding tensor with respect to the size of the simulation cell used. This convergence is equivalent to the convergence with respect to the sampling of the phonon BZ. 

For the case of MgO, we have used a simulation cell with $16$ atoms containing $2\times2\times2$ primitive cells. In the case of the molecular crystals, the primitive cells of L-alanine and bDA contain $52$ and $104$ atoms respectively, and a simulation cell size convergence study would require a prohibitive amount of computational resources. Although such a study is not necessary for the arguments put forward in this work, we refer the reader to the work of Robinson and Haynes.\cite{nmr_force_fields} They found a significant effect of the simulation cell size on the calculation of chemical shifts in L-alanine using classical force fields fitted to quantum mechanical forces in molecular dynamics simulations. The larger size of the primitive cell of bDA suggests that simulation size effects might be smaller in this system than in L-alanine.

\subsubsection{Convergence of couplings with normal mode amplitude} \label{subsubsec:coupling_convergence}

\begin{figure}
\centering
\includegraphics[scale=0.35]{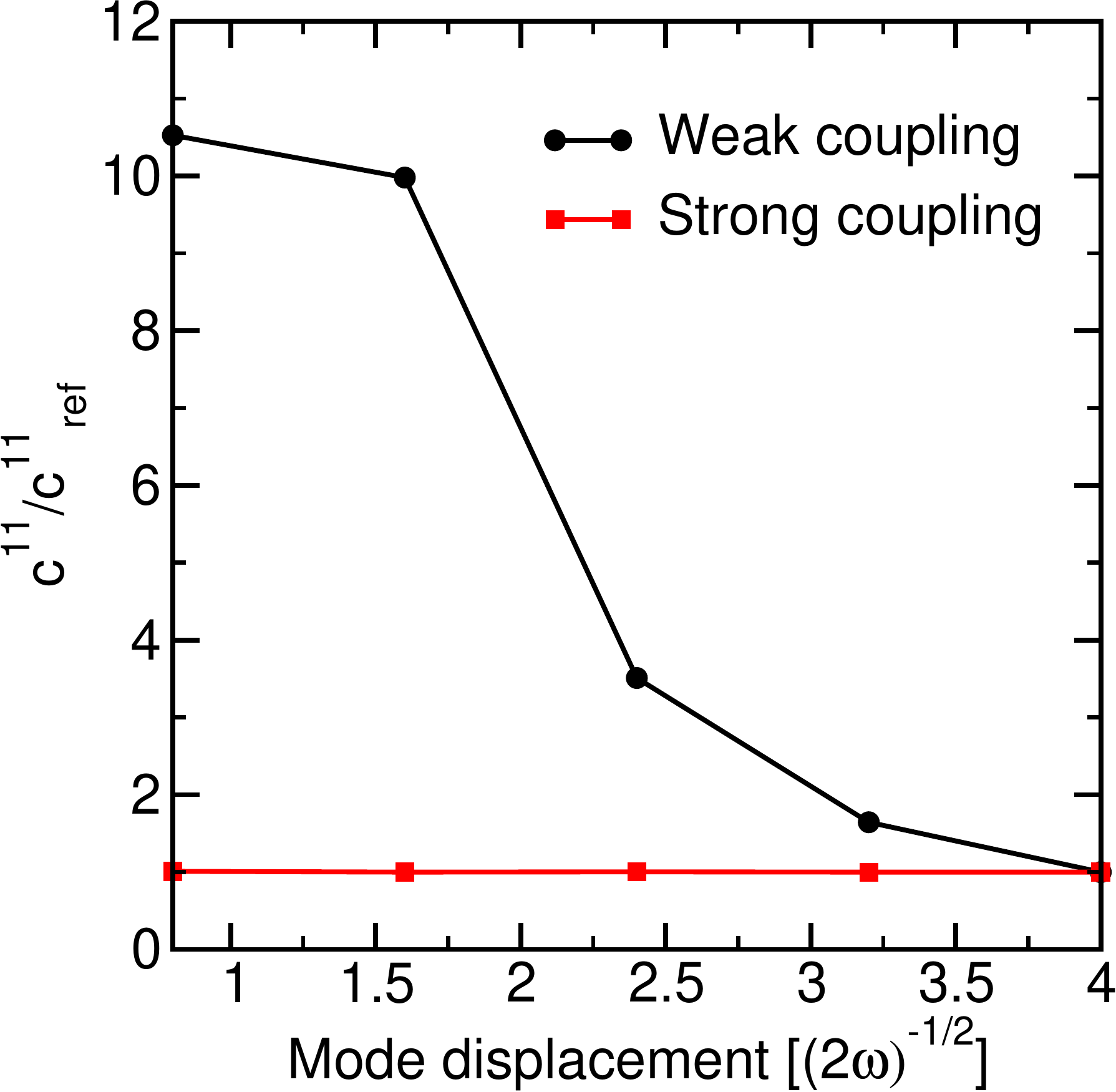}
\caption{Convergence of the coupling to the chemical shielding tensor $(1,1)$ component for a vibrational mode with large coupling (red squares) and small coupling (black circles). The reference $c^{11}_{\mathrm{ref}}$ is the coupling for mode amplitude $4/\sqrt{2\omega}$.}
\label{fig:convergence}
\end{figure}

The use of the quadratic expansion for calculating the vibrational coupling to the chemical shielding tensor requires a careful analysis of the amplitude of the vibrational mode at which the sampling points are taken. In this section we give some details of this using bDA as an example.

For all atoms, the coupling strength is large for a small number of vibrational modes (about $10$), and is small for the rest. The small size of the coupling strength for the vast majority of vibrational modes can lead to problems with numerical noise. In Fig.~\ref{fig:convergence} we plot the coupling to the $(1,1)$ component of the chemical shielding tensor of a representative hydrogen atom as a function of the normal mode amplitude at which the coupling is calculated. The data is normalised with respect to the largest amplitude considered, corresponding to $4/\sqrt{2\omega}$. The red squares correspond to data for one of the modes that has a strong coupling to $c^{11}$, and we can see that the coupling strength is independent of the amplitude at which it is calculated for the range considered. In contrast, the black circles, which correspond to a mode with weak coupling, show a strong dependence of the value of the coupling on the amplitude. Although this coupling is weak, in bDA there are about $300$ such modes, all with similar behaviour, which means that the error in the coupling of one mode increases by two orders of magnitude in the final result, and this leads to a significant error in the final result. As the coupling strength of the modes with strong coupling is largely independent of the sampling amplitude, the strategy we follow is to choose the sampling amplitude as that at which the weak-coupling modes are converged. In the case of bDA, an amplitude of $4/\sqrt{2\omega}$ leads to converged results (see Fig.~\ref{fig:convergence}).

This behaviour could be expected in most molecular crystals with a large number of atoms in the primitive cell, as these systems are expected to have a small range of modes with strong coupling while the majority of modes have weak coupling. Therefore, any calculation using the quadratic expansion should be preceded by a convergence test to estimate the sampling amplitude required for converged results. For this convergence test, a weak-coupling and a strong-coupling mode should suffice. For any given atom, these can be identified by investigating the atomic motion of the vibrational normal modes, and selecting a mode for which the atom of interest moves significantly, and one in which it does not.

\section{Anharmonic vibrations} \label{sec:anharmonic}

The results presented in Sec.~\ref{sec:harmonic} show very good agreement between the quadratic approximation and the Monte Carlo sampling when the vibrational wave function is treated within the harmonic approximation. This suggests that quartic terms in the expansion in Eq.~(\ref{eq:expansion}) are not important for the description of the coupling of the chemical shielding tensor with the vibrational state of the solid. However, when the vibrations of the solid have an anharmonic asymmetric component, odd terms in the expansion in Eq.~(\ref{eq:expansion}) may become important. Recent work has made the incorporation of anharmonic vibrations in first-principles calculations of solids possible.\cite{PhysRevLett.100.095901,PhysRevLett.106.165501,PhysRevB.84.180301,PhysRevB.86.054119,PhysRevB.87.144302,errea_prl,helium,errea_prb,prl_dissociation_hydrogen} In this section we use the method described in Ref.~\onlinecite{PhysRevB.87.144302} to obtain an anharmonic vibrational wave function for L-alanine, and use it to evaluate the coupling of the anharmonic vibrational state to the chemical shielding tensor.

\subsection{Anharmonic vibrations}

The harmonic vibrational energy of L-alanine, calculated using only $\Gamma$-point vibrations, is $\epsilon_{\mathrm{har}}=228.2$~meV/atom. The anharmonic energy is only slightly larger, at $\epsilon_{\mathrm{har}}=228.8$~meV/atom, leading to an anharmonic correction to the ZP vibrational energy of only $0.6$~meV/atom. The vibrational modes that dominate the anharmonic correction to the energy correspond to bond stretching vibrations of individual hydrogen atoms.  


\begin{figure}
\centering
\includegraphics[scale=0.7]{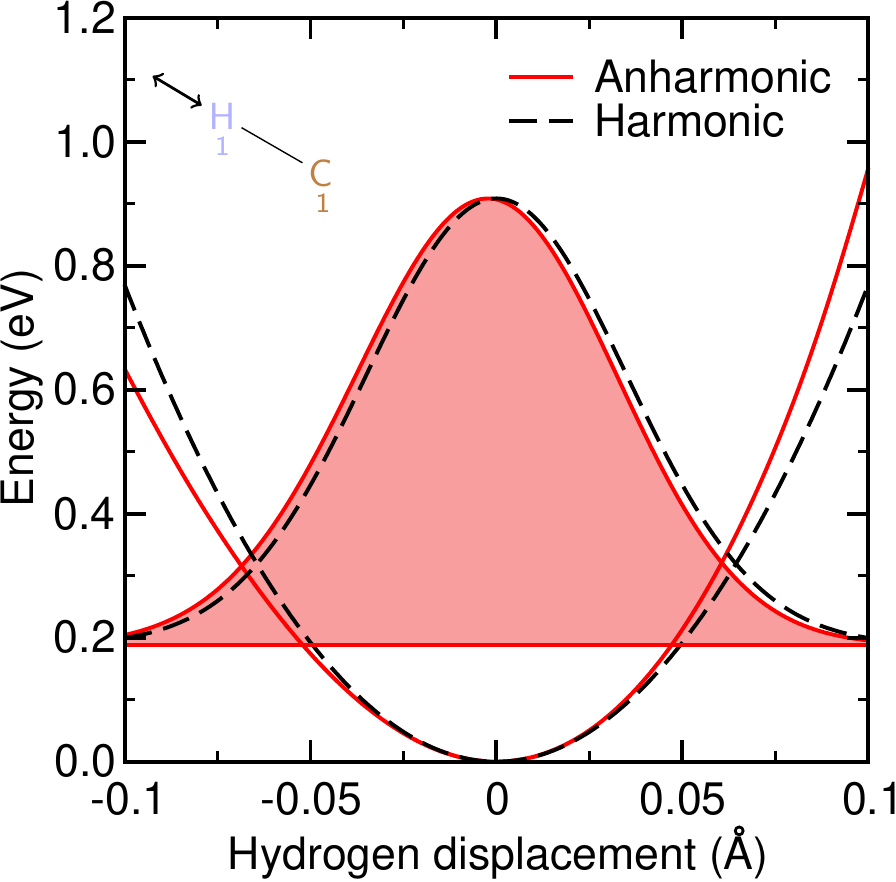}
\caption{Harmonic (black dashed lines) and anharmonic (solid red lines) potentials and ground state densities for the stretching mode of hydrogen atom $1$ (shown schematically in the inset). The harmonic and anharmonic energies are very similar, at $188$~meV.}
\label{fig:anh_pot}
\end{figure}

It is interesting to note that, although the anharmonic contribution to the vibrational energy is small, the anharmonic vibrational wave function of some atoms deviates from the corresponding harmonic wave function by acquiring an asymmetric component. An example of such a feature is shown in Fig.~\ref{fig:anh_pot}, in which we plot the harmonic and anharmonic potentials and densities. This particular example corresponds to the bond-stretching vibrations of hydrogen atom $1$. When the hydrogen atom approaches the carbon atom (positive displacements), the anharmonic potential is steeper than the harmonic one, whereas when the hydrogen atom moves away from the carbon atom, the anharmonic potential becomes shallower. The anharmonic vibrational density shifts towards the shallower part of the potential, acquiring an asymmetric component.

\subsection{Anharmonic coupling to the chemical shielding tensor}

In order to calculate the renormalized shielding tensor using an anharmonic vibrational wave function, we reuse the data points calculated in Sec.~\ref{sec:harmonic} for the Monte Carlo calculation sampled from a harmonic vibrational wave function. For the ZP correction, the expectation value reads 
\begin{eqnarray}
\langle\bm{\sigma}\rangle&=&\int d\mathbf{q}\,|\Phi_{\mathrm{anh}}(\mathbf{q})|^2\bm{\sigma}(\mathbf{q}) \nonumber \\ 
&=&\int d\mathbf{q}\,\frac{|\Phi_{\mathrm{anh}}(\mathbf{q})|^2}{|\Phi_{\mathrm{har}}(\mathbf{q})|^2}|\Phi_{\mathrm{har}}(\mathbf{q})|^2\bm{\sigma}(\mathbf{q}), \label{eq:anh}
\end{eqnarray}
and within a Monte Carlo integration scheme it may be evaluated as
\begin{equation}
\langle\bm{\sigma}\rangle_{\mathrm{MC}}\simeq\frac{1}{M}\sum_{i=1}^{M}\frac{|\Phi_{\mathrm{anh}}(\mathbf{q}_i)|^2}{|\Phi_{\mathrm{har}}(\mathbf{q}_i)|^2}\bm{\sigma}(\mathbf{q}_i),
\end{equation}
where the data points $\mathbf{q}_i$ are distributed according to the harmonic density $|\Phi_{\mathrm{har}}|^2$.
Therefore, the data points calculated in Sec.~\ref{sec:harmonic} that were distributed according to $|\Phi_{\mathrm{har}}|^2$ may be reused with the inclusion of the appropriate weighting $|\Phi_{\mathrm{anh}}|^2/|\Phi_{\mathrm{har}}|^2$ to obtain the expectation value in Eq.~(\ref{eq:anh}). This weighting procedure leads to an increase in the statistical uncertainty of the weighted integral, and this is the reason why we used a very large number of sampling points in the case of L-alanine. The large number of sampling points is also necessary to ensure that the sampling of the tails of the anharmonic distribution is correct.

\begin{figure}
\centering
\includegraphics[scale=0.35]{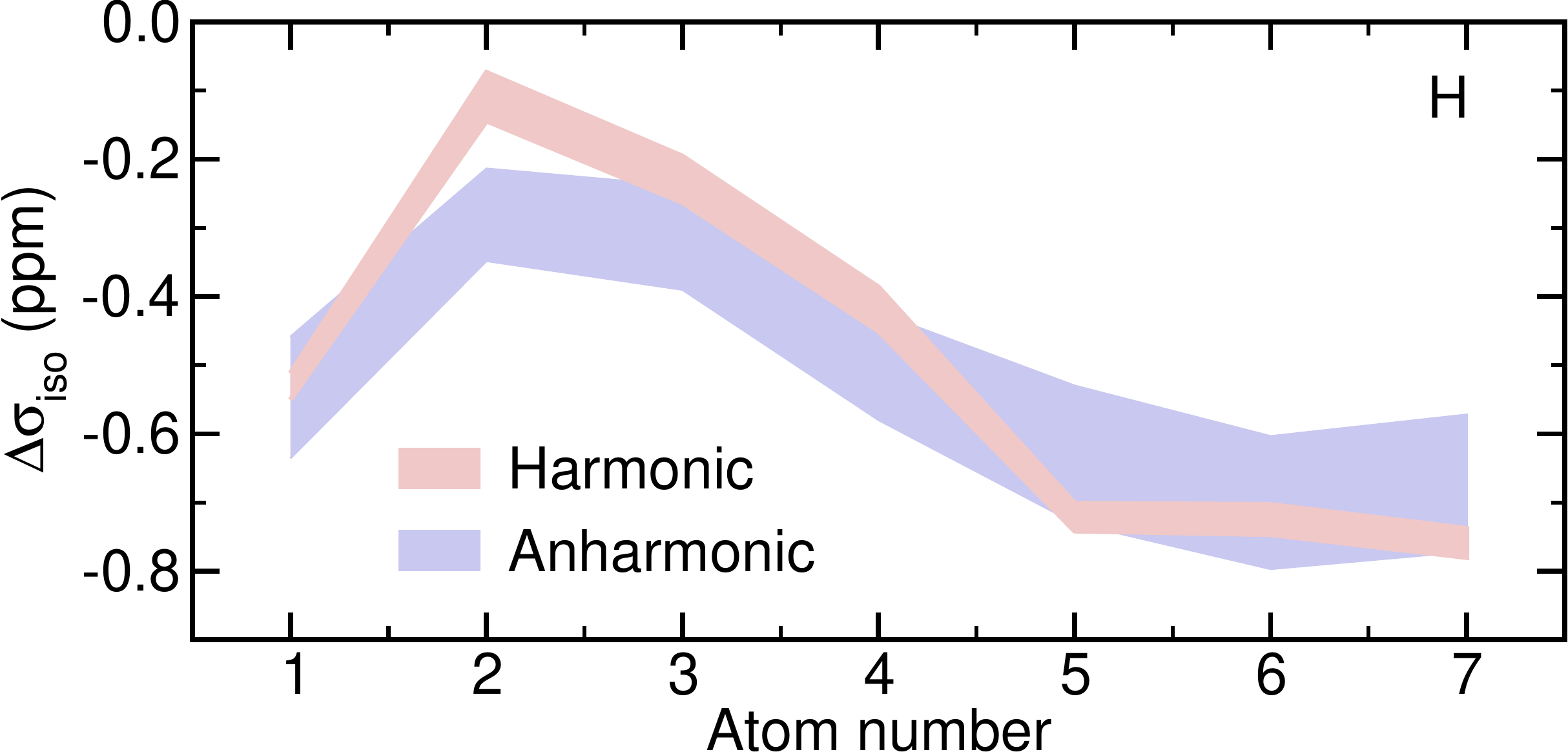}
\includegraphics[scale=0.35]{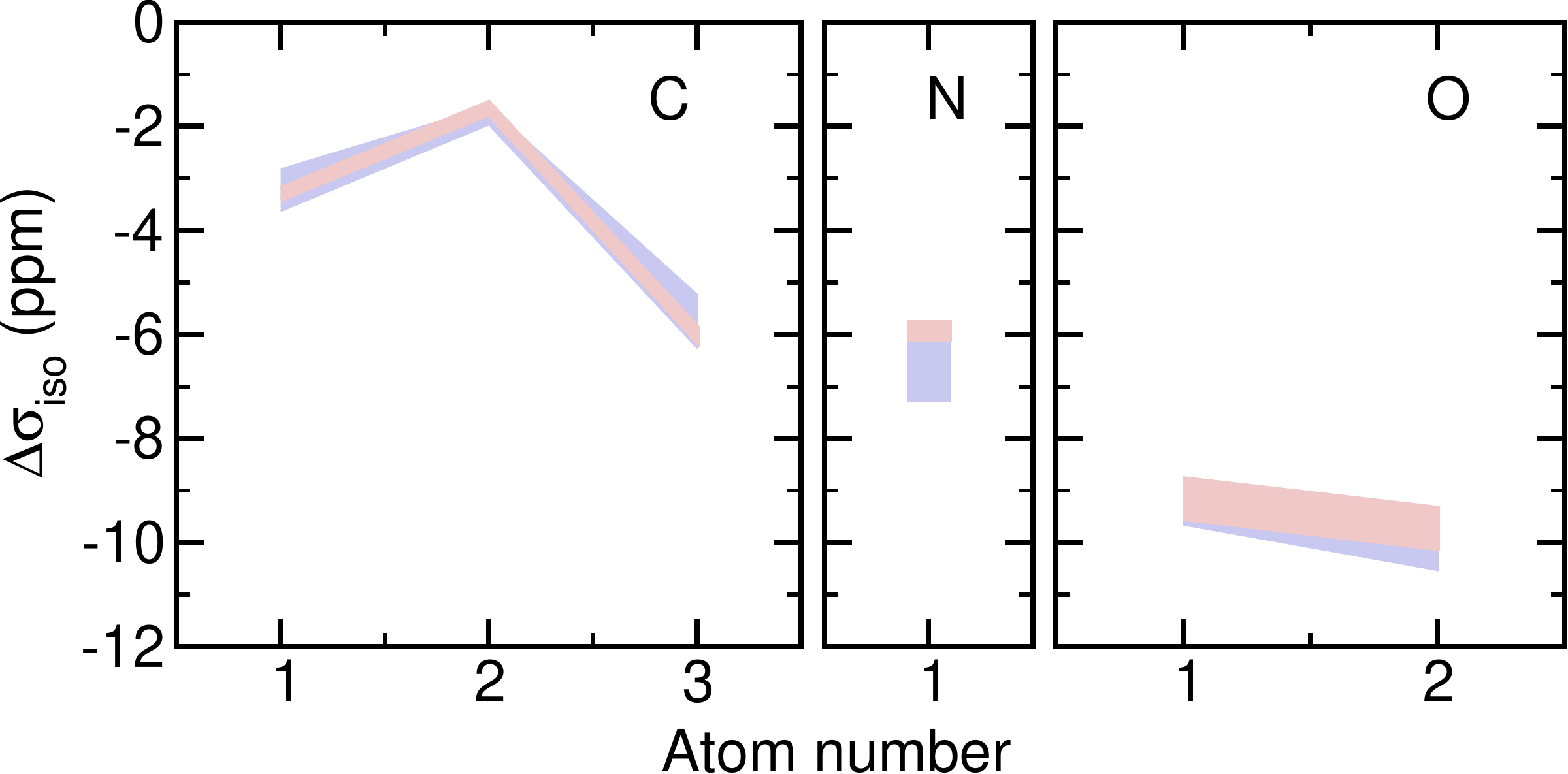}
\caption{ZP correction to the isotropic chemical shift from the static lattice value for L-alanine. The light red bands correspond to the use of a harmonic vibrational wave function, and the light blue bands to the use of an anharmonic vibrational wave function. The links between atom numbers are only an aid to the eye.}
\label{fig:alanine_iso_anh}
\end{figure}

In Fig.~\ref{fig:alanine_iso_anh} we show the ZP correction to the isotropic chemical shift of L-alanine calculated using the Monte Carlo sampling approach with harmonic (light red bands) and anharmonic (light blue bands) vibrational wave functions. The use of a more accurate anharmonic vibrational wave function leads to small differences in the isotropic shift. 
We can observe a small increase in the ZP correction to the isotropic shift of the hydrogen atoms belonging to the CH$_3$ methyl group, and a small decrease in those belonging to the NH$_3$ group, when an anharmonic wave function is used. However, these differences are very small, therefore the overall picture is that anharmonic corrections are small, even for the lightest elements, and may therefore be safely neglected.

\begin{figure}
\centering
\includegraphics[scale=0.35]{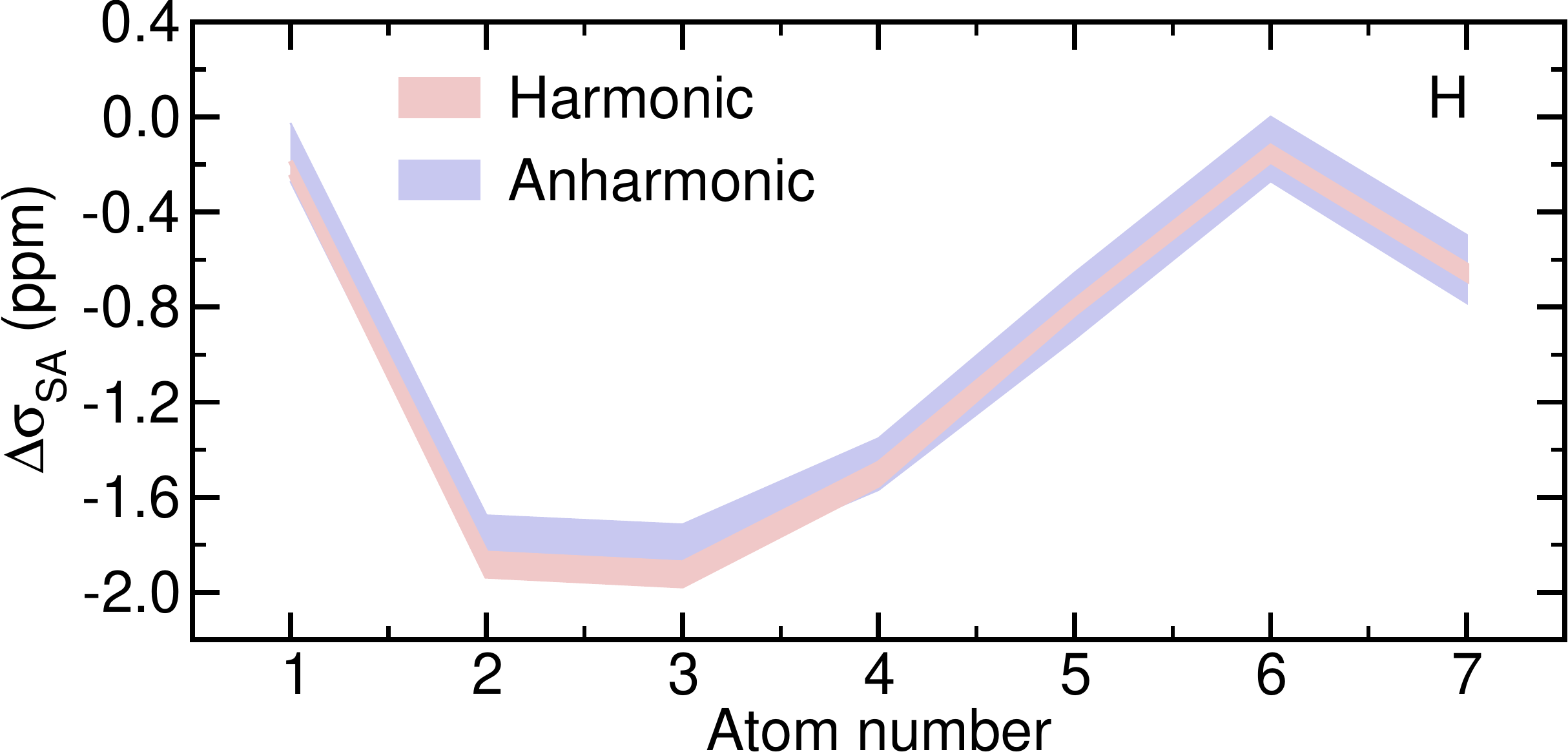}
\includegraphics[scale=0.35]{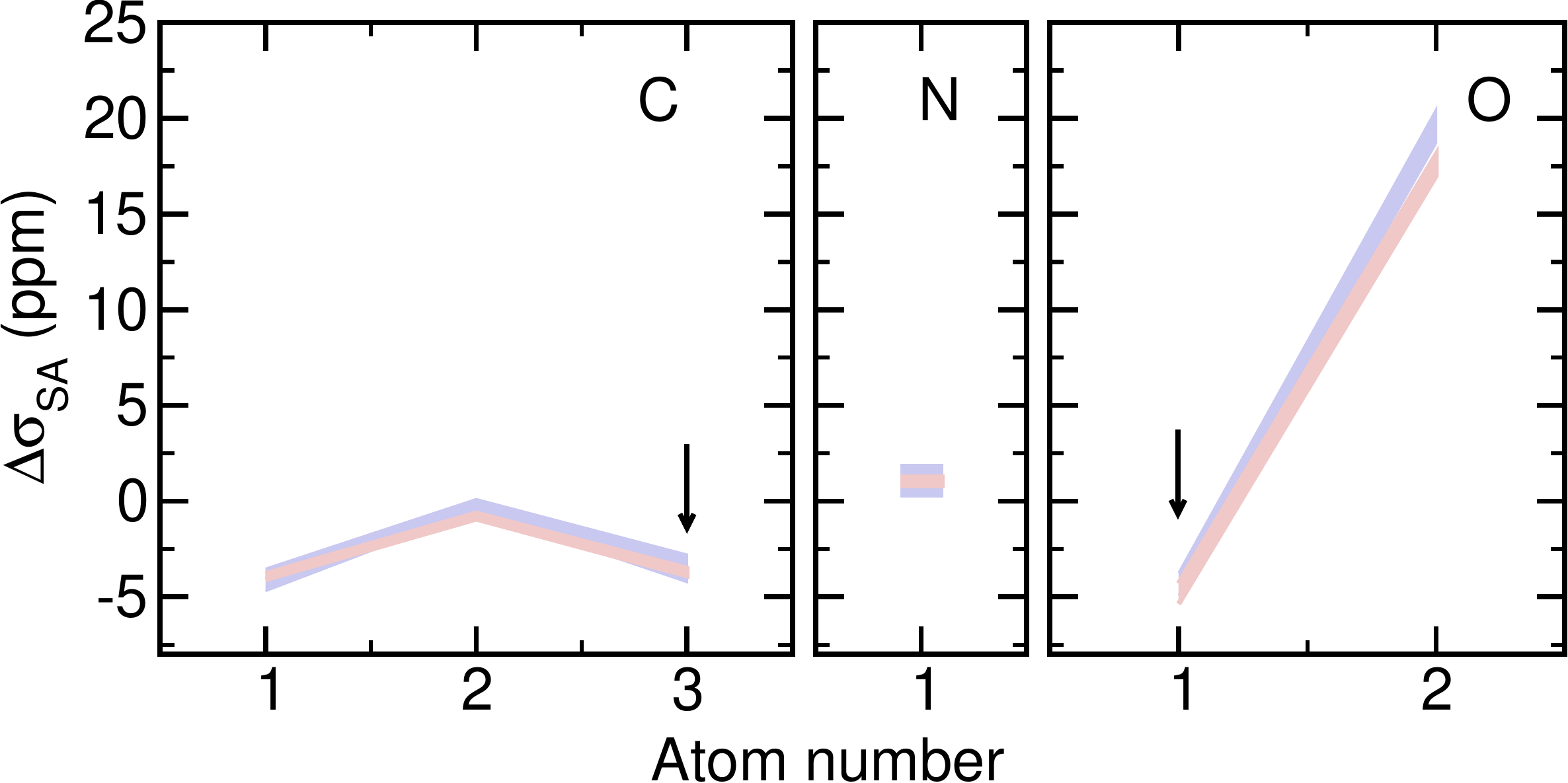}
\caption{ZP correction to the shielding anisotropy from the static lattice value of L-alanine. The light red bands correspond to the use of a harmonic vibrational wave function, and the light blue bands to the use of an anharmonic vibrational wave function. The arrows indicate the atoms for which the shielding anisotropy changes sign from the static lattice value to the vibrationally averaged value. The links between atom numbers are only an aid to the eye.}
\label{fig:alanine_ani_anh}
\end{figure}

In Fig.~\ref{fig:alanine_ani_anh} we show the ZP correction to the shielding anisotropy of L-alanine calculated with a harmonic (light red bands) and an anharmonic (light blue bands) wave function. The inclusion of anharmonic terms in the description of vibrations also has a small effect on the off-diagonal part of the chemical shielding tensor.


\section{Discussion and conclusions} \label{sec:conclusions}

\begin{figure}
\centering
\includegraphics[scale=0.35]{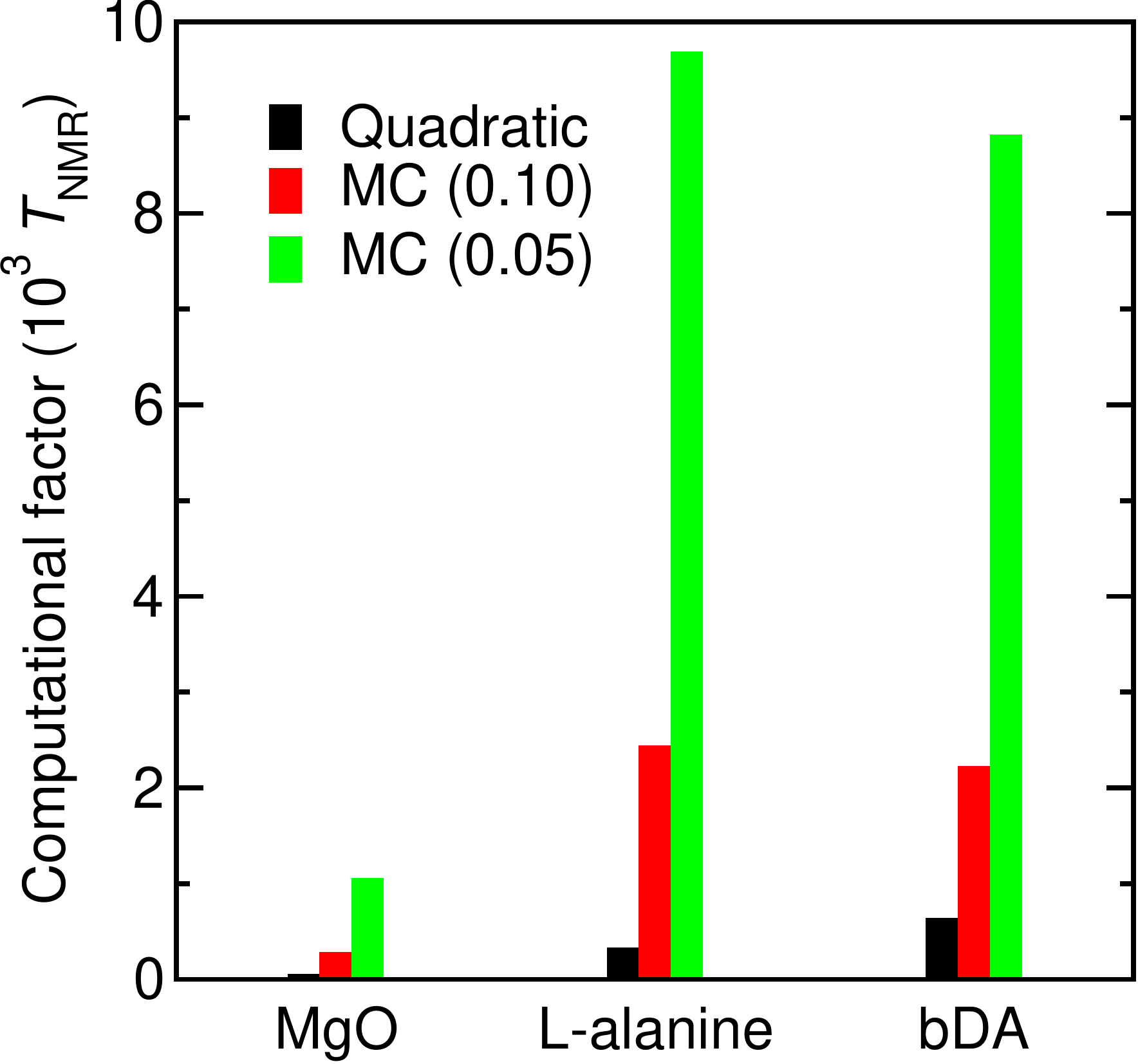}
\caption{Computational cost of the quadratic and Monte Carlo methods in terms of the computational cost $T_{\mathrm{NMR}}$ of a single NMR calculation. Two values for the Monte Carlo sampling approach are reported, corresponding to uncertainties of $0.1$ (red) and $0.05$ (green) of the overall ZP correction.}
\label{fig:speedup}
\end{figure}

We have investigated two approaches for studying the effects of nuclear vibrations on the chemical shielding tensor of solids. The first is based on the parametrising the quadratic approximation to the coupling of the chemical shielding tensor to the vibrational state of a solid, which leads to a scheme requiring moderate computational resources compared to other approaches. The second consists on calculating the expectation value of the chemical shielding tensor with respect to the vibrational wave function by means of Monte Carlo integration. We have tested them using MgO, L-alanine, and bDA as model systems. 

At the harmonic level, we have found excellent agreement between the quadratic approximation and Monte Carlo sampling, demonstrating the practical usefulness of the former. In all cases for which experimental data was available, we have found an improved agreement between theory and experiment when the effects of temperature are included in the calculations. We have also shown that anharmonic vibrations have a small influence on the effects of vibrations on the chemical shielding tensor, at least for the systems studied here. It would be interesting to investigate a wider range of materials to discover systems with important anharmonic contributions. 

The quadratic and Monte Carlo approaches are computationally trivially parallelizable as the sampling points are independent, unlike the correlated paths of the dynamical methods that require a serial evaluation. This means that the computational time of the quadratic and Monte Carlo methods can be expressed as a multiple of the computational time $T_{\mathrm{NMR}}$ for a single chemical shielding tensor calculation. The calculation times for MgO, L-alanine, and bDA are shown in Fig.~\ref{fig:speedup}. 
As the system size increases, the dimensionality of the vibrational phase space to be explored also increases, and Monte Carlo sampling should eventually become the most efficient method. However, the results in Fig.~\ref{fig:speedup} show that for the system sizes studied (up to $104$ atoms in the simulation cell) the benefits of using the quadratic approximation are significant.

The results reported in this work suggest the following scenario. For an accurate treatment of the effects of vibrations (and therefore of temperature) on the chemical shielding tensor it is necessary to include quantum zero-point effects as these are of a similar size to thermal effects even at temperatures as high as $500$~K. The quadratic approach is capable of including these effects while treating vibrations at the harmonic level, which has been shown to provide very accurate results. Furthermore, the computational benefits of using the quadratic method allow us to propose it as the method of choice for a systematic inclusion of the effects of vibrations in calculations of the chemical shielding tensor from first principles in solids.


\acknowledgements

We acknowledge enlightening discussions with Pascal Bugnion, Jonathan Lloyd-Williams, Neil Drummond, and Jonathan Yates. Financial support was provided by the Engineering and Physical Sciences Research Council (UK). The calculations were performed on the Cambridge High Performance Computing Service facility and the Archer facility of the UK's national high-performance computing service (for which access was obtained via the UKCP consortium).

\bibliography{manuscript.bbl}

\begin{thebibliography}{48}%
\makeatletter
\providecommand \@ifxundefined [1]{%
 \@ifx{#1\undefined}
}%
\providecommand \@ifnum [1]{%
 \ifnum #1\expandafter \@firstoftwo
 \else \expandafter \@secondoftwo
 \fi
}%
\providecommand \@ifx [1]{%
 \ifx #1\expandafter \@firstoftwo
 \else \expandafter \@secondoftwo
 \fi
}%
\providecommand \natexlab [1]{#1}%
\providecommand \enquote  [1]{``#1''}%
\providecommand \bibnamefont  [1]{#1}%
\providecommand \bibfnamefont [1]{#1}%
\providecommand \citenamefont [1]{#1}%
\providecommand \href@noop [0]{\@secondoftwo}%
\providecommand \href [0]{\begingroup \@sanitize@url \@href}%
\providecommand \@href[1]{\@@startlink{#1}\@@href}%
\providecommand \@@href[1]{\endgroup#1\@@endlink}%
\providecommand \@sanitize@url [0]{\catcode `\\12\catcode `\$12\catcode
  `\&12\catcode `\#12\catcode `\^12\catcode `\_12\catcode `\%12\relax}%
\providecommand \@@startlink[1]{}%
\providecommand \@@endlink[0]{}%
\providecommand \url  [0]{\begingroup\@sanitize@url \@url }%
\providecommand \@url [1]{\endgroup\@href {#1}{\urlprefix }}%
\providecommand \urlprefix  [0]{URL }%
\providecommand \Eprint [0]{\href }%
\@ifxundefined \urlstyle {%
  \providecommand \doi  [0]{\begingroup \@sanitize@url \@doi}%
  \providecommand \@doi [1]{\endgroup \@@startlink {\doibase
  #1}doi:\discretionary {}{}{}#1\@@endlink }%
}{%
  \providecommand \doi  [0]{doi:\discretionary{}{}{}\begingroup
  \urlstyle{rm}\Url }%
}%
\providecommand \doibase [0]{http://dx.doi.org/}%
\providecommand \Doi [0]{\begingroup \@sanitize@url \@Doi }%
\providecommand \@Doi  [1]{\endgroup\@@startlink{\doibase#1}\@@Doi}%
\providecommand \@@Doi [1]{#1\@@endlink}%
\providecommand \selectlanguage [0]{\@gobble}%
\providecommand \bibinfo  [0]{\@secondoftwo}%
\providecommand \bibfield  [0]{\@secondoftwo}%
\providecommand \translation [1]{[#1]}%
\providecommand \BibitemOpen [0]{}%
\providecommand \bibitemStop [0]{}%
\providecommand \bibitemNoStop [0]{.\EOS\space}%
\providecommand \EOS [0]{\spacefactor3000\relax}%
\providecommand \BibitemShut  [1]{\csname bibitem#1\endcsname}%
\bibitem [{\citenamefont {Lesage}(2009)}]{exp_nmr_review}%
  \BibitemOpen
  \bibfield  {author} {\bibinfo {author} {\bibfnamefont {A.}~\bibnamefont
  {Lesage}},\ }\bibfield  {title} {\enquote {\bibinfo {title} {Recent advances
  in solid-state nmr spectroscopy of spin i = 1/2 nuclei},}\ }\href
  {http://dx.doi.org/10.1039/B907733M} {\bibfield  {journal} {\bibinfo
  {journal} {Phys. Chem. Chem. Phys.},\ }\textbf {\bibinfo {volume} {11}},\
  \bibinfo {pages} {6876} (\bibinfo {year} {2009})}\BibitemShut {NoStop}%
\bibitem [{\citenamefont {Ashbrook}(2009)}]{exp_nmr_review2}%
  \BibitemOpen
  \bibfield  {author} {\bibinfo {author} {\bibfnamefont {S.~E.}\ \bibnamefont
  {Ashbrook}},\ }\bibfield  {title} {\enquote {\bibinfo {title} {Recent
  advances in solid-state nmr spectroscopy of quadrupolar nuclei},}\ }\href
  {http://dx.doi.org/10.1039/B907183K} {\bibfield  {journal} {\bibinfo
  {journal} {Phys. Chem. Chem. Phys.},\ }\textbf {\bibinfo {volume} {11}},\
  \bibinfo {pages} {6892} (\bibinfo {year} {2009})}\BibitemShut {NoStop}%
\bibitem [{\citenamefont {Bonhomme}\ \emph {et~al.}(2012)\citenamefont
  {Bonhomme}, \citenamefont {Gervais}, \citenamefont {Babonneau}, \citenamefont
  {Coelho}, \citenamefont {Pourpoint}, \citenamefont {Azaïs}, \citenamefont
  {Ashbrook}, \citenamefont {Griffin}, \citenamefont {Yates}, \citenamefont
  {Mauri},\ and\ \citenamefont {Pickard}}]{gipaw_review}%
  \BibitemOpen
  \bibfield  {author} {\bibinfo {author} {\bibfnamefont {C.}~\bibnamefont
  {Bonhomme}}, \bibinfo {author} {\bibfnamefont {C.}~\bibnamefont {Gervais}},
  \bibinfo {author} {\bibfnamefont {F.}~\bibnamefont {Babonneau}}, \bibinfo
  {author} {\bibfnamefont {C.}~\bibnamefont {Coelho}}, \bibinfo {author}
  {\bibfnamefont {F.}~\bibnamefont {Pourpoint}}, \bibinfo {author}
  {\bibfnamefont {T.}~\bibnamefont {Azaïs}}, \bibinfo {author} {\bibfnamefont
  {S.~E.}\ \bibnamefont {Ashbrook}}, \bibinfo {author} {\bibfnamefont {J.~M.}\
  \bibnamefont {Griffin}}, \bibinfo {author} {\bibfnamefont {J.~R.}\
  \bibnamefont {Yates}}, \bibinfo {author} {\bibfnamefont {F.}~\bibnamefont
  {Mauri}}, \ and\ \bibinfo {author} {\bibfnamefont {C.~J.}\ \bibnamefont
  {Pickard}},\ }\bibfield  {title} {\enquote {\bibinfo {title}
  {First-principles calculation of nmr parameters using the gauge including
  projector augmented wave method: A chemist’s point of view},}\ }\href
  {http://pubs.acs.org/doi/abs/10.1021/cr300108a} {\bibfield  {journal}
  {\bibinfo  {journal} {Chem. Rev.},\ }\textbf {\bibinfo {volume} {112}},\
  \bibinfo {pages} {5733} (\bibinfo {year} {2012})}\BibitemShut {NoStop}%
\bibitem [{\citenamefont {Florian}\ and\ \citenamefont
  {Massiot}(2013)}]{exp_nmr_review_disorder}%
  \BibitemOpen
  \bibfield  {author} {\bibinfo {author} {\bibfnamefont {P.}~\bibnamefont
  {Florian}}\ and\ \bibinfo {author} {\bibfnamefont {D.}~\bibnamefont
  {Massiot}},\ }\bibfield  {title} {\enquote {\bibinfo {title} {Beyond
  periodicity: probing disorder in crystalline materials by solid-state nuclear
  magnetic resonance spectroscopy},}\ }\href
  {http://dx.doi.org/10.1039/C3CE40982A} {\bibfield  {journal} {\bibinfo
  {journal} {Cryst. Eng. Comm.},\ }\textbf {\bibinfo {volume} {15}},\ \bibinfo
  {pages} {8623} (\bibinfo {year} {2013})}\BibitemShut {NoStop}%
\bibitem [{\citenamefont {Hohenberg}\ and\ \citenamefont
  {Kohn}(1964)}]{PhysRev.136.B864}%
  \BibitemOpen
  \bibfield  {author} {\bibinfo {author} {\bibfnamefont {P.}~\bibnamefont
  {Hohenberg}}\ and\ \bibinfo {author} {\bibfnamefont {W.}~\bibnamefont
  {Kohn}},\ }\bibfield  {title} {\enquote {\bibinfo {title} {Inhomogeneous
  electron gas},}\ }\href {http://link.aps.org/doi/10.1103/PhysRev.136.B864}
  {\bibfield  {journal} {\bibinfo  {journal} {Phys. Rev.},\ }\textbf {\bibinfo
  {volume} {136}},\ \bibinfo {pages} {B864} (\bibinfo {year}
  {1964})}\BibitemShut {NoStop}%
\bibitem [{\citenamefont {Kohn}\ and\ \citenamefont
  {Sham}(1965)}]{PhysRev.140.A1133}%
  \BibitemOpen
  \bibfield  {author} {\bibinfo {author} {\bibfnamefont {W.}~\bibnamefont
  {Kohn}}\ and\ \bibinfo {author} {\bibfnamefont {L.~J.}\ \bibnamefont
  {Sham}},\ }\bibfield  {title} {\enquote {\bibinfo {title} {Self-consistent
  equations including exchange and correlation effects},}\ }\href
  {http://link.aps.org/doi/10.1103/PhysRev.140.A1133} {\bibfield  {journal}
  {\bibinfo  {journal} {Phys. Rev.},\ }\textbf {\bibinfo {volume} {140}},\
  \bibinfo {pages} {A1133} (\bibinfo {year} {1965})}\BibitemShut {NoStop}%
\bibitem [{\citenamefont {Payne}\ \emph {et~al.}(1992)\citenamefont {Payne},
  \citenamefont {Teter}, \citenamefont {Allan}, \citenamefont {Arias},\ and\
  \citenamefont {Joannopoulos}}]{dft_rev_mod_phys}%
  \BibitemOpen
  \bibfield  {author} {\bibinfo {author} {\bibfnamefont {M.~C.}\ \bibnamefont
  {Payne}}, \bibinfo {author} {\bibfnamefont {M.~P.}\ \bibnamefont {Teter}},
  \bibinfo {author} {\bibfnamefont {D.~C.}\ \bibnamefont {Allan}}, \bibinfo
  {author} {\bibfnamefont {T.~A.}\ \bibnamefont {Arias}}, \ and\ \bibinfo
  {author} {\bibfnamefont {J.~D.}\ \bibnamefont {Joannopoulos}},\ }\bibfield
  {title} {\enquote {\bibinfo {title} {Iterative minimization techniques for ab
  initio total-energy calculations: molecular dynamics and conjugate
  gradients},}\ }\href {http://link.aps.org/doi/10.1103/RevModPhys.64.1045}
  {\bibfield  {journal} {\bibinfo  {journal} {Rev. Mod. Phys.},\ }\textbf
  {\bibinfo {volume} {64}},\ \bibinfo {pages} {1045} (\bibinfo {year}
  {1992})}\BibitemShut {NoStop}%
\bibitem [{\citenamefont {Pickard}\ and\ \citenamefont {Mauri}(2001)}]{gipaw}%
  \BibitemOpen
  \bibfield  {author} {\bibinfo {author} {\bibfnamefont {C.~J.}\ \bibnamefont
  {Pickard}}\ and\ \bibinfo {author} {\bibfnamefont {F.}~\bibnamefont
  {Mauri}},\ }\bibfield  {title} {\enquote {\bibinfo {title} {All-electron
  magnetic response with pseudopotentials: Nmr chemical shifts},}\ }\href
  {http://link.aps.org/doi/10.1103/PhysRevB.63.245101} {\bibfield  {journal}
  {\bibinfo  {journal} {Phys. Rev. B},\ }\textbf {\bibinfo {volume} {63}},\
  \bibinfo {pages} {245101} (\bibinfo {year} {2001})}\BibitemShut {NoStop}%
\bibitem [{\citenamefont {Yates}\ \emph {et~al.}(2007)\citenamefont {Yates},
  \citenamefont {Pickard},\ and\ \citenamefont {Mauri}}]{gipaw_ultrasoft}%
  \BibitemOpen
  \bibfield  {author} {\bibinfo {author} {\bibfnamefont {J.~R.}\ \bibnamefont
  {Yates}}, \bibinfo {author} {\bibfnamefont {C.~J.}\ \bibnamefont {Pickard}},
  \ and\ \bibinfo {author} {\bibfnamefont {F.}~\bibnamefont {Mauri}},\
  }\bibfield  {title} {\enquote {\bibinfo {title} {Calculation of nmr chemical
  shifts for extended systems using ultrasoft pseudopotentials},}\ }\href
  {http://link.aps.org/doi/10.1103/PhysRevB.76.024401} {\bibfield  {journal}
  {\bibinfo  {journal} {Phys. Rev. B},\ }\textbf {\bibinfo {volume} {76}},\
  \bibinfo {pages} {024401} (\bibinfo {year} {2007})}\BibitemShut {NoStop}%
\bibitem [{\citenamefont {Gervais}\ \emph {et~al.}(2004)\citenamefont
  {Gervais}, \citenamefont {Profeta}, \citenamefont {Lafond}, \citenamefont
  {Bonhomme}, \citenamefont {Azaïs}, \citenamefont {Mutin}, \citenamefont
  {Pickard}, \citenamefont {Mauri},\ and\ \citenamefont
  {Babonneau}}]{gipaw_example_organic}%
  \BibitemOpen
  \bibfield  {author} {\bibinfo {author} {\bibfnamefont {C.}~\bibnamefont
  {Gervais}}, \bibinfo {author} {\bibfnamefont {M.}~\bibnamefont {Profeta}},
  \bibinfo {author} {\bibfnamefont {V.}~\bibnamefont {Lafond}}, \bibinfo
  {author} {\bibfnamefont {C.}~\bibnamefont {Bonhomme}}, \bibinfo {author}
  {\bibfnamefont {T.}~\bibnamefont {Azaïs}}, \bibinfo {author} {\bibfnamefont
  {H.}~\bibnamefont {Mutin}}, \bibinfo {author} {\bibfnamefont {C.~J.}\
  \bibnamefont {Pickard}}, \bibinfo {author} {\bibfnamefont {F.}~\bibnamefont
  {Mauri}}, \ and\ \bibinfo {author} {\bibfnamefont {F.}~\bibnamefont
  {Babonneau}},\ }\bibfield  {title} {\enquote {\bibinfo {title} {Combined ab
  initio computational and experimental multinuclear solid-state magnetic
  resonance study of phenylphosphonic acid},}\ }\href
  {http://dx.doi.org/10.1002/mrc.1360} {\bibfield  {journal} {\bibinfo
  {journal} {Mag. Res. Chem.},\ }\textbf {\bibinfo {volume} {42}},\ \bibinfo
  {pages} {445} (\bibinfo {year} {2004})},\ ISSN \bibinfo {issn}
  {1097-458X}\BibitemShut {NoStop}%
\bibitem [{\citenamefont {Profeta}\ \emph {et~al.}(2003)\citenamefont
  {Profeta}, \citenamefont {Mauri},\ and\ \citenamefont
  {Pickard}}]{gipaw_example_inorganic}%
  \BibitemOpen
  \bibfield  {author} {\bibinfo {author} {\bibfnamefont {M.}~\bibnamefont
  {Profeta}}, \bibinfo {author} {\bibfnamefont {F.}~\bibnamefont {Mauri}}, \
  and\ \bibinfo {author} {\bibfnamefont {C.~J.}\ \bibnamefont {Pickard}},\
  }\bibfield  {title} {\enquote {\bibinfo {title} {Accurate first principles
  prediction of 17o nmr parameters in sio2:  assignment of the zeolite
  ferrierite spectrum},}\ }\href
  {http://pubs.acs.org/doi/abs/10.1021/ja027124r} {\bibfield  {journal}
  {\bibinfo  {journal} {J. Am. Chem. Soc.},\ }\textbf {\bibinfo {volume}
  {125}},\ \bibinfo {pages} {541} (\bibinfo {year} {2003})}\BibitemShut
  {NoStop}%
\bibitem [{\citenamefont {Charpentier}\ \emph {et~al.}(2004)\citenamefont
  {Charpentier}, \citenamefont {Ispas}, \citenamefont {Profeta}, \citenamefont
  {Mauri},\ and\ \citenamefont {Pickard}}]{gipaw_example_glass}%
  \BibitemOpen
  \bibfield  {author} {\bibinfo {author} {\bibfnamefont {T.}~\bibnamefont
  {Charpentier}}, \bibinfo {author} {\bibfnamefont {S.}~\bibnamefont {Ispas}},
  \bibinfo {author} {\bibfnamefont {M.}~\bibnamefont {Profeta}}, \bibinfo
  {author} {\bibfnamefont {F.}~\bibnamefont {Mauri}}, \ and\ \bibinfo {author}
  {\bibfnamefont {C.~J.}\ \bibnamefont {Pickard}},\ }\bibfield  {title}
  {\enquote {\bibinfo {title} {First-principles calculation of 17o, 29si, and
  23na nmr spectra of sodium silicate crystals and glasses},}\ }\href
  {http://pubs.acs.org/doi/abs/10.1021/jp0367225} {\bibfield  {journal}
  {\bibinfo  {journal} {J. Phys. Chem. B},\ }\textbf {\bibinfo {volume}
  {108}},\ \bibinfo {pages} {4147} (\bibinfo {year} {2004})}\BibitemShut
  {NoStop}%
\bibitem [{\citenamefont {Fr\"{u}chtl}\ \emph {et~al.}(2009)\citenamefont
  {Fr\"{u}chtl}, \citenamefont {van Mourik}, \citenamefont {Pickard},\ and\
  \citenamefont {Woollins}}]{gipaw_example_polymer1}%
  \BibitemOpen
  \bibfield  {author} {\bibinfo {author} {\bibfnamefont {H.}~\bibnamefont
  {Fr\"{u}chtl}}, \bibinfo {author} {\bibfnamefont {T.}~\bibnamefont {van
  Mourik}}, \bibinfo {author} {\bibfnamefont {C.}~\bibnamefont {Pickard}}, \
  and\ \bibinfo {author} {\bibfnamefont {J.}~\bibnamefont {Woollins}},\
  }\bibfield  {title} {\enquote {\bibinfo {title} {The structure of (scn)x: A
  study using molecular and solid-state density functional theory
  calculations},}\ }\href {http://dx.doi.org/10.1002/chem.200802075} {\bibfield
   {journal} {\bibinfo  {journal} {Chem. Eur. J.},\ }\textbf {\bibinfo {volume}
  {15}},\ \bibinfo {pages} {2687} (\bibinfo {year} {2009})},\ ISSN \bibinfo
  {issn} {1521-3765}\BibitemShut {NoStop}%
\bibitem [{\citenamefont {Filhol}\ \emph {et~al.}(2009)\citenamefont {Filhol},
  \citenamefont {Deschamps}, \citenamefont {Dutremez}, \citenamefont {Boury},
  \citenamefont {Barisien}, \citenamefont {Legrand},\ and\ \citenamefont
  {Schott}}]{gipaw_example_polymer2}%
  \BibitemOpen
  \bibfield  {author} {\bibinfo {author} {\bibfnamefont {J.-S.}\ \bibnamefont
  {Filhol}}, \bibinfo {author} {\bibfnamefont {J.}~\bibnamefont {Deschamps}},
  \bibinfo {author} {\bibfnamefont {S.~G.}\ \bibnamefont {Dutremez}}, \bibinfo
  {author} {\bibfnamefont {B.}~\bibnamefont {Boury}}, \bibinfo {author}
  {\bibfnamefont {T.}~\bibnamefont {Barisien}}, \bibinfo {author}
  {\bibfnamefont {L.}~\bibnamefont {Legrand}}, \ and\ \bibinfo {author}
  {\bibfnamefont {M.}~\bibnamefont {Schott}},\ }\bibfield  {title} {\enquote
  {\bibinfo {title} {Polymorphs and colors of polydiacetylenes: A first
  principles study},}\ }\href {http://pubs.acs.org/doi/abs/10.1021/ja803768u}
  {\bibfield  {journal} {\bibinfo  {journal} {J. Am. Chem. Soc.},\ }\textbf
  {\bibinfo {volume} {131}},\ \bibinfo {pages} {6976} (\bibinfo {year}
  {2009})}\BibitemShut {NoStop}%
\bibitem [{\citenamefont {Chappell}\ \emph {et~al.}(2008)\citenamefont
  {Chappell}, \citenamefont {Duer}, \citenamefont {Groom}, \citenamefont
  {Pickard},\ and\ \citenamefont {Bristowe}}]{gipaw_example_surface}%
  \BibitemOpen
  \bibfield  {author} {\bibinfo {author} {\bibfnamefont {H.}~\bibnamefont
  {Chappell}}, \bibinfo {author} {\bibfnamefont {M.}~\bibnamefont {Duer}},
  \bibinfo {author} {\bibfnamefont {N.}~\bibnamefont {Groom}}, \bibinfo
  {author} {\bibfnamefont {C.~J.}\ \bibnamefont {Pickard}}, \ and\ \bibinfo
  {author} {\bibfnamefont {P.}~\bibnamefont {Bristowe}},\ }\bibfield  {title}
  {\enquote {\bibinfo {title} {Probing the surface structure of hydroxyapatite
  using nmr spectroscopy and first principles calculations},}\ }\href
  {http://dx.doi.org/10.1039/B714512H} {\bibfield  {journal} {\bibinfo
  {journal} {Phys. Chem. Chem. Phys.},\ }\textbf {\bibinfo {volume} {10}},\
  \bibinfo {pages} {600} (\bibinfo {year} {2008})}\BibitemShut {NoStop}%
\bibitem [{\citenamefont {Gee}\ and\ \citenamefont
  {Raynes}(2000)}]{perturbative_shift}%
  \BibitemOpen
  \bibfield  {author} {\bibinfo {author} {\bibfnamefont {C.~H.}\ \bibnamefont
  {Gee}}\ and\ \bibinfo {author} {\bibfnamefont {W.~T.}\ \bibnamefont
  {Raynes}},\ }\bibfield  {title} {\enquote {\bibinfo {title} {Nuclear motion
  effects on the 13c, 19f and 1h shielding in methyl fluoride},}\ }\href
  {http://www.sciencedirect.com/science/article/pii/S0009261400011301}
  {\bibfield  {journal} {\bibinfo  {journal} {Chem. Phys. Lett.},\ }\textbf
  {\bibinfo {volume} {330}},\ \bibinfo {pages} {595 } (\bibinfo {year}
  {2000})}\BibitemShut {NoStop}%
\bibitem [{\citenamefont {B\"{o}hm}\ \emph {et~al.}(2000)\citenamefont
  {B\"{o}hm}, \citenamefont {Schulte},\ and\ \citenamefont
  {Ram\'{i}rez}}]{schulte_pimd_nmr_2}%
  \BibitemOpen
  \bibfield  {author} {\bibinfo {author} {\bibfnamefont {M.}~\bibnamefont
  {B\"{o}hm}}, \bibinfo {author} {\bibfnamefont {J.}~\bibnamefont {Schulte}}, \
  and\ \bibinfo {author} {\bibfnamefont {R.}~\bibnamefont {Ram\'{i}rez}},\
  }\bibfield  {title} {\enquote {\bibinfo {title} {Nuclear quantum effects in
  calculated nmr shieldings of ethylene; a feynman path integral - ab initio
  study},}\ }\href
  {http://www.ingentaconnect.com/content/els/00092614/2000/00000332/00000001/art01232}
  {\bibfield  {journal} {\bibinfo  {journal} {Chem. Phys. Lett.},\ }\textbf
  {\bibinfo {volume} {332}},\ \bibinfo {pages} {117} (\bibinfo {year}
  {2000})}\BibitemShut {NoStop}%
\bibitem [{\citenamefont {Schulte}\ \emph {et~al.}(2001)\citenamefont
  {Schulte}, \citenamefont {Ram\'{i}rez},\ and\ \citenamefont
  {B\"{o}hm}}]{schulte_pimd_nmr_1}%
  \BibitemOpen
  \bibfield  {author} {\bibinfo {author} {\bibfnamefont {J.}~\bibnamefont
  {Schulte}}, \bibinfo {author} {\bibfnamefont {R.}~\bibnamefont
  {Ram\'{i}rez}}, \ and\ \bibinfo {author} {\bibfnamefont {M.~C.}\ \bibnamefont
  {B\"{o}hm}},\ }\bibfield  {title} {\enquote {\bibinfo {title} {Nuclear
  quantum effects in calculated nmr shieldings of benzene; a feynman path
  integral study},}\ }\href
  {http://www.tandfonline.com/doi/abs/10.1080/00268970110043048} {\bibfield
  {journal} {\bibinfo  {journal} {Mol. Phys.},\ }\textbf {\bibinfo {volume}
  {99}},\ \bibinfo {pages} {1155} (\bibinfo {year} {2001})}\BibitemShut
  {NoStop}%
\bibitem [{\citenamefont {Ruden}\ \emph {et~al.}(2003)\citenamefont {Ruden},
  \citenamefont {Lutn{\ae}s}, \citenamefont {Helgaker},\ and\ \citenamefont
  {Ruud}}]{perturbative_jj}%
  \BibitemOpen
  \bibfield  {author} {\bibinfo {author} {\bibfnamefont {T.~A.}\ \bibnamefont
  {Ruden}}, \bibinfo {author} {\bibfnamefont {O.~B.}\ \bibnamefont
  {Lutn{\ae}s}}, \bibinfo {author} {\bibfnamefont {T.}~\bibnamefont
  {Helgaker}}, \ and\ \bibinfo {author} {\bibfnamefont {K.}~\bibnamefont
  {Ruud}},\ }\bibfield  {title} {\enquote {\bibinfo {title} {Vibrational
  corrections to indirect nuclear spin–spin coupling constants calculated by
  density-functional theory},}\ }\href
  {http://scitation.aip.org/content/aip/journal/jcp/118/21/10.1063/1.1569846}
  {\bibfield  {journal} {\bibinfo  {journal} {J. Chem. Phys.},\ }\textbf
  {\bibinfo {volume} {118}},\ \bibinfo {pages} {9572} (\bibinfo {year}
  {2003})}\BibitemShut {NoStop}%
\bibitem [{\citenamefont {Dra\v{c}\'{i}nsk\'{y}}\ \emph
  {et~al.}(2009)\citenamefont {Dra\v{c}\'{i}nsk\'{y}}, \citenamefont
  {Kaminsk\'{y}},\ and\ \citenamefont {Bou\v{r}}}]{perturbative_shift_jj}%
  \BibitemOpen
  \bibfield  {author} {\bibinfo {author} {\bibfnamefont {M.}~\bibnamefont
  {Dra\v{c}\'{i}nsk\'{y}}}, \bibinfo {author} {\bibfnamefont {J.}~\bibnamefont
  {Kaminsk\'{y}}}, \ and\ \bibinfo {author} {\bibfnamefont {P.}~\bibnamefont
  {Bou\v{r}}},\ }\bibfield  {title} {\enquote {\bibinfo {title} {Relative
  importance of first and second derivatives of nuclear magnetic resonance
  chemical shifts and spin-spin coupling constants for vibrational
  averaging},}\ }\href
  {http://scitation.aip.org/content/aip/journal/jcp/130/9/10.1063/1.3081317}
  {\bibfield  {journal} {\bibinfo  {journal} {J. Chem. Phys.},\ }\textbf
  {\bibinfo {volume} {130}},\ \bibinfo {pages} {094106} (\bibinfo {year}
  {2009})}\BibitemShut {NoStop}%
\bibitem [{\citenamefont {Rossano}\ \emph {et~al.}(2005)\citenamefont
  {Rossano}, \citenamefont {Mauri}, \citenamefont {Pickard},\ and\
  \citenamefont {Farnan}}]{pickard_original_nmr_vibrational}%
  \BibitemOpen
  \bibfield  {author} {\bibinfo {author} {\bibfnamefont {S.}~\bibnamefont
  {Rossano}}, \bibinfo {author} {\bibfnamefont {F.}~\bibnamefont {Mauri}},
  \bibinfo {author} {\bibfnamefont {C.~J.}\ \bibnamefont {Pickard}}, \ and\
  \bibinfo {author} {\bibfnamefont {I.}~\bibnamefont {Farnan}},\ }\bibfield
  {title} {\enquote {\bibinfo {title} {First-principles calculation of 17o and
  25mg nmr shieldings in mgo at finite temperature: Rovibrational effect in
  solids},}\ }\href {http://pubs.acs.org/doi/abs/10.1021/jp044251w} {\bibfield
  {journal} {\bibinfo  {journal} {J. Phys. Chem. B},\ }\textbf {\bibinfo
  {volume} {109}},\ \bibinfo {pages} {7245} (\bibinfo {year}
  {2005})}\BibitemShut {NoStop}%
\bibitem [{\citenamefont {Dumez}\ and\ \citenamefont
  {Pickard}(2009)}]{pickard_nmr_vibrations_organic}%
  \BibitemOpen
  \bibfield  {author} {\bibinfo {author} {\bibfnamefont {J.-N.}\ \bibnamefont
  {Dumez}}\ and\ \bibinfo {author} {\bibfnamefont {C.~J.}\ \bibnamefont
  {Pickard}},\ }\bibfield  {title} {\enquote {\bibinfo {title} {Calculation of
  nmr chemical shifts in organic solids: Accounting for motional effects},}\
  }\href
  {http://scitation.aip.org/content/aip/journal/jcp/130/10/10.1063/1.3081630}
  {\bibfield  {journal} {\bibinfo  {journal} {J. Chem. Phys.},\ }\textbf
  {\bibinfo {volume} {130}},\ \bibinfo {pages} {104701} (\bibinfo {year}
  {2009})}\BibitemShut {NoStop}%
\bibitem [{\citenamefont {Schmidt}\ and\ \citenamefont
  {Sebastiani}(2005)}]{md_quadrupolar_coupling}%
  \BibitemOpen
  \bibfield  {author} {\bibinfo {author} {\bibfnamefont {J.}~\bibnamefont
  {Schmidt}}\ and\ \bibinfo {author} {\bibfnamefont {D.}~\bibnamefont
  {Sebastiani}},\ }\bibfield  {title} {\enquote {\bibinfo {title} {Anomalous
  temperature dependence of nuclear quadrupole interactions in strongly
  hydrogen-bonded systems from first principles},}\ }\href
  {http://scitation.aip.org/content/aip/journal/jcp/123/7/10.1063/1.2000241}
  {\bibfield  {journal} {\bibinfo  {journal} {J. Chem. Phys.},\ }\textbf
  {\bibinfo {volume} {123}},\ \bibinfo {pages} {074501} (\bibinfo {year}
  {2005})}\BibitemShut {NoStop}%
\bibitem [{\citenamefont {Lee}\ \emph {et~al.}(2007)\citenamefont {Lee},
  \citenamefont {Bingöl}, \citenamefont {Murakhtina}, \citenamefont
  {Sebastiani}, \citenamefont {Meyer}, \citenamefont {Wegner},\ and\
  \citenamefont {Spiess}}]{md_nmr_lee}%
  \BibitemOpen
  \bibfield  {author} {\bibinfo {author} {\bibfnamefont {Y.~J.}\ \bibnamefont
  {Lee}}, \bibinfo {author} {\bibfnamefont {B.}~\bibnamefont {Bingöl}},
  \bibinfo {author} {\bibfnamefont {T.}~\bibnamefont {Murakhtina}}, \bibinfo
  {author} {\bibfnamefont {D.}~\bibnamefont {Sebastiani}}, \bibinfo {author}
  {\bibfnamefont {W.~H.}\ \bibnamefont {Meyer}}, \bibinfo {author}
  {\bibfnamefont {G.}~\bibnamefont {Wegner}}, \ and\ \bibinfo {author}
  {\bibfnamefont {H.~W.}\ \bibnamefont {Spiess}},\ }\bibfield  {title}
  {\enquote {\bibinfo {title} {High-resolution solid-state nmr studies of
  poly(vinyl phosphonic acid) proton-conducting polymer:  molecular structure
  and proton dynamics},}\ }\href
  {http://pubs.acs.org/doi/abs/10.1021/jp072112j} {\bibfield  {journal}
  {\bibinfo  {journal} {J. Phys. Chem. B},\ }\textbf {\bibinfo {volume}
  {111}},\ \bibinfo {pages} {9711} (\bibinfo {year} {2007})}\BibitemShut
  {NoStop}%
\bibitem [{\citenamefont {Robinson}\ and\ \citenamefont
  {Haynes}(2010)}]{nmr_force_fields}%
  \BibitemOpen
  \bibfield  {author} {\bibinfo {author} {\bibfnamefont {M.}~\bibnamefont
  {Robinson}}\ and\ \bibinfo {author} {\bibfnamefont {P.~D.}\ \bibnamefont
  {Haynes}},\ }\bibfield  {title} {\enquote {\bibinfo {title} {Dynamical
  effects in ab initio nmr calculations: Classical force fields fitted to
  quantum forces},}\ }\href
  {http://scitation.aip.org/content/aip/journal/jcp/133/8/10.1063/1.3474573}
  {\bibfield  {journal} {\bibinfo  {journal} {J. Chem. Phys.},\ }\textbf
  {\bibinfo {volume} {133}},\ \bibinfo {eid} {084109} (\bibinfo {year}
  {2010})}\BibitemShut {NoStop}%
\bibitem [{\citenamefont {Gortari}\ \emph {et~al.}(2010)\citenamefont
  {Gortari}, \citenamefont {Portella}, \citenamefont {Salvatella},
  \citenamefont {Bajaj}, \citenamefont {van~der Wel}, \citenamefont {Yates},
  \citenamefont {Segall}, \citenamefont {Pickard}, \citenamefont {Payne},\ and\
  \citenamefont {Vendruscolo}}]{long_time_md}%
  \BibitemOpen
  \bibfield  {author} {\bibinfo {author} {\bibfnamefont {I.~D.}\ \bibnamefont
  {Gortari}}, \bibinfo {author} {\bibfnamefont {G.}~\bibnamefont {Portella}},
  \bibinfo {author} {\bibfnamefont {X.}~\bibnamefont {Salvatella}}, \bibinfo
  {author} {\bibfnamefont {V.~S.}\ \bibnamefont {Bajaj}}, \bibinfo {author}
  {\bibfnamefont {P.~C.~A.}\ \bibnamefont {van~der Wel}}, \bibinfo {author}
  {\bibfnamefont {J.~R.}\ \bibnamefont {Yates}}, \bibinfo {author}
  {\bibfnamefont {M.~D.}\ \bibnamefont {Segall}}, \bibinfo {author}
  {\bibfnamefont {C.~J.}\ \bibnamefont {Pickard}}, \bibinfo {author}
  {\bibfnamefont {M.~C.}\ \bibnamefont {Payne}}, \ and\ \bibinfo {author}
  {\bibfnamefont {M.}~\bibnamefont {Vendruscolo}},\ }\bibfield  {title}
  {\enquote {\bibinfo {title} {Time averaging of nmr chemical shifts in the mlf
  peptide in the solid state},}\ }\href
  {http://pubs.acs.org/doi/abs/10.1021/ja9062629} {\bibfield  {journal}
  {\bibinfo  {journal} {J. Am. Chem. Soc.},\ }\textbf {\bibinfo {volume}
  {132}},\ \bibinfo {pages} {5993} (\bibinfo {year} {2010})}\BibitemShut
  {NoStop}%
\bibitem [{\citenamefont {Dra\v{c}\'{i}nsk\'{y}}\ and\ \citenamefont
  {Hodgkinson}(2013)}]{dracinsky_md_nmr}%
  \BibitemOpen
  \bibfield  {author} {\bibinfo {author} {\bibfnamefont {M.}~\bibnamefont
  {Dra\v{c}\'{i}nsk\'{y}}}\ and\ \bibinfo {author} {\bibfnamefont
  {P.}~\bibnamefont {Hodgkinson}},\ }\bibfield  {title} {\enquote {\bibinfo
  {title} {A molecular dynamics study of the effects of fast molecular motions
  on solid-state nmr parameters},}\ }\href
  {http://dx.doi.org/10.1039/C3CE40612A} {\bibfield  {journal} {\bibinfo
  {journal} {Cryst. Eng. Comm.},\ }\textbf {\bibinfo {volume} {15}},\ \bibinfo
  {pages} {8705} (\bibinfo {year} {2013})}\BibitemShut {NoStop}%
\bibitem [{\citenamefont {Dra\v{c}\'{i}nsk\'{y}}\ and\ \citenamefont
  {Hodgkinson}(2014)}]{dracinsky_pimd_nmr}%
  \BibitemOpen
  \bibfield  {author} {\bibinfo {author} {\bibfnamefont {M.}~\bibnamefont
  {Dra\v{c}\'{i}nsk\'{y}}}\ and\ \bibinfo {author} {\bibfnamefont
  {P.}~\bibnamefont {Hodgkinson}},\ }\bibfield  {title} {\enquote {\bibinfo
  {title} {Effects of quantum nuclear delocalisation on nmr parameters from
  path integral molecular dynamics},}\ }\href
  {http://dx.doi.org/10.1002/chem.201303496} {\bibfield  {journal} {\bibinfo
  {journal} {Chem. Eur. J.},\ }\textbf {\bibinfo {volume} {20}},\ \bibinfo
  {pages} {2201} (\bibinfo {year} {2014})},\ ISSN \bibinfo {issn}
  {1521-3765}\BibitemShut {NoStop}%
\bibitem [{\citenamefont {Dra\v{c}\'{i}nsk\'{y}}\ and\ \citenamefont
  {Bou\v{r}}(2012)}]{dracinsky_md_quadratic_nmr}%
  \BibitemOpen
  \bibfield  {author} {\bibinfo {author} {\bibfnamefont {M.}~\bibnamefont
  {Dra\v{c}\'{i}nsk\'{y}}}\ and\ \bibinfo {author} {\bibfnamefont
  {P.}~\bibnamefont {Bou\v{r}}},\ }\bibfield  {title} {\enquote {\bibinfo
  {title} {Vibrational averaging of the chemical shift in crystalline
  $\alpha$-glycine},}\ }\href {http://dx.doi.org/10.1002/jcc.22940} {\bibfield
  {journal} {\bibinfo  {journal} {J. Comp. Chem.},\ }\textbf {\bibinfo {volume}
  {33}},\ \bibinfo {pages} {1080} (\bibinfo {year} {2012})},\ ISSN \bibinfo
  {issn} {1096-987X}\BibitemShut {NoStop}%
\bibitem [{\citenamefont {Monserrat}\ \emph {et~al.}(2013)\citenamefont
  {Monserrat}, \citenamefont {Drummond},\ and\ \citenamefont
  {Needs}}]{PhysRevB.87.144302}%
  \BibitemOpen
  \bibfield  {author} {\bibinfo {author} {\bibfnamefont {B.}~\bibnamefont
  {Monserrat}}, \bibinfo {author} {\bibfnamefont {N.~D.}\ \bibnamefont
  {Drummond}}, \ and\ \bibinfo {author} {\bibfnamefont {R.~J.}\ \bibnamefont
  {Needs}},\ }\bibfield  {title} {\enquote {\bibinfo {title} {Anharmonic
  vibrational properties in periodic systems: energy, electron-phonon coupling,
  and stress},}\ }\href {http://link.aps.org/doi/10.1103/PhysRevB.87.144302}
  {\bibfield  {journal} {\bibinfo  {journal} {Phys. Rev. B},\ }\textbf
  {\bibinfo {volume} {87}},\ \bibinfo {pages} {144302} (\bibinfo {year}
  {2013})}\BibitemShut {NoStop}%
\bibitem [{\citenamefont {Monserrat}\ \emph {et~al.}(2014)\citenamefont
  {Monserrat}, \citenamefont {Drummond}, \citenamefont {Pickard},\ and\
  \citenamefont {Needs}}]{helium}%
  \BibitemOpen
  \bibfield  {author} {\bibinfo {author} {\bibfnamefont {B.}~\bibnamefont
  {Monserrat}}, \bibinfo {author} {\bibfnamefont {N.~D.}\ \bibnamefont
  {Drummond}}, \bibinfo {author} {\bibfnamefont {C.~J.}\ \bibnamefont
  {Pickard}}, \ and\ \bibinfo {author} {\bibfnamefont {R.~J.}\ \bibnamefont
  {Needs}},\ }\bibfield  {title} {\enquote {\bibinfo {title} {Electron-phonon
  coupling and the metallization of solid helium at terapascal pressures},}\
  }\href {http://link.aps.org/doi/10.1103/PhysRevLett.112.055504} {\bibfield
  {journal} {\bibinfo  {journal} {Phys. Rev. Lett.},\ }\textbf {\bibinfo
  {volume} {112}},\ \bibinfo {pages} {055504} (\bibinfo {year}
  {2014})}\BibitemShut {NoStop}%
\bibitem [{\citenamefont {Azadi}\ \emph {et~al.}(2014)\citenamefont {Azadi},
  \citenamefont {Monserrat}, \citenamefont {Foulkes},\ and\ \citenamefont
  {Needs}}]{prl_dissociation_hydrogen}%
  \BibitemOpen
  \bibfield  {author} {\bibinfo {author} {\bibfnamefont {S.}~\bibnamefont
  {Azadi}}, \bibinfo {author} {\bibfnamefont {B.}~\bibnamefont {Monserrat}},
  \bibinfo {author} {\bibfnamefont {W.~M.~C.}\ \bibnamefont {Foulkes}}, \ and\
  \bibinfo {author} {\bibfnamefont {R.~J.}\ \bibnamefont {Needs}},\ }\bibfield
  {title} {\enquote {\bibinfo {title} {Dissociation of high-pressure solid
  molecular hydrogen: A quantum monte~carlo and anharmonic vibrational
  study},}\ }\href {http://link.aps.org/doi/10.1103/PhysRevLett.112.165501}
  {\bibfield  {journal} {\bibinfo  {journal} {Phys. Rev. Lett.},\ }\textbf
  {\bibinfo {volume} {112}},\ \bibinfo {pages} {165501} (\bibinfo {year}
  {2014})}\BibitemShut {NoStop}%
\bibitem [{\citenamefont {Clark}\ \emph {et~al.}(2005)\citenamefont {Clark},
  \citenamefont {Segall}, \citenamefont {Pickard}, \citenamefont {Hasnip},
  \citenamefont {Probert}, \citenamefont {Refson},\ and\ \citenamefont
  {Payne}}]{CASTEP}%
  \BibitemOpen
  \bibfield  {author} {\bibinfo {author} {\bibfnamefont {S.~J.}\ \bibnamefont
  {Clark}}, \bibinfo {author} {\bibfnamefont {M.~D.}\ \bibnamefont {Segall}},
  \bibinfo {author} {\bibfnamefont {C.~J.}\ \bibnamefont {Pickard}}, \bibinfo
  {author} {\bibfnamefont {P.~J.}\ \bibnamefont {Hasnip}}, \bibinfo {author}
  {\bibfnamefont {M.~I.~J.}\ \bibnamefont {Probert}}, \bibinfo {author}
  {\bibfnamefont {K.}~\bibnamefont {Refson}}, \ and\ \bibinfo {author}
  {\bibfnamefont {M.~C.}\ \bibnamefont {Payne}},\ }\bibfield  {title} {\enquote
  {\bibinfo {title} {First principles methods using castep},}\ }\href
  {http://www.oldenbourg-link.com/doi/abs/10.1524/zkri.220.5.567.65075}
  {\bibfield  {journal} {\bibinfo  {journal} {Z. Kristallogr.},\ }\textbf
  {\bibinfo {volume} {220}},\ \bibinfo {pages} {567} (\bibinfo {year}
  {2005})}\BibitemShut {NoStop}%
\bibitem [{\citenamefont {Perdew}\ \emph {et~al.}(1996)\citenamefont {Perdew},
  \citenamefont {Burke},\ and\ \citenamefont
  {Ernzerhof}}]{PhysRevLett.77.3865}%
  \BibitemOpen
  \bibfield  {author} {\bibinfo {author} {\bibfnamefont {J.~P.}\ \bibnamefont
  {Perdew}}, \bibinfo {author} {\bibfnamefont {K.}~\bibnamefont {Burke}}, \
  and\ \bibinfo {author} {\bibfnamefont {M.}~\bibnamefont {Ernzerhof}},\
  }\bibfield  {title} {\enquote {\bibinfo {title} {Generalized gradient
  approximation made simple},}\ }\href
  {http://link.aps.org/doi/10.1103/PhysRevLett.77.3865} {\bibfield  {journal}
  {\bibinfo  {journal} {Phys. Rev. Lett.},\ }\textbf {\bibinfo {volume} {77}},\
  \bibinfo {pages} {3865} (\bibinfo {year} {1996})}\BibitemShut {NoStop}%
\bibitem [{\citenamefont {Vanderbilt}(1990)}]{PhysRevB.41.7892}%
  \BibitemOpen
  \bibfield  {author} {\bibinfo {author} {\bibfnamefont {D.}~\bibnamefont
  {Vanderbilt}},\ }\bibfield  {title} {\enquote {\bibinfo {title} {Soft
  self-consistent pseudopotentials in a generalized eigenvalue formalism},}\
  }\href {http://link.aps.org/doi/10.1103/PhysRevB.41.7892} {\bibfield
  {journal} {\bibinfo  {journal} {Phys. Rev. B},\ }\textbf {\bibinfo {volume}
  {41}},\ \bibinfo {pages} {7892} (\bibinfo {year} {1990})}\BibitemShut
  {NoStop}%
\bibitem [{\citenamefont {Monkhorst}\ and\ \citenamefont
  {Pack}(1976)}]{PhysRevB.13.5188}%
  \BibitemOpen
  \bibfield  {author} {\bibinfo {author} {\bibfnamefont {H.~J.}\ \bibnamefont
  {Monkhorst}}\ and\ \bibinfo {author} {\bibfnamefont {J.~D.}\ \bibnamefont
  {Pack}},\ }\bibfield  {title} {\enquote {\bibinfo {title} {Special points for
  brillouin-zone integrations},}\ }\href
  {http://link.aps.org/doi/10.1103/PhysRevB.13.5188} {\bibfield  {journal}
  {\bibinfo  {journal} {Phys. Rev. B},\ }\textbf {\bibinfo {volume} {13}},\
  \bibinfo {pages} {5188} (\bibinfo {year} {1976})}\BibitemShut {NoStop}%
\bibitem [{\citenamefont {Wilson}\ \emph {et~al.}(2005)\citenamefont {Wilson},
  \citenamefont {Myles}, \citenamefont {Ghosh}, \citenamefont {Johnson},\ and\
  \citenamefont {Wang}}]{alanine_crystal}%
  \BibitemOpen
  \bibfield  {author} {\bibinfo {author} {\bibfnamefont {C.~C.}\ \bibnamefont
  {Wilson}}, \bibinfo {author} {\bibfnamefont {D.}~\bibnamefont {Myles}},
  \bibinfo {author} {\bibfnamefont {M.}~\bibnamefont {Ghosh}}, \bibinfo
  {author} {\bibfnamefont {L.~N.}\ \bibnamefont {Johnson}}, \ and\ \bibinfo
  {author} {\bibfnamefont {W.}~\bibnamefont {Wang}},\ }\bibfield  {title}
  {\enquote {\bibinfo {title} {Neutron diffraction investigations of l- and
  d-alanine at different temperatures: the search for structural evidence for
  parity violation},}\ }\href {http://dx.doi.org/10.1039/B419295H} {\bibfield
  {journal} {\bibinfo  {journal} {New J. Chem.},\ }\textbf {\bibinfo {volume}
  {29}},\ \bibinfo {pages} {1318} (\bibinfo {year} {2005})}\BibitemShut
  {NoStop}%
\bibitem [{\citenamefont {G\"{o}rbitz}(1987)}]{bDA_crystal}%
  \BibitemOpen
  \bibfield  {author} {\bibinfo {author} {\bibfnamefont {C.~H.}\ \bibnamefont
  {G\"{o}rbitz}},\ }\bibfield  {title} {\enquote {\bibinfo {title} {Crystal and
  molecular structures of the isomeric dipeptides $\alpha$-l-aspartyl-l-alanine
  and $\beta$-l-aspartyl-l-alanine},}\ }\href@noop {} {\bibfield  {journal}
  {\bibinfo  {journal} {Acta Chem. Scand. B},\ }\textbf {\bibinfo {volume}
  {41}},\ \bibinfo {pages} {679} (\bibinfo {year} {1987})}\BibitemShut
  {NoStop}%
\bibitem [{\citenamefont {Kunc}\ and\ \citenamefont
  {Martin}(1982)}]{phonon_finite_displacement}%
  \BibitemOpen
  \bibfield  {author} {\bibinfo {author} {\bibfnamefont {K.}~\bibnamefont
  {Kunc}}\ and\ \bibinfo {author} {\bibfnamefont {R.~M.}\ \bibnamefont
  {Martin}},\ }\bibfield  {title} {\enquote {\bibinfo {title} {\textit{Ab
  Initio} force constants of gaas: A new approach to calculation of phonons and
  dielectric properties},}\ }\href
  {http://link.aps.org/doi/10.1103/PhysRevLett.48.406} {\bibfield  {journal}
  {\bibinfo  {journal} {Phys. Rev. Lett.},\ }\textbf {\bibinfo {volume} {48}},\
  \bibinfo {pages} {406} (\bibinfo {year} {1982})}\BibitemShut {NoStop}%
\bibitem [{Note1()}]{Note1}%
  \BibitemOpen
  \bibinfo {note} {This is true only if the variance of the function being
  sampled does not change when increasing the dimension of the
  integral.}\BibitemShut {Stop}%
\bibitem [{\citenamefont {Naito}\ \emph {et~al.}(1981)\citenamefont {Naito},
  \citenamefont {Ganapathy}, \citenamefont {Akasaka},\ and\ \citenamefont
  {McDowell}}]{alanine_nmr_experiment}%
  \BibitemOpen
  \bibfield  {author} {\bibinfo {author} {\bibfnamefont {A.}~\bibnamefont
  {Naito}}, \bibinfo {author} {\bibfnamefont {S.}~\bibnamefont {Ganapathy}},
  \bibinfo {author} {\bibfnamefont {K.}~\bibnamefont {Akasaka}}, \ and\
  \bibinfo {author} {\bibfnamefont {C.~A.}\ \bibnamefont {McDowell}},\
  }\bibfield  {title} {\enquote {\bibinfo {title} {Chemical shielding tensor
  and 13c–14n dipolar splitting in single crystals of l‐alanine},}\ }\href
  {http://scitation.aip.org/content/aip/journal/jcp/74/6/10.1063/1.441513}
  {\bibfield  {journal} {\bibinfo  {journal} {J. Chem. Phys.},\ }\textbf
  {\bibinfo {volume} {74}},\ \bibinfo {pages} {3190} (\bibinfo {year}
  {1981})}\BibitemShut {NoStop}%
\bibitem [{\citenamefont {Pickard}\ \emph {et~al.}(2007)\citenamefont
  {Pickard}, \citenamefont {Salager}, \citenamefont {Pintacuda}, \citenamefont
  {Elena},\ and\ \citenamefont {Emsley}}]{bDA_nmr_experiment}%
  \BibitemOpen
  \bibfield  {author} {\bibinfo {author} {\bibfnamefont {C.~J.}\ \bibnamefont
  {Pickard}}, \bibinfo {author} {\bibfnamefont {E.}~\bibnamefont {Salager}},
  \bibinfo {author} {\bibfnamefont {G.}~\bibnamefont {Pintacuda}}, \bibinfo
  {author} {\bibfnamefont {B.}~\bibnamefont {Elena}}, \ and\ \bibinfo {author}
  {\bibfnamefont {L.}~\bibnamefont {Emsley}},\ }\bibfield  {title} {\enquote
  {\bibinfo {title} {Resolving structures from powders by nmr crystallography
  using combined proton spin diffusion and plane wave dft calculations},}\
  }\href {http://pubs.acs.org/doi/abs/10.1021/ja071829h} {\bibfield  {journal}
  {\bibinfo  {journal} {J. Am. Chem. Soc.},\ }\textbf {\bibinfo {volume}
  {129}},\ \bibinfo {pages} {8932} (\bibinfo {year} {2007})}\BibitemShut
  {NoStop}%
\bibitem [{\citenamefont {Souvatzis}\ \emph {et~al.}(2008)\citenamefont
  {Souvatzis}, \citenamefont {Eriksson}, \citenamefont {Katsnelson},\ and\
  \citenamefont {Rudin}}]{PhysRevLett.100.095901}%
  \BibitemOpen
  \bibfield  {author} {\bibinfo {author} {\bibfnamefont {P.}~\bibnamefont
  {Souvatzis}}, \bibinfo {author} {\bibfnamefont {O.}~\bibnamefont {Eriksson}},
  \bibinfo {author} {\bibfnamefont {M.~I.}\ \bibnamefont {Katsnelson}}, \ and\
  \bibinfo {author} {\bibfnamefont {S.~P.}\ \bibnamefont {Rudin}},\ }\bibfield
  {title} {\enquote {\bibinfo {title} {Entropy driven stabilization of
  energetically unstable crystal structures explained from first principles
  theory},}\ }\href {http://link.aps.org/doi/10.1103/PhysRevLett.100.095901}
  {\bibfield  {journal} {\bibinfo  {journal} {Phys. Rev. Lett.},\ }\textbf
  {\bibinfo {volume} {100}},\ \bibinfo {pages} {095901} (\bibinfo {year}
  {2008})}\BibitemShut {NoStop}%
\bibitem [{\citenamefont {Errea}\ \emph {et~al.}(2011)\citenamefont {Errea},
  \citenamefont {Rousseau},\ and\ \citenamefont
  {Bergara}}]{PhysRevLett.106.165501}%
  \BibitemOpen
  \bibfield  {author} {\bibinfo {author} {\bibfnamefont {I.}~\bibnamefont
  {Errea}}, \bibinfo {author} {\bibfnamefont {B.}~\bibnamefont {Rousseau}}, \
  and\ \bibinfo {author} {\bibfnamefont {A.}~\bibnamefont {Bergara}},\
  }\bibfield  {title} {\enquote {\bibinfo {title} {Anharmonic stabilization of
  the high-pressure simple cubic phase of calcium},}\ }\href
  {http://link.aps.org/doi/10.1103/PhysRevLett.106.165501} {\bibfield
  {journal} {\bibinfo  {journal} {Phys. Rev. Lett.},\ }\textbf {\bibinfo
  {volume} {106}},\ \bibinfo {pages} {165501} (\bibinfo {year}
  {2011})}\BibitemShut {NoStop}%
\bibitem [{\citenamefont {Hellman}\ \emph {et~al.}(2011)\citenamefont
  {Hellman}, \citenamefont {Abrikosov},\ and\ \citenamefont
  {Simak}}]{PhysRevB.84.180301}%
  \BibitemOpen
  \bibfield  {author} {\bibinfo {author} {\bibfnamefont {O.}~\bibnamefont
  {Hellman}}, \bibinfo {author} {\bibfnamefont {I.~A.}\ \bibnamefont
  {Abrikosov}}, \ and\ \bibinfo {author} {\bibfnamefont {S.~I.}\ \bibnamefont
  {Simak}},\ }\bibfield  {title} {\enquote {\bibinfo {title} {Lattice dynamics
  of anharmonic solids from first principles},}\ }\href
  {http://link.aps.org/doi/10.1103/PhysRevB.84.180301} {\bibfield  {journal}
  {\bibinfo  {journal} {Phys. Rev. B},\ }\textbf {\bibinfo {volume} {84}},\
  \bibinfo {pages} {180301} (\bibinfo {year} {2011})}\BibitemShut {NoStop}%
\bibitem [{\citenamefont {Antolin}\ \emph {et~al.}(2012)\citenamefont
  {Antolin}, \citenamefont {Restrepo},\ and\ \citenamefont
  {Windl}}]{PhysRevB.86.054119}%
  \BibitemOpen
  \bibfield  {author} {\bibinfo {author} {\bibfnamefont {N.}~\bibnamefont
  {Antolin}}, \bibinfo {author} {\bibfnamefont {O.~D.}\ \bibnamefont
  {Restrepo}}, \ and\ \bibinfo {author} {\bibfnamefont {W.}~\bibnamefont
  {Windl}},\ }\bibfield  {title} {\enquote {\bibinfo {title} {Fast free-energy
  calculations for unstable high-temperature phases},}\ }\href
  {http://link.aps.org/doi/10.1103/PhysRevB.86.054119} {\bibfield  {journal}
  {\bibinfo  {journal} {Phys. Rev. B},\ }\textbf {\bibinfo {volume} {86}},\
  \bibinfo {pages} {054119} (\bibinfo {year} {2012})}\BibitemShut {NoStop}%
\bibitem [{\citenamefont {Errea}\ \emph {et~al.}(2013)\citenamefont {Errea},
  \citenamefont {Calandra},\ and\ \citenamefont {Mauri}}]{errea_prl}%
  \BibitemOpen
  \bibfield  {author} {\bibinfo {author} {\bibfnamefont {I.}~\bibnamefont
  {Errea}}, \bibinfo {author} {\bibfnamefont {M.}~\bibnamefont {Calandra}}, \
  and\ \bibinfo {author} {\bibfnamefont {F.}~\bibnamefont {Mauri}},\ }\bibfield
   {title} {\enquote {\bibinfo {title} {First-principles theory of
  anharmonicity and the inverse isotope effect in superconducting
  palladium-hydride compounds},}\ }\href
  {http://link.aps.org/doi/10.1103/PhysRevLett.111.177002} {\bibfield
  {journal} {\bibinfo  {journal} {Phys. Rev. Lett.},\ }\textbf {\bibinfo
  {volume} {111}},\ \bibinfo {pages} {177002} (\bibinfo {year}
  {2013})}\BibitemShut {NoStop}%
\bibitem [{\citenamefont {Errea}\ \emph {et~al.}(2014)\citenamefont {Errea},
  \citenamefont {Calandra},\ and\ \citenamefont {Mauri}}]{errea_prb}%
  \BibitemOpen
  \bibfield  {author} {\bibinfo {author} {\bibfnamefont {I.}~\bibnamefont
  {Errea}}, \bibinfo {author} {\bibfnamefont {M.}~\bibnamefont {Calandra}}, \
  and\ \bibinfo {author} {\bibfnamefont {F.}~\bibnamefont {Mauri}},\ }\bibfield
   {title} {\enquote {\bibinfo {title} {Anharmonic free energies and phonon
  dispersions from the stochastic self-consistent harmonic approximation:
  Application to platinum and palladium hydrides},}\ }\href
  {http://link.aps.org/doi/10.1103/PhysRevB.89.064302} {\bibfield  {journal}
  {\bibinfo  {journal} {Phys. Rev. B},\ }\textbf {\bibinfo {volume} {89}},\
  \bibinfo {pages} {064302} (\bibinfo {year} {2014})}\BibitemShut {NoStop}%
\end{thebibliography}%


\end{document}